\documentclass[12pt]{article}
\usepackage{amsmath,amssymb,epsfig,amsfonts}
\usepackage{graphicx,floatflt,subfigure}
\usepackage{cite}

\makeatletter

\newcommand{\cb}{c_{1,B}}
\newcommand{\cs}{c_{1,S}}
\newcommand{\bh}{\hat{b}}
\newcommand{\ah}{\hat{a}}
\newcommand{\eh}{\hat{e}}
\newcommand{\dhat}{\hat{d}}
\newcommand{\hh}{\hat{h}}
\newcommand{\ch}{\hat{c}}
\newcommand{\alphah}{\hat{\alpha}}
\newcommand{\cloc}{ {\cal{C}}_{\text{Higgs,loc}} }
\newcommand{\ploc}{ {p_{\text{loc}}} }

\newcommand{\ctate}{ {\cal{C}}_{\text{Tate}} }
\newcommand{\tencurve}{ \Sigma_{\mathbf{10},\uparrow}^{(4)} }
\newcommand{\fiveonecurve} { \Sigma_{\mathbf{\overline{5}},\uparrow}^{(\bar{4}1)} }
\newcommand{\fivefourcurve}{ \Sigma_{\mathbf{\overline{5}},\uparrow}^{(44)} }
\newcommand{\zenriques}{\mathbb{Z}_2^{(\text{Enriques})}}
\newcommand{\ztau}{\mathbb{Z}_2^{(\tau)}}

\def\unit{{1\kern-.65ex {\rm l}}}
\def\1{{1\kern-.65ex {\rm l}}}

\def\CA{{\cal A}}
\def\CB{{\cal B}}

\def\CD{{\cal D}}
\def\CE{{\cal E}}

\def\CG{{\cal G}}

\def\CI{{\cal I}}

\def\CL{{\cal L}}

\def\CO{{\cal O}}

\def\CS{{\cal S}}

\newcount\hour \newcount\minute
\hour=\time \divide \hour by 60
\minute=\time
\def\now{%
\ifnum \hour<13
  \ifnum \hour=0 \advance \hour by 12 \number\hour:\else \number\hour:\fi%
     \ifnum \minute<10 0\fi%
     \number\minute%
\ A.M.%
\else \advance \hour by -12 \number\hour:%
  \ifnum \minute<10 0\fi%
  \number\minute%
  \ P.M.%
\fi%
}

\makeatother

\begin{document}

\baselineskip=18pt  
\numberwithin{equation}{section}  
\allowdisplaybreaks  



%
%


\thispagestyle{empty}

\vspace*{-2cm}
\begin{flushright}
{EFI-12-14$\quad$OHSTPY-HEP-T-12-002}
\end{flushright}

\vspace*{0.8cm}
\begin{center}
 {\LARGE A Global $SU(5)$ F-theory model with Wilson line breaking\\}
 \vspace*{1.5cm}
J.~Marsano $^1$, H.~Clemens$^2$,  T.~Pantev $^3$, S.~Raby$^4$, H-H.~Tseng $^2$ \\
\vspace*{1.0cm}
$^1$  Enrico Fermi Institute, University of Chicago,\\
 5640 S Ellis Ave, Chicago, IL 60637, USA \\
$^2$ Department of Mathematics, The Ohio State University, \\
100 Math Tower, 231 West 18th Avenue, Columbus, OH 43210  \\
$^3$ Department of Mathematics, University of Pennsylvania,\\
209 S 33rd Street, Philadelphia, PA 19104, USA \\
$^4$ Department of Physics, The Ohio State University, \\
191 W. Woodruff Ave,  Columbus, OH 43210 \\
 \vspace*{0.8cm}
\end{center}
\vspace*{.5cm}

\noindent
We engineer compact $SU(5)$ Grand Unified Theories in F-theory in which GUT-breaking is achieved by a discrete Wilson line.
Because the internal gauge field is flat, these models avoid the high scale threshold corrections associated with hypercharge flux.
Along the  way, we exemplify the `local-to-global' approach in F-theory model building and demonstrate how the Tate divisor formalism can be used to address several challenges of extending local models to global ones.  These include in particular the construction of $G$-fluxes that extend non-inherited bundles and the engineering of $U(1)$ symmetries.
We go beyond chirality computations and determine the precise (charged) massless spectrum, finding exactly three families of quarks and leptons but excessive doublet and/or triplet pairs in the Higgs sector (depending on the example) and vector-like exotics descending from the adjoint of $SU(5)_{\rm GUT}$.  Understanding why vector-like pairs persist in the Higgs sector without an obvious symmetry to protect them may shed light on new solutions to the $\mu$ problem in F-theory GUTs.

\newpage
\setcounter{page}{1} 



\tableofcontents


\section{Introduction}
Supersymmetric grand unified theories \cite{Dimopoulos:1981yj,Dimopoulos:1981zb,Ibanez:1981yh} 
have many desirable features, including
(1) addressing the gauge hierarchy problem, i.e.  why  $M_Z \ll M_G$ is ``natural,"
(2)  explaining charge quantization and families;
(3)  predicting the unification of gauge couplings;
(4)  predicting Yukawa coupling unification,
(5)  and with broken family symmetries easily accommodating the fermion mass hierarchy.
(6)  They are consistent with the observed neutrino masses via the See-Saw mechanism;
(7)  provide a ``natural" dark matter candidate;
(8)  and enable baryogenesis via leptogenesis.
A SUSY desert with no new physics between the weak and GUT scales puts experiments at the LHC in the enviable position
of probing physics at the highest scales in Nature.  Finally, SUSY GUTs  are a  ``natural extension of the Standard Model."

That said, SUSY GUTs in four dimensions have two serious problems - (1) Higgs doublet - triplet splitting and (2) GUT symmetry breaking.
Although both of these problems have four dimensional solutions requiring large GUT representations and extensive symmetry breaking sectors,
a very elegant solution is found in the context of GUTs in higher dimensions.  They are broken to the Standard Model in four dimensions via
non-trivial GUT boundary conditions on orbifolds,  non-trivial gauge field strengths and/or Wilson lines.  In particular,
such theories can naturally be found in the string landscape and they retain many, or all, of the desirable features of SUSY GUTs mentioned above.  In fact,  it has been shown that by demanding SUSY GUTs in string constructions one can find many models with features much like that of the minimal supersymmetric Standard Model [MSSM] \cite{Lebedev:2006kn,Lebedev:2007hv,Kim:2007mt,Lebedev:2008un,Blumenhagen:2008zz,Anderson:2011ns,Anderson:2012yf}.

The past several years have seen significant attention devoted to the study of supersymmetric GUTs in F-theory \cite{Donagi:2008ca,Beasley:2008dc,Hayashi:2008ba,Beasley:2008kw,Donagi:2008kj}, which is a framework for studying nonperturbative configurations of $(p,q)$ 7-branes in type IIB strings.  F-theory compactifications are described by specifying an elliptically fibered Calabi-Yau 4-fold, $Y_4$, with section along with a suitable choice of $G$-flux.  The base $B_3$ of $Y_4$ corresponds to the physical IIB compactification space while the elliptic fibration determines the configuration of the axio-dilaton which, in turn, encodes the locations of the $(p,q)$ 7-branes.  The GUT sector of F-theory models is realized on a single stack of 7-branes that wrap a complex surface $S_2$ inside $B_3$.  In this way, essential GUT physics is captured by an 8-dimensional gauge theory on the brane worldvolume that provides an explicit realization of a higher dimensional GUT.  The stringy origin of this GUT adds a very nice feature: the question of UV completing to gravity is in principle well-defined.  As the GUT model is determined by the local behavior of an F-theory geometry and a certain type of flux on that geometry, the task of adding gravity amounts to nothing more than embedding these structures into a globally consistent F-theory compactification.  This suggests a `local-to-global' approach to the study of F-theory models that we pursue throughout this paper.

\subsection{Local-to-Global Model Building}

Unlike the complicated space of compact Calabi-Yau 4-folds, the local geometries that control most GUT physics are well-understood and exhibit a rigid structure that places tight constraints on model-building.  By studying this structure, we can ascertain our ability to solve phenomenological problems and understand the mechanisms available for doing so in F-theory.  In a sense, then, local F-theory models provide a restrictive class of effective field theories that serve as a starting point for building concrete string constructions.

Theoretically, the task of embedding `local models' into global completions is highly nontrivial.   The local model is defined in part by specifying a Higgs bundle, which consists of nontrivial vacuum solutions for an adjoint scalar and gauge field along the internal brane worldvolume $S_2$ \cite{Donagi:2009ra,Hayashi:2009ge,Tatar:2009jk}.  To embed this into a global completion, one must identify the local geometric data that determine the Higgs bundle on the brane and understand how that local data can be extended to appropriate global quantities.  The scalar part of the Higgs bundle is determined by complex structure moduli of the local geometry which are easy to understand \cite{Donagi:2008ca,Donagi:2009ra}.  The gauge field, on the other hand, is specified by a local `$G$-flux' that can be trickier.  Techniques for dealing with this exist in the context of Heterotic/F-theory duality \cite{Curio:1998bva,Donagi:2008ca} and have recently been adapted to more general settings through the `Tate divisor' formalism \cite{Marsano:2010ix,Marsano:2011nn,Marsano:2011hv}.  Recent related work on this formalism in $SO(10)$ and $E_6$ models includes \cite{Kuntzler:2012bu,Tatar:2012tm} Using the Tate divisor as a bridge, we can build global completions of local models that extend both the geometric and flux data thereby realizing the `bottom-up' approach to string model building in a very explicit way.

\subsection{Summary}

In this paper, we provide a complete demonstration of the `local-to-global' paradigm for F-theory model-building by starting with a restrictive set of requirements on a set of local models and applying the Tate divisor approach to construct global completions from the ground up that extend both the geometric data and the flux data.  
We intend our treatment to be self-contained in the sense that almost all aspects of local and global model building are worked out in complete detail, from the requisite structures of local models and their ability to be tuned, to the global extension of those structures and explicit computations of global quantities.  Our starting point is a set of local models that are chosen to implement specific solutions to the phenomenological problems of GUT-breaking and proton decay.  A number of compact geometries for global models have been constructed in recent years that use different methods than we do \cite{Andreas:2009uf,Marsano:2009ym,Collinucci:2009uh,Blumenhagen:2009up,Marsano:2009gv,Blumenhagen:2009yv,Marsano:2009wr,Grimm:2009yu,Cvetic:2010rq,Chen:2010ts,Chen:2010tp,Chung:2010bn,Chen:2010tg,Knapp:2011wk,Knapp:2011ip} with varying level of detail concerning the subtle issues of $U(1)$'s \cite{Hayashi:2010zp,Grimm:2010ez,Marsano:2010ix,Grimm:2011tb,Marsano:2011nn,Marsano:2011hv,Krause:2011xj} and fluxes \cite{Marsano:2010ix,Collinucci:2010gz,Grimm:2011tb,Braun:2011zm,Marsano:2011nn,Marsano:2011hv,Krause:2011xj,Grimm:2011sk,Grimm:2011fx,Krause:2012yh,Grimm:2012rg,Collinucci:2012as}.  Ours is the first example of a global model that utilizes the Tate divisor description of $U(1)$ symmetries and extends a local model with non-inherited bundle data.  Understanding models with non-inherited bundles is crucial because they will play an important role in stabilizing complex structure moduli.

\subsubsection{Setup}

In general, GUT-breaking can be accomplished in F-theory models by turning on a nonzero expectation value for a chiral adjoint along the internal surface $S_2$ or by introducing a nontrivial configuration for the GUT gauge field on that surface.  The first can be achieved `by hand' in F-theory by explicitly separating some of the branes in the stack (or, equivalently, unfolding the GUT singularity) but the dynamics of this process is connected to the problem of moduli stabilization and will generally result in GUT-breaking near the Planck scale.  For this reason, the most popular approach in the literature is to break the GUT with an internal gauge field configuration \cite{Beasley:2008kw,Donagi:2008kj}, in which case the GUT scale is associated with the volume of $S_2$ and can be separated from $M_{\text{Planck}}$.  In all current F-theory models, the GUT-breaking gauge field generates a nontrivial flux along $S_2$ that introduces potentially problematic high scale threshold corrections to the 4-dimensional gauge couplings.

Though breaking to the Standard Model is simpler in F-theory models than ordinary extra-dimensional GUTs,Êwe see that it doesn't come problem-free.  One encounters similar issues in Heterotic constructions, where the use of chiral adjoints or internal flux is also difficult, if not impossible.
Indeed, it is difficult to obtain chiral adjoints in heterotic constructions;  in fact, impossible at level 1 Kac-Moody.   It is not possible to break the GUT group $SU(5)$ to the Standard Model group using non-trivial hypercharge field strength on the heterotic side,  since this would give mass to the hypercharge gauge boson, and thus the photon.  Heterotic models are able to deal with these problems, however, by using a gauge configuration that doesn't introduce any nontrivial flux, namely a Wilson line.  This has several nice features.  It breaks $SU(5)$ to the Standard Model,  splits Higgs doublets and triplets and is consistent with gauge coupling unification.   The latter virtue is non-trivial.   The two low energy gauge couplings, $\alpha_{1,2}$, unify at a scale of order $M_G \sim 3 \times 10^{16}$ GeV.  Depending on the nature of the SUSY particle spectrum at the weak scale it turns out that $\alpha_3(M_G)$ is a few percent smaller than $\alpha_{1,2}(M_G) = \alpha_G$.   In addition the string scale is typically of order $M_S \sim 5 \times 10^{17}$ GeV and gauge couplings are expected to unify at the string scale.   This discrepancy can be resolved by allowing for one large extra dimension of order $M_C \leq M_G$ such that Kaluza-Klein modes between $M_C$ and $M_S$ fix this apparent discrepancy \cite{Hebecker:2004ce,Dundee:2008ts,Dundee:2008gr}.

In this paper, we attempt to borrow these Heterotic successes by building F-theory models that use Wilson lines to break $SU(5)_{\rm GUT}\rightarrow SU(3)\times SU(2)\times U(1)_Y$.  We further require some structure to suppress rapid proton decay and choose to engineer $U(1)_{B-L}$  for this purpose
{\footnote{More properly, we engineer the unique linear combination $U(1)_{\chi}$ of $B-L$ and hypercharge that commutes with $SU(5)_{\rm GUT}$.}}.  To that end, we develop the Tate divisor approach to $U(1)$ symmetries in an explicit setting.  This gives us an opportunity
to comment on the relation to the $U(1)$ restricted Tate models of \cite{Grimm:2010ez}.

The use of Wilson lines for GUT-breaking was first proposed in \cite{Donagi:2008ca,Beasley:2008kw} but no global models have been produced thus far.  To build such a model we must start with a surface $S_2$ that admits Wilson lines, which is to say a surface with nontrivial fundamental group $\pi_1(S_2)\ne 0$.
To keep things simple, we will take $S_2$ to be perhaps the best-known (non-toroidal) surface of this type in physics, namely the Enriques surface for which $\pi_1(S_2)=\mathbb{Z}_2$.  It should be noted that finding other candidate surfaces that satisfy the requirements of housing $A_4$ singularities in a Calabi-Yau 4-fold is very difficult and we do not know of any other examples.
Note that the presentation of the Enriques surface that we use cannot be arbitrary but must be carefully chosen with an eye toward the eventual embedding into a global compactification.  In particular, the Enriques $S_2$ must sit inside a 3-fold $B_3$ in such a way that the requisite elliptic fibration $Y_4$ over $B_3$ can be constructed.   This amounts to requiring that certain line bundles on $B_3$ admit holomorphic sections and greatly restricts the freedom in choosing $S_2$.  We were able to find only one class of examples, namely surfaces given by the intersection of three quadrics in $\mathbb{P}^5$ along with their resolutions when some of the quadrics are singular{\footnote{Geometries that support Wilson-line breaking in principle were constructed in \cite{Braun:2010hr} but they exhibited additional singularities that we prefer to avoid.  These singularities should correspond to orientifold 3-planes that intersect the GUT branes \cite{Denef:2005mm,Collinucci:2008zs,Collinucci:2009uh,Blumenhagen:2009up} so in principle their physics can be understood.  In this paper, however, we prefer to build models that do not require O3 planes on top of the GUT branes.}}.  The difficulty in finding a suitable presentation of the Enriques surface exemplifies the general principle that local model-building is never completely separated from global concerns.  Nevertheless, it is easier to engineer details of the model, such as choices for spectral and bundle data that yield three chiral generations, in the local setting where complications are fewer.  We proceed in this way, uncovering new structure that must be introduced at the level of the spectral data, and use the Tate divisor approach to lift this data to the global completion.

\subsubsection{Our Models}

With our choice of $S_2$, we construct explicit local 3-generation GUT models with $U(1)_{B-L}$ that break $SU(5)_{\rm GUT}$ with a discrete Wilson line.  We find a two-parameter family of models that yield 3 chiral generations and, in addition, go beyond the typical chirality computations to explicitly determine the precise spectrum of all light (charged) fields in two cases using recently developed techniques for computing line bundle cohomologies on toric spaces \cite{Blumenhagen:2010pv,Jow,Blumenhagen:2010ed,cohomCalg:Implementation}.  From there, we proceed to construct explicit global completions that show the full power of the `Tate divisor' formalism \cite{Marsano:2010ix,Marsano:2011nn,Marsano:2011hv}.  In particular, our local model construction requires two features whose global extension can be tricky: a $U(1)$ symmetry and a so-called non-inherited bundle \cite{Donagi:2009ra,Marsano:2009gv,Marsano:2009wr} which amounts to a gauge field configuration of the local model that can only be switched on for special choices of the Higgs bundle spectral data.  We use the `Tate divisor' formalism \cite{Marsano:2010ix,Marsano:2011nn,Marsano:2011hv} to construct global extensions of both features in a very explicit way in the fully resolved Calabi-Yau 4-fold.
The explicit global embedding allows a direct computation of the D3-brane charge induced by the various branes and fluxes of the models.  Within the full 2-parameter family of bundles that yield 3 chiral generations in the local model, we find precisely one choice in the global extension that does not require the introduction of anti-D3-branes to cancel this charge.  Interestingly, we find a second choice that only requires a single anti-D3-brane, a situation that we view favorably in light of the need to break supersymmetry.  

The models that we construct have several positive features.   Phenomenologically, each has exactly 3 families of quarks and leptons with no extra vector-like generations.  By construction, GUT-breaking is triggered by a discrete Wilson line so we do not have to worry about large GUT scale gauge threshold corrections \cite{Donagi:2008kj,Blumenhagen:2008aw}.  Also by construction, the models exhibit a $U(1)_{B-L}$ symmetry that eliminates rapid proton decay processes involving dimension 4 operators.
The fluxes that we require fix many (but not all) of the complex structure moduli of $Y_4${\footnote{More specifically, the fluxes are not inherited from the ambient 5-fold in which $Y_4$ sits as an anti-canonical divisor and can only be switched on for special choices of the complex structure moduli.  In this sense, they are similar in character to those of \cite{Anderson:2010mh,Anderson:2011cza,Anderson:2011ty}.}}.
Mathematically the models are appealing because the base 3-fold $B_3$ is toric.  This means that the Calabi-Yau 4-fold $Y_4$ can be described as a hypersurface inside a 5-dimensional toric variety and it is this property that simplifies our spectrum computations.

Unfortunately, our models exhibit undesirable features as well.  For starters, neither has an acceptable Higgs sector.  As we do not introduce any symmetry to distinguish the up- and down- Higgs doublets, we might be surprised to find any Higgs sector at all.
Quite surprisingly, our spectrum computations reveal that the problem is not too few Higgs fields but \emph{too many}!
The supersymmetric model exhibits 4 pairs of Higgs doublets while the model that requires the introduction of a single anti-D3-brane exhibits 2 pairs of doublets and 4 pairs of triplets.  We do not have a good understanding for the mechanism responsible for retaining so many massless vector-like pairs of fields.  Developing this further may shed light on new ways to deal with the $\mu$ problem in F-theory.

The models also suffer from the existence of vector-like exotics from the adjoint of $SU(5)_{\rm GUT}$.  We provide a simple argument along the lines of those in \cite{Beasley:2008kw,Hayashi:2011aa} that it is never possible to lift all chiral matter on $S_2$ without a nontrivial internal flux, so exotics like this are a general feature of Wilson line models.  Since the exotics are vector-like, it is possible that they may be lifted dynamically perhaps through nonperturbative physics associated to M5 instantons.  In the absence of such effects, one possible remedy is to start with a larger GUT group like $SO(10)$ or $SU(6)$, perhaps along the lines of \cite{Donagi:2011dv}, and use a combination of Wilson line and gauge flux breaking.  This will almost certainly lead to large threshold corrections to gauge coupling unification, though, destroying some of the motivation for considering Wilson lines in the first place.  Heterotic models with discrete Wilson lines manage to avoid similar problems through the introduction of extra singularities (for which the analog in our models would be singularities of the GUT surface $S_2$ as in the geometries of \cite{Braun:2010hr}) or the use of discrete Wilson lines associated to non-holomorphic involutions.  It may be possible to apply similar ideas to F-theory constructions as well, though we do not pursue either in the present paper.

 Finally, our model suffers from unacceptable proton decay rates due to dimension 5 operators that are suppressed only by the compactification scale, $M_{KK} \sim 10^{16}$ GeV.  $U(1)$-based solutions to this problem tend to be problematic \cite{Marsano:2009gv,Marsano:2009wr,Marsano:2010sq,Dolan:2011iu,Dolan:2011aq} so the best way to address this issue is likely with a discrete symmetry such as the $\mathbb{Z}_4^R$ of Ref. \cite{Lee:2010gv,Lee:2011dya,Kappl:2010yu}.

 \subsection{Outline}

 The rest of the paper is organized as follows.  In section \ref{sec:generalsetup}, we provide a general description of F-theory models, the local-to-global framework, and the `Tate divisor' formalism.  We proceed in section \ref{sec:FEnriquesLocal} to construct local $SU(5)_{\rm GUT}$ models that admit Wilson lines and $U(1)$ symmetries and compute the precise spectrum in a few interesting examples.  We then study global extensions in section \ref{sec:globalgeneral} including very explicit constructions of $U(1)$'s and fluxes with the Tate divisor in resolved Calabi-Yau's.  We combine the two in section \ref{sec:globalmodel} to construct our two globally consistent F-theory models with $U(1)_{\chi}$ and Wilson line breaking.  A number of computational details are relegated to the appendices.

\section{General Setup}
\label{sec:generalsetup}

We begin in this section with an overview of the basic structures of F-theory models and the strategy we will adopt to build F-theory GUTs that utilize Wilson lines to break $SU(5)_{\rm GUT}$.

\subsection{Global Models}

To specify an F-theory compactification, we need two pieces of data: an elliptically fibered Calabi-Yau 4-fold $Y_4$ with section and a $(2,2)$-form $G_4$ on $Y_4$ that defines a $G$-flux.  The geometry of $Y_4$ controls the structure of non-Abelian gauge groups and charged matter while the $G$-flux couples to the internal profiles of these degrees of freedom and, in so doing, determines the actual 4-dimensional spectrum.

The connection between geometry and physics is most manifest when F-theory is viewed as a limit of M-theory.  For this, we consider M-theory on $\mathbb{R}^{2,1}\times Y_4$ and take the volume of the elliptic fiber to zero while keeping the complex structure modulus fixed.  In this limit, we pass to type IIB with varying axio-dilaton and the $\mathbb{R}^{2,1}$ grows to $\mathbb{R}^{3,1}$ as a result of $T$-duality.  The basic objects of M-theory that we must follow through this limit are the 3-form potential $C_3$ and the M2-branes to which it couples.  We are most interested in M2-branes that wrap holomorphic curves in $Y_4$ and reductions of $C_3$ on $(1,1)$-forms{\footnote{More generally we would like to reduce $C_3$ on 2-forms but, on $Y_4$, $h^{2,0}=h^{0,2}=0$ so all 2-forms are of $(1,1)$ type and, correspondingly, all 6-forms are of $(3,3)$ type.  
}} since these give rise to particles and gauge fields in 4-dimensions.  Any M2 that wraps the elliptic fiber or a curve in the base $B_3$, though, maps to a fundamental string or wrapped D3-brane that carries momentum or extends along the new direction of $\mathbb{R}^{3,1}$.  Correspondingly, any reduction of $C_3$ on a $(1,1)$-form that integrates over the fiber or a curve in the base passes to a field configuration that breaks 4-dimensional Lorentz invariance.  Consequently, we are interested in $(1,1)$-forms with `one leg on the fiber' and holomorphic curves that do not sit inside or meet the section .  If $Y_4$ is smooth then no such objects exist.  When $Y_4$ has a singular Weierstrass model, some will appear when the singularities are resolved.

\subsubsection{Codimension 1}
\label{subsubsec:cod1}

The appearance of gauge groups is correlated with singular fibers over codimension 1 loci in $B_3$.  Singularities of this type admit an ADE classification due to Kodaira \cite{Kodaira} and the geometric origin of (part of) the gauge symmetry is standard.  Upon resolution, the singular fiber over a surface $S_2$ will admit several components $C_i$.  Each $C_i$ that is contracted in the singular limit is a holomorphic curve that satisfies the `one leg on the fiber' condition and, consequently, gives rise to light particle states from wrapped M2 branes.  In addition, each $C_i$ yields a divisor obtained by fibering it over $S_2${\footnote{More specifically, we resolve $Y_4$ by embedding it in an ambient 5-fold, blowing up the 5-fold, and passing to the proper transform.  The new divisor is the restriction of an exceptional divisor of the blow-up \cite{Marsano:2011hv}.}}.  This determines a $(1,1)$-form $\omega_i$ that satisfies the `one leg on the fiber' condition and produces a 4-dimensional gauge field through the reduction of $C_3$.  States from wrapped M2's couple to this gauge field with a charge that is determined by the intersection number of relevant 2-cycles in the singular fiber
\begin{equation}q\sim \int_{C_i}\omega_j = C_i\cdot_{\text{fiber}}C_j\end{equation}
As Kodaira fibers take the form of (extended) Dynkin diagrams, the $U(1)$'s are naturally interpreted as Cartan generators and the wrapped M2's as adjoint states corresponding to simple roots.  While we cannot directly quantize zero volume M2's, we anticipate that a full non-Abelian gauge group corresponding to the ADE type of the fiber emerges and Heterotic/F-theory duality corroborates this expectation.

In this paper we are interested in models with an $SU(5)_{\rm GUT}$ gauge group so we seek $Y_4$'s that exhibit $A_4$ singular fibers over a distinguished surface $S_2$ in $B_3$.  We can build such a $Y_4$ as a hypersurface inside the 5-fold
\begin{equation}W_5 = \mathbb{P}\left(\CO\oplus K_{B_3}^{-2}\oplus K_{B_3}^{-3}\right)\end{equation}
whose defining equation is written in `Tate form'
\begin{equation}vy^2 = x^3 + \bh_0 z^5v^3 + \bh_2 z^3 v^2x + \bh_3 z^2v^2 y + \bh_4 zv x^2 + \bh_5 vxy\label{genericTate}\end{equation}
The various objects that appear here are sections of the indicated bundles
\begin{equation}\begin{array}{c|c}
\text{Section} & \text{Bundle} \\ \hline
v & \CO(\sigma) \\
x & \CO(\sigma + 2\cb) \\
y & \CO(\sigma + 3\cb) \\
z & \CO(S_2) \\
\bh_m & \CO( [6-m]\cb - [5-m]S_2)
\end{array}\label{bmbundles}\end{equation}
Here $\sigma$ denotes a hyperplane of the $\mathbb{P}^2$ fiber of $W_5$ and we have introduced the notation $\cb$ for the anti-canonical divisor of $B_3$, roughly $\cb\sim c_1(B_3)$.  Note that we do not notationally distinguish between divisors on $B_3$ and their pullbacks to $W_5$.

\subsubsection{Adding $U(1)$'s}

We could proceed with geometries of the type \eqref{genericTate} but one typically encounters phenomenological problems if additional structure is not imposed.  The most general superpotential of the Minimal Supersymmetric Standard Model (MSSM) that is allowed by gauge invariance, for instance, leads to rapid proton decay.  To gain some level of control over this we need additional symmetry, which can descend from either a discrete isometry of the Calabi-Yau 4-fold or the presence of additional $U(1)$ gauge fields{\footnote{These $U(1)$'s are typically lifted by a Stuckelberg mechanism.}}.  We focus on the latter in this paper, which requires specializing the structure of \eqref{genericTate} so that the resulting 4-fold exhibits an additional harmonic $(1,1)$-form satisfying the `one leg on the fiber' condition.

Adding $U(1)$'s is somewhat subtle \cite{Hayashi:2010zp} but by now a procedure for doing so is well-understood \cite{Grimm:2010ez,Marsano:2010ix,Marsano:2011nn,Grimm:2011tb,Krause:2011xj},.  To proceed, we consider a distinguished divisor of \eqref{genericTate} that was given the name `Tate divisor' in \cite{Dolan:2011iu}
\begin{equation}\ctate :\quad \bh_0z^5v^2 + \bh_2z^3vx + \bh_3z^2vy + \bh_4zx^2 + \bh_5xy=0\label{TateDivisor}\end{equation}
As written, this divisor has two components with one being the section at $v=x=0$.  We will usually not worry about this additional component since most objects of interest satisfy `one leg on the fiber' conditions that make them orthogonal to objects in the section.

The divisor $\ctate$, apart from the zero section component, is a 5-sheeted covering of $B_3$ inside $Y_4$ that is singular at $z=0$ where all of the sheets come together.  When the $A_4$ singularity is resolved, the sheets of $\ctate$ are separated so that, if the $\bh_m$ are suitably generic, the resulting divisor is smooth.  A useful way to think about $\ctate$ is to write its defining equation in terms of the meromorphic section $t=y/x$, which is holomorphic along $\ctate$.  In that case, $\ctate$ is described by a homogeneous polynomial of degree 5 in $z$ and $t$
\begin{equation}\bh_0 z^5 + \bh_2 z^3 t^2 + \bh_3 z^2t^3 + \bh_4 zt^4 + \bh_5 t^5\label{tatets}\end{equation}
If we choose the $\bh_m$'s so that this polynomial factors, $\ctate$ becomes reducible and the Calabi-Yau develops an additional singularity along a curve that sits inside the intersection of its components.  The two components of $\ctate$ are in fact Weil divisors of the partially resolved 4-fold obtained by resolving only the $A_4$ singular fibers and we complete the resolution of the 4-fold by blowing up the ambient 5-fold along one of the components as in \cite{Krause:2011xj}.  In this paper, we will be interested in a special case of this where the split of \eqref{tatets} is into a quartic and linear piece
\begin{equation}\ctate\rightarrow \ctate^{(4)}+\ctate^{(1)}\end{equation}
with the special property that we can move the linear component to $t=0$ by a change of variables
\begin{equation}\bh_0 z^5 + \bh_2 z^3 t^2 + \bh_3 z^2t^3 + \bh_4 zt^4 + \bh_5 t^5 = (\ah_0 z^4 + \ah_1 z^3 t + \ah_2 z^2t^2 + \ah_3 zt^3 + \ah_4 t^4)(\eh_0 z + \eh_1 t)\label{splitspecdiv}\end{equation}
where $\eh_1$ is a nonzero constant so that the shift $t\rightarrow t - \eh_0 z / \eh_1$ is holomorphic.  Performing this change of variables carefully at the level of $x$ and $y$ recovers the $U(1)$-restricted Tate models of \cite{Grimm:2010ez}.  In the fully resolved Calabi-Yau, the splitting of $\ctate$ corresponds to the appearance of a new $(1,1)$-form satisfying the `one leg on the fiber' condition.  We can get it by starting with the $(1,1)$-form corresponding to the divisor
\begin{equation}\ctate^{(4)}-4\ctate^{(1)}\label{U1divgen}\end{equation}
and adding a correction term to ensure that it is orthogonal to all horizontal and vertical curves.  We do this explicitly in section \ref{subsubsec:splittatedivisorandU1} and, in this way, recover the $(1,1)$-form of \cite{Krause:2011xj}.

\subsubsection{Codimension 2 and 3}
\label{subsubsec:cod23}

Above loci of codimension 2 and 3 in $B_3$, the singularity type of the fiber can enhance.  In \eqref{genericTate}, the codimension 2 enhancements occur above the `matter curves'
\begin{equation}\begin{split}SO(10):&\quad z = \bh_5=0 \\
SU(6):&\quad z = \bh_0\bh_5^2 - \bh_2\bh_3\bh_5+\bh_3^2\bh_4=0
\end{split}\label{mattercurvesgen}\end{equation}
The hallmark of a matter curve $\Sigma_{R,\downarrow}$ is that some irreducible components $C_i$ of the $A_4$ singular fiber degenerate into reducible collections of curves over generic points in $\Sigma_{R,\downarrow}$.
This yields a set of new effective curves $C_{R,a}$ in the singular fiber above $\Sigma_{R,\downarrow}$ that, through wrapped M2 branes, yield new light degrees of freedom that localize along $\Sigma_{R,\downarrow}$.  The homology classes of the $C_{R,a}$ are determined by their intersections with the Cartan divisors and hence by the $U(1)$ charges of the corresponding state.  As the notation indicates, they will correspond to weights of some representation $R$ of $SU(5)_{\rm GUT}$ or its conjugate.  Indeed, the holomorphic curves in $\pi^*\Sigma_{\mathbf{10},\downarrow}$ generate a sublattice of $H^2(Y_4,\mathbb{Z})\cap H^{1,1}(Y_4,\mathbb{C})$ that is isomorphic to the $\mathbf{10}$ weight lattice of $SU(5)_{\rm GUT}$ while the holomorphic curves in $\pi^*\Sigma_{\mathbf{\overline{5}},\downarrow}$ generate a sublattice that is isomorphic to the $\mathbf{\overline{5}}$ weight lattice of $SU(5)_{\rm GUT}$.
The singular fibers themselves are not necessarily precise Dynkin diagrams \cite{Esole:2011sm} but this is not important \cite{Marsano:2011hv}.

Once we have charged matter representations we need to know the precise spectrum and the pattern of couplings.  The generation of couplings from this perspective is via M2-brane instantons along 3-chains that connect homologously trivial combinations of wrapped M2's.  Because the homology class of each 2-cycle is determined by its $U(1)$ charges, any gauge invariant combination of 2-cycles must be connected by a 3-chain.  The coupling induced by an M2-brane instanton that wraps such a 3-chain can sometimes be zero and, even when it is nonzero, will be exponentially suppressed by the instanton volume.  Dominant contributions to Yukawas therefore come from places where the corresponding 3-chain degenerates and this can happen only above the isolated points where matter curves intersect.  The singularity type worsens considerably above these points and the cone of effective curves in the fiber includes all $\mathbf{10}$ and $\mathbf{5}$ weights.  The algebra of effective curves can vary, though, depending on the nature of the singular fiber and, correspondingly, the set of 3-chains that degenerate can vary also.  In general the `top-type' $\mathbf{10}\times\mathbf{10}\times\mathbf{5}$ couplings descend from `$E_6$' points where $\bh_5=\bh_4=z=0$ while `bottom-type' $\mathbf{10}\times\mathbf{\overline{5}}\times\mathbf{\overline{5}}$ couplings descend from `$SO(12)$' points where $\bh_5=\bh_3=0$.  The singular fibers themselves can be quite complicated \cite{Esole:2011sm} and do not correspond to ordinary Dynkin diagrams.

The spectrum of charged fields is determined by the $G$-flux, which is the field strength associated to $C_3$.  As is familiar, the only $G$-fluxes we can introduce without violating Lorentz invariance in the F-theory limit are those that have `one leg on the fiber'.  Since the coupling of an M2 to $C_3$ is simply
\begin{equation}\int_{M2\text{ worldvolume}}C_3\end{equation}
the state that we get from an M2 wrapping a degenerate 2-cycle $C_{R,a}$ above a matter curve $\Sigma_{R,\downarrow}$ couples to a line bundle $L_{R,a}$ on $\Sigma_{R,\downarrow}$ that is induced by integrating $C_3$ over $C_{R,a}$.
The number of chiral multiplets in the representation $R$ and its conjugate are then determined by
\begin{equation}n_{R} = h^0(\Sigma_{R,\downarrow},K_{\Sigma_{R,\downarrow}}^{1/2}\otimes L_{R,a})\qquad n_{\overline{R}} = h^1(\Sigma_{R,\downarrow},K_{\Sigma_{R,\downarrow}}^{1/2}\otimes L_{R,a})\end{equation}
where $K_{\Sigma_{R,\downarrow}}^{1/2}$ is a spin bundle on $\Sigma_{R,\downarrow}$ that is inherited from $B_3$.  The operation of inducing a bundle $L_{R,a}$ on $\Sigma_{R,\downarrow}$ is naturally described by introducing the notion of a `matter surface' $\CS_{R,a}$ that is essentially the fibration of a new 2-cycle $C_{R,a}$ over the matter curve $\Sigma_{R,\downarrow}$ above which it appears
{\footnote{More specifically, $\CS_{R,a}$ is the irreducible component of $\pi^*\Sigma_{R,\downarrow}$ that contains the 2-cycle $C_{R,a}$ above generic points on $\Sigma_{R,\downarrow}$.}}.  With this notation, the net chirality is given by an index that takes a very natural form{\footnote{This formula was recently derived from 11-dimensional supergravity by moving to the Coulomb branch (i.e. giving finite volume to the resolved cycles) in \cite{Grimm:2011fx}.}}
\begin{equation}n_{R}-n_{\overline{R}} = \int_{\CS_{R,a}}G_4\label{chiralityfromGflux}\end{equation}
We typically specify $G_4$ by its Poincare dual holomorphic surface so this amounts to an intersection of two surfaces in $Y_4$ and is, in principle, easily computable.

\subsection{Local Models}
\label{subsec:localmodelsgen}

At this point, we could proceed to build models directly at a global level by constructing a suitable $Y_4$ and defining $G$-fluxes on it.  The degrees of freedom that carry $SU(5)_{\rm GUT}$ charge, though, all descend from M2's that wrap holomorphic curves in singular fibers over $S_2$.
Their physics should depend only on the local geometry near $S_2$ and the local behavior of the flux.  Indeed, the local geometric and flux data combine to specify the holomorphic physics of the worldvolume theory on a stack of 7-branes wrapping $\mathbb{R}^{3,1}\times S_2$.  In practice, we will start by building our model in this setting, that is to say specifying the local data needed to achieve desired properties like 3 chiral generations.
We will then extend this data to construct a full global model complete with a proper description of the $G$-flux.

A local to global approach like this has advantages and drawbacks.  The principal advantage is that local models are significantly simpler than global ones.  We do not have to work with global Calabi-Yau's or their resolutions which can be quite cumbersome.  This will be especially helpful for the models that we build in this paper because the fluxes that are available for generic choices of local data will be unable to yield 3 chiral generations.  Dealing with this will require a very strong specialization of the local data that would be difficult to uncover within the complexities of the global model.  The drawback to working with local models, however, is that they are insensitive to important global features like tadpoles and $D$-terms.  Our local models will be globally extendable but we will see that all but one choice of flux will generate a 3-brane tadpole that requires the introduction of anti-D3-branes to cancel.  We will also investigate a second choice of flux that requires the introduction of only one anti-D3-brane.

\subsubsection{Spectral Cover Generalities}
\label{subsubsec:specgen}

The connection between local and global models can be made explicit through the Tate divisor \cite{Marsano:2010ix,Dolan:2011iu,Marsano:2011nn}.  The restriction of the Tate divisor to $\pi^*S_2$ is a 5-sheeted covering of $S_2$ that specifies the local model Higgs bundle spectral cover \cite{Marsano:2011hv}
\begin{equation}\ctate|_{\pi^*S_2}\supset \cloc\label{Tate2Higgs}\end{equation}
This object lives in the total space of $K_{S_2}$ which we can compactify to $\mathbb{P}(\CO\oplus K_{S_2})$,
and is defined by
\begin{equation}b_0 U^5 + b_2U^3V^2 + b_3U^2V^3+b_4UV^4+b_5V^5\label{Clocalgeneric}\end{equation}
where
\begin{equation}b_m = \bh_m|_{S_2}\end{equation}
For clarity, the objects in \eqref{Clocalgeneric} are sections of the indicated bundles
\begin{equation}\begin{array}{c|c}
\text{Section} & \text{Bundle} \\ \hline
U & \CO(\sigma_{loc}) \\
V & \CO(\sigma_{loc} + \cs) \\
b_m & \CO(\eta-m\cs)
\end{array}\end{equation}
Here, $\sigma_{loc}$ denotes a hyperplane in the $\mathbb{P}^1$ fiber of $\mathbb{P}(\CO\oplus K_{S_2})$ and $\cs$ is shorthand for the anti-canonical divisor class in $S_2$, roughly $\cs\sim c_1(S_2)$.  We also introduce the standard local model notation $\eta$ for the class
\begin{equation}\eta = 6\cs - t\end{equation}
with $t$ determined by the normal bundle $N_{S_2/B_3}$ of $S_2$ in $B_3$ according to
\begin{equation}\CO(-t) = N_{S_2/B_3}\end{equation}

We construct $G$-fluxes from holomorphic surfaces $\Gamma$ inside $\ctate$ as in \cite{Marsano:2010ix,Marsano:2011nn,Marsano:2011hv}.  The restriction of $\Gamma$ under \eqref{Tate2Higgs} yields a divisor $\gamma$ in $\cloc$ that, together with $\cloc$ itself, specifies an $SU(5)_{\perp}$ Higgs bundle on $S_2$.  Defining $p_{\text{loc}}$ as the projection
\begin{equation}p_{\text{loc}}:\cloc\rightarrow S_2\end{equation}
the adjoint Higgs field $\Phi$ and gauge bundle $V$ of this Higgs bundle are given by \cite{Donagi:2009ra,Donagi:2011jy}
\begin{equation}\Phi \sim p_{\text{loc},\ast}U\qquad V\sim p_{\text{loc},\ast}{\cal{L}}_{\gamma}\end{equation}
where
\begin{equation}{\cal{L}}_{\gamma}=\CO\left(\gamma+\frac{r}{2}\right)\end{equation}
with $r$ the ramification divisor of the covering $p_{\text{loc}}${\footnote{That ${\cal{L}}_{\gamma}$ is an integer bundle is connected to the quantization condition of $G$-flux \cite{Witten:1996md}.}}.

The worldvolume theory on $\mathbb{R}^{3,1}\times S_2$ is a twisted supersymmetric Yang-Mills theory \cite{Donagi:2008ca,Beasley:2008dc,Hayashi:2009ge,Donagi:2009ra} with gauge group $E_8$ in the presence of the (in general meromorphic) $SU(5)_{\perp}$ Higgs bundle, which breaks $E_8\rightarrow SU(5)_{\rm GUT}$.  Bifundamentals under the breaking
\begin{equation}\mathbf{248}\rightarrow (\mathbf{24},\mathbf{1})\oplus (\mathbf{1},\mathbf{24})\oplus \left[(\mathbf{10},\mathbf{5})\oplus\text{cc}\right]\oplus\left[(\mathbf{\overline{5}},\mathbf{10})\oplus\text{cc}\right]\end{equation}
acquire position-dependent masses along $S_2$ and become locally massless along curves in $S_2$ where suitable linear combinations of $\Phi$ eigenvalues $t_i$ vanish.  These define matter curves `downstairs' in $S_2$ and in this case take the form
\begin{equation}\begin{split}\Sigma_{\mathbf{10},\downarrow}:&\,\, 0=b_5\sim \prod_{i=1}^5t_i\\
\Sigma_{\mathbf{\overline{5}},\downarrow}:&\,\,0 = b_0b_5^2-b_2b_3b_5+b_3^2b_4\sim \prod_{i<j}(t_i+t_j)
\end{split}\end{equation}
where we recall that the $b_m$ are related to symmetric polynomials in the eigenvalues according to $b_m\sim b_0 e_n(t_i)$.  Zero modes for $\mathbf{10}$'s and $\mathbf{\overline{5}}$'s are directly counted by cohomology groups of matter curves `upstairs', that is distinguished curves in $\cloc$ that project to the $\Sigma_{R\,\downarrow}$ under $p_{\text{loc}}$.  These are given by
\begin{equation}\begin{split}\Sigma_{\mathbf{10},\uparrow}:&\,\, \cloc\cap [U=0] \\
\Sigma_{\mathbf{\overline{5}},\uparrow}:&\,\, \cloc\cap \tau \cloc - [\text{curves fixed by $\tau$}]\end{split}\end{equation}
where $\tau$ is the involution that takes $V\rightarrow -V$.  The zero mode counting problem can be expressed in terms of `upstairs' cohomologies or reformulated in terms of `downstairs' cohomologies following \cite{Hayashi:2008ba}
\begin{equation}\label{simplecohomologies}\begin{array}{c|cc}
\text{Representation} & \text{Upstairs cohomology} & \text{Downstairs cohomology} \\ \hline
\mathbf{10} & h^0(\Sigma_{\mathbf{10},\uparrow},p_{\text{loc}}^*K_{S_2}\otimes {\cal{L}}_{\gamma}|_{\Sigma_{\mathbf{10},\uparrow}}) &
h^0(\Sigma_{\mathbf{10},\downarrow},K_{\Sigma_{\mathbf{10},\downarrow}}^{1/2}\otimes \CO(p_{\text{loc}*}\gamma|_{\Sigma_{\mathbf{10},\uparrow}})) \\
\mathbf{\overline{10}}& h^1(\Sigma_{\mathbf{10},\uparrow},p_{\text{loc}}^*K_{S_2}\otimes {\cal{L}}_{\gamma}|_{\Sigma_{\mathbf{10},\uparrow}}) &
h^1(\Sigma_{\mathbf{10},\downarrow},K_{\Sigma_{\mathbf{10},\downarrow}}^{1/2}\otimes \CO(p_{\text{loc}*}\gamma|_{\Sigma_{\mathbf{10},\uparrow}})) \\ \hline
\mathbf{\overline{5}} & h^0_-(\Sigma_{\mathbf{\overline{5}},\uparrow},p_{\text{loc}}^*K_{S_2}\otimes {\cal{L}}_{\gamma}|_{\Sigma_{\mathbf{\overline{5}},\uparrow}}\otimes \tau {\cal{L}}_{\gamma}|_{\Sigma_{\mathbf{\overline{5}},\uparrow}}) &
h^0(\Sigma_{\mathbf{\overline{5}},\downarrow},K_{\Sigma_{\mathbf{\overline{5}},\downarrow}}^{1/2}\otimes \CO(p_{\text{loc}*}\gamma|_{\Sigma_{\mathbf{\overline{5}},\uparrow}})) \\
\mathbf{5} &h^1_-(\Sigma_{\mathbf{\overline{5}},\uparrow},p_{\text{loc}}^*K_{S_2}\otimes {\cal{L}}_{\gamma}|_{\Sigma_{\mathbf{\overline{5}},\uparrow}}\otimes \tau {\cal{L}}_{\gamma}|_{\Sigma_{\mathbf{\overline{5}},\uparrow}}) &
 h^1(\Sigma_{\mathbf{\overline{5}},\downarrow},K_{\Sigma_{\mathbf{\overline{5}},\downarrow}}^{1/2}\otimes \CO(p_{\text{loc}*}\gamma|_{\Sigma_{\mathbf{\overline{5}},\uparrow}})) \\
\end{array}\end{equation}
where the $`-'$ for the $\mathbf{\overline{5}}/\mathbf{5}$ `upstairs' cohomology indicates to take the antisymmetric cohomology with respect to the $\ztau$ involution $\tau$, which interchanges the two sheets of $\Sigma_{\mathbf{\overline{5}},\uparrow}$.  Net chiralities are determined by indices which, in turn, can be computed by simple intersections
\begin{equation}\begin{split}n_{\mathbf{10}}-n_{\mathbf{\overline{10}}} &= \gamma\cdot_{ \cloc}\Sigma_{\mathbf{10},\uparrow} \\
n_{\mathbf{\overline{5}}}-n_{\mathbf{5}} &= \gamma\cdot_{ \cloc}\Sigma_{\mathbf{\overline{5}},\uparrow}
\end{split}\end{equation}
If we are interested in going beyond index computations, the upstairs cohomologies are easier to work with because the downstairs picture requires one to be careful about precisely which $\theta$ characteristic, $K_{\Sigma,\downarrow}^{1/2}$, is the `right' one.  The downstairs picture, on the other hand, is more helpful for seeing the connection with the global description of $G$-flux \cite{Marsano:2011nn,Marsano:2011hv}.  If we construct $G$ from a holomorphic surface $\Gamma$ inside $\ctate$, the intersection of $\Gamma$ with the $\mathbf{10}$ ($\mathbf{\overline{5}}$) matter surface is a collection of points specified by the intersection of $\gamma$ with the corresponding `upstairs' matter curve $\Sigma_{\mathbf{10},\uparrow}$ ($\Sigma_{\mathbf{\overline{5}},\uparrow}$).  Integrating $G$ over the fiber induces a bundle on the `downstairs' matter curve $\Sigma_{\mathbf{10},\downarrow}$ ($\Sigma_{\mathbf{\overline{5}},\downarrow}$) given by $\CO(p_{\text{loc}*}\gamma|_{\Sigma_{\mathbf{10},\uparrow}})$ ($\CO(p_{\text{loc}*}\gamma|_{\Sigma_{\mathbf{\overline{5}},\uparrow}})$).

Before moving to split spectral covers, let us give a more explicit description of the matter curves and recall the general chirality formulae.  The defining equations of each `upstairs' matter curve and the corresponding class in $\mathbb{P}(\CO\oplus K_{S_2})$ are easily determined
\begin{equation}\begin{array}{c|cc}
\text{Matter Curve} & \text{Equations} & \text{Class} \\ \hline
\Sigma_{\mathbf{10},\uparrow} & U=0 & \sigma_{loc}\cdot (\eta-5\cs) \\
& b_5=0 & \\ \hline
\Sigma_{\mathbf{\overline{5}},\uparrow} & b_0 U^4 + b_2 V^2U^2 + b_4 V^4=0 & 2\sigma_{loc}\cdot (3\eta-10\cs)  \\
& b_3 U^2 + b_5 V^2=0 & + \eta\cdot (\eta - 3\cs)
\end{array}\label{genericmattcurves}\end{equation}
The chirality on each matter curve depends on $\gamma$ and our choices for this are limited.  For suitably generic $\cloc$, we can only construct $\gamma$ from curves that are inherited from $\mathbb{P}(\CO\oplus K_{S_2})$ namely $\sigma_{loc}\cdot \cloc$ and $\ploc^*\Sigma_{\downarrow}$ for curves $\Sigma_{\downarrow}$ in $S_2$.  We must make sure that the gauge bundle $V$ satisfies $c_1(V)=0$ to ensure that its structure group is truly $SU(5)_{\perp}$ in order to prevent the bundle from breaking $SU(5)_{\rm GUT}$.  In terms of $\gamma$ this leads to the condition{\footnote{It is easy to see that $G$-fluxes constructed from $\gamma$ in the Tate divisor formalism break $SU(5)_{\rm GUT}$ if this condition is violated.}}
\begin{equation}\ploc_*\gamma=0\end{equation}
and there is only one combination of inherited curves that satisfies it
\begin{equation}\gamma_u = 5\sigma_{loc}\cdot \cloc - \ploc^*(\eta-5\cs)\label{gammaugeneric}\end{equation}
The impact of $\gamma_u$ on the net chirality of each matter curve is now easy to compute.  We find
\begin{equation}\gamma_u\cdot \Sigma_{\mathbf{10},\uparrow} = \gamma_u\cdot \Sigma_{\mathbf{\overline{5}},\uparrow} = -\eta\cdot_{S_2}(\eta-5\cs)\label{genericspec}\end{equation}
which is the standard result for generic $SU(5)_{\rm GUT}$ local models \cite{Donagi:2009ra}.  Reproducing this will give us a sanity check on our first global extensions of local model data to $G$-flux in section \ref{subsubsec:extensionofinherited}.

\subsubsection{Split Spectral Cover}

To build models with an extra $U(1)$ symmetry, we will be interested in situations where the Tate divisor $\ctate$ splits.  This will induce a splitting of the Higgs bundle spectral cover ${\cal{C}}_{\text{Higgs,loc}}$ and, consequently, a naive reduction in the rank of the Higgs bundle spectral cover.
For our 4+1 split, the Higgs bundle spectral cover will become
\begin{equation}(a_4V^4+a_3V^3U+a_2V^2U^2+\alpha U^3[e_1V - e_0U] ) (e_1V+e_0 U)=0\label{Clocsplit}\end{equation}
where the various objects that appear are sections of the indicated bundles
\begin{equation}\begin{array}{c|c}\text{Section} & \text{Bundle} \\ \hline
a_m & \CO(\eta_4- m \cs) \\
\alpha & \CO(\eta_4 - \cs) \\
e_0 & \CO(\cs) \\
e_1 & \CO
\end{array}\end{equation}
and
\begin{equation}\eta_4 = 5\cs-t\end{equation}
For the bundle, we will specify independently a $\gamma_4$ and $\gamma_1$ on the two components $\cloc^{(4)}$ and $\cloc^{(1)}$ that are quantized so that the corresponding line bundles ${\cal{L}}_{\gamma}^{(4)}$ and ${\cal{L}}_{\gamma}^{(1)}$
\begin{equation}\begin{split}{\cal{L}}_{\gamma}^{(4)} &= \CO_{\cloc^{(4)}}\left(\gamma_4+\frac{r_4}{2}\right) \\
{\cal{L}}_{\gamma}^{(1)} &= \CO_{\cloc^{(1)}}\left(\gamma_1+\frac{r_1}{2}\right)
\end{split}\end{equation}
are well-defined.  Here $r_m$ is the ramification divisor of the covering
\begin{equation}p_m:\cloc^{(m)}\rightarrow S_2\end{equation}
with $r_1=0$ since $p_1$ is a smooth 1-1 cover.

Taken together, this data naively specifies an $S[U(4)\times U(1)]$ bundle.  The reduced structure group has the virtue that it will break $E_8$ not just to $SU(5)_{\rm GUT}$ but to the product $SU(5)_{\rm GUT}\times U(1)$, explicitly displaying the extra $U(1)$ symmetry that we wish to retain{\footnote{When $\cloc$ splits it becomes singular along the locus where its components intersect and it is necessary to specify how the bundles on the two different components are glued together there \cite{Cecotti:2010bp,Donagi:2011jy,Donagi:2011dv}.  We insist on a global split of the Tate divisors which engineers an honest $U(1)$ and hence corresponds to a trivial choice of this gluing data wherein the sheaf on $\cloc$ is just ${\cal{L}}_{\gamma}^{(m)}$ on each component $\cloc^{(m)}$ away from the intersection locus and a direct sum ${\cal{L}}_{\gamma}^{(1)}\oplus {\cal{L}}_{\gamma}^{(4)}$ on the intersection  $\cloc^{(1)}\cap \cloc^{(4)}$.}}.

The types of $\mathbf{10}$ and $\mathbf{\overline{5}}$ representation that can descend from $E_8$ are graded by their $U(1)$ charges and, correspondingly, the matter curves split.  For the simple factorization structure \eqref{Clocsplit} the $\mathbf{10}$ matter curve remains unchanged
\begin{equation}\Sigma_{\mathbf{10},\uparrow}\rightarrow \Sigma_{\mathbf{10},\uparrow}^{(4)} = \cloc^{(4)}\cdot [U=0]\end{equation}
but the $\mathbf{\overline{5}}$ curve splits in two
\begin{equation}\Sigma_{\mathbf{\overline{5}},\uparrow}\rightarrow\left\{\begin{array}{cl}\Sigma_{\mathbf{\overline{5}},\uparrow}^{(44)}= & \cloc^{(4)}\cap \tau \cloc^{(4)} - [\text{curves in }\cloc^{(4)}\text{ fixed by }\tau] \\
\Sigma_{\mathbf{\overline{5}},\uparrow}^{(41)}= & \cloc^{(4)}\cap \tau \cloc^{(1)}+\cloc^{(1)}\cap \tau \cloc^{(4)}- [\text{curves in }\cloc^{(4)}\cap \cloc^{(1)}\text{ fixed by }\tau]
\end{array}\right.\end{equation}
Correspondingly, $\Sigma_{\mathbf{\overline{5}},\downarrow}$ splits into two components that are projections of the upstairs curves
\begin{equation}\Sigma_{\mathbf{\overline{5}},\downarrow}\rightarrow \Sigma_{\mathbf{\overline{5}},\downarrow}^{(44)}\Sigma_{\mathbf{\overline{5}},\downarrow}^{(41)}\end{equation}
with
\begin{equation}\begin{split} \Sigma_{\mathbf{\overline{5}},\downarrow}^{(44)} &= a_4\alpha e_1^2 - a_3(a_2e_1+a_3e_0) \\
 \Sigma_{\mathbf{\overline{5}},\downarrow}^{(41)} &= a_4e_0^2+a_3e_0e_1+a_2e_1^2
 \end{split}\end{equation}
 The curve $\Sigma_{\mathbf{\overline{5}},\uparrow}^{(44)}$ is a smooth 2-sheeted cover of $\Sigma_{\mathbf{\overline{5}},\downarrow}^{(44)}$ whose sheets are permuted by the involution $\tau$.  The curve $\Sigma_{\mathbf{\overline{5}},\uparrow}^{(41)}$, on the other hand, is a reducible union of two 1-sheeted covers of $\Sigma_{\mathbf{\overline{5}},\downarrow}$ one of which sits inside $\cloc^{(4)}$ and the other inside $\cloc^{(1)}$.  It will be convenient to introduce separate notation for each of these sheets.  We take
 \begin{equation}\begin{split}\Sigma_{\mathbf{\overline{5}},\uparrow}^{(\bar{4}1)} & \subset \cloc^{(4)} \\
 \Sigma_{\mathbf{\overline{5}},\uparrow}^{(4\bar{1})} & \subset \cloc^{(1)}
\end{split}\end{equation}

 The cohomologies \eqref{simplecohomologies} that determine the 4-dimensional spectrum now split in the natural way.  We summarize this as follows
 \begin{equation}\begin{array}{c|cc}
\text{Rep} & \text{Upstairs cohomology} & \text{Downstairs cohomology} \\ \hline
\mathbf{10}^{(4)}_{+1} & h^0(\Sigma_{\mathbf{10},\uparrow}^{(4)},p_{\text{loc}\ast}^*K_{S_2}\otimes {\cal{L}}_{\gamma}^{(4)}|_{\Sigma_{\mathbf{10},\uparrow}}) & h^0(\Sigma_{\mathbf{10},\downarrow}^{(44)},K_{\Sigma_{\mathbf{10},\downarrow}^{(44)}}^{1/2}\otimes \CO(p_{\text{loc}*}\gamma_4|_{\Sigma_{\mathbf{10},\uparrow}^{(44)}})) \\
\mathbf{\overline{10}}^{(4)}_{-1} & h^1(\Sigma_{\mathbf{10},\uparrow}^{(4)},p_{\text{loc}\ast}^*K_{S_2}\otimes {\cal{L}}_{\gamma}^{(4)}|_{\Sigma_{\mathbf{10},\uparrow}}) & h^1(\Sigma_{\mathbf{10},\downarrow}^{(44)},K_{\Sigma_{\mathbf{10},\downarrow}^{(44)}}^{1/2}\otimes \CO(p_{\text{loc}*}\gamma_4|_{\Sigma_{\mathbf{10},\uparrow}^{(44)}})) \\ \hline
\mathbf{\overline{5}}^{(44)}_{+2} & h^0_-(\Sigma_{\mathbf{\overline{5}},\uparrow}^{(44)},p_{\text{loc}}^*K_{S_2}\otimes {\cal{L}}_{\gamma}^{(4)}|_{\Sigma_{\mathbf{\overline{5}},\uparrow}^{(44)}}\otimes \tau {\cal{L}}_{\gamma}^{(4)}|_{\Sigma_{\mathbf{\overline{5}},\uparrow}^{(44)}}) &
h^0(\Sigma_{\mathbf{\overline{5}},\downarrow}^{(44)},K_{\Sigma_{\mathbf{\overline{5}},\downarrow}^{(44)}}^{1/2}\otimes \CO(p_{\text{loc}*}\gamma_4|_{\Sigma_{\mathbf{\overline{5}},\uparrow}^{(44)}})) \\
\mathbf{5}^{(44)}_{-2} & h^1_-(\Sigma_{\mathbf{\overline{5}},\uparrow}^{(44)},p_{\text{loc}}^*K_{S_2}\otimes {\cal{L}}_{\gamma}^{(4)}|_{\Sigma_{\mathbf{\overline{5}},\uparrow}^{(44)}}\otimes \tau {\cal{L}}_{\gamma}^{(4)}|_{\Sigma_{\mathbf{\overline{5}},\uparrow}^{(44)}}) &
h^1(\Sigma_{\mathbf{\overline{5}},\downarrow}^{(44)},K_{\Sigma_{\mathbf{\overline{5}},\downarrow}^{(44)}}^{1/2}\otimes \CO(p_{\text{loc}*}\gamma_4|_{\Sigma_{\mathbf{\overline{5}},\uparrow}^{(44)}})) \\ \hline
\mathbf{\overline{5}}^{(41)}_{-3} & h^0(\Sigma_{\mathbf{\overline{5}},\uparrow}^{(\bar{4}1)},p_{\text{loc}}^*K_{S_2}\otimes {\cal{L}}_{\gamma}^{(4)}|_{\Sigma_{\mathbf{\overline{5}},\uparrow}^{(\bar{4}1)}}\otimes \tau ({\cal{L}}_{\gamma}^{(1)}|_{\Sigma_{\mathbf{\overline{5}},\uparrow}^{(4\bar{1})}})) & h^0(\Sigma_{\mathbf{\overline{5}},\downarrow}^{(41)},K_{\Sigma_{\mathbf{\overline{5}},\downarrow}^{(41)}}^{1/2}\otimes \CO(p_{\text{loc}*}[\gamma_4|_{\Sigma_{\mathbf{\overline{5}},\uparrow}^{(\bar{4}1)}}-\gamma_1|_{\Sigma_{\mathbf{\overline{5}},\uparrow}^{(4\bar{1})}}])) \\
\mathbf{5}^{(41)}_{+3}  & h^1(\Sigma_{\mathbf{\overline{5}},\uparrow}^{(\bar{4}1)},p_{\text{loc}}^*K_{S_2}\otimes {\cal{L}}_{\gamma}^{(4)}|_{\Sigma_{\mathbf{\overline{5}},\uparrow}^{(\bar{4}1)}}\otimes \tau ({\cal{L}}_{\gamma}^{(1)}|_{\Sigma_{\mathbf{\overline{5}},\uparrow}^{(4\bar{1})}})) & h^1(\Sigma_{\mathbf{\overline{5}},\downarrow}^{(41)},K_{\Sigma_{\mathbf{\overline{5}},\downarrow}^{(41)}}^{1/2}\otimes \CO(p_{\text{loc}*}[\gamma_4|_{\Sigma_{\mathbf{\overline{5}},\uparrow}^{(\bar{4}1)}}-\gamma_1|_{\Sigma_{\mathbf{\overline{5}},\uparrow}^{(4\bar{1})}}])) \\
\end{array}\label{cohomsforspectrum}\end{equation}
Net chiralities can again be computed by simple intersections
\begin{equation}\begin{split}n_{\mathbf{10}^{(4)}_{+1}}-n_{\mathbf{\overline{10}}^{(4)}_{-1}} &= \gamma_4\cdot_{\cloc^{(4)}}\Sigma_{\mathbf{10},\uparrow}^{(4)} \\
n_{\mathbf{\overline{5}}^{(44)}_{+2}}-n_{\mathbf{5}^{(44)}_{-2}} &= \gamma_4\cdot_{\cloc^{(4)}}\Sigma_{\mathbf{\overline{5}},\uparrow}^{(44)} \\
n_{\mathbf{\overline{5}}^{(41)}_{-3}}-n_{\mathbf{5}^{(41)}_{+3}} &= \gamma_4\cdot_{\cloc^{(4)}}\Sigma_{\mathbf{\overline{5}},\uparrow}^{(\bar{4}1)} - \gamma_1\cdot_{\cloc^{(1)}}\Sigma_{\mathbf{\overline{5}},\uparrow}^{(4\bar{1})}
\end{split}\end{equation}

For the split \eqref{Clocsplit} we can give a more explicit description of the matter curves that will be useful for the chirality computations of the next section.  We provide the defining equations as well as the class of each curve in the ambient space $\mathbb{P}(\CO\oplus K_{S_2})$
\begin{equation}\begin{array}{c|cc}
\text{Matter Curve} & \text{Equations} & \text{Class} \\ \hline
\Sigma_{\mathbf{10},\uparrow}^{(4)} & U=0 & \sigma_{loc}\cdot(\eta_4-4\cs) \\
& a_4=0 & \\ \hline
\Sigma_{\mathbf{\overline{5}},\uparrow}^{(44)} & a_3 V^2 + \alpha e_1U^2 = 0 & 2\sigma_{loc}\cdot (2\eta_4-5\cs) \\
& a_4V^2 + (a_2+a_3e_0)U^2=0 & +\eta_4\cdot \eta_4 - 3\eta_4\cdot \cs + 2\cs\cdot \cs \\ \hline
\Sigma_{\mathbf{\overline{5}},\uparrow}^{(41)} & e_1V\pm e_0 U = 0 & 2\times \left(\sigma_{loc}+\cs\right)\cdot\left(\eta_4-2\cs\right) \\
& a_4V^2 \mp a_3 VU + a_2U^2 &
\end{array}\label{upstairsmattercurves}\end{equation}

\subsubsection{Chirality from Inherited and Noninherited Fluxes}
\label{subsubsec:chiralitygen}

We close this section by describing the construction of the line bundles $\CL_{\gamma}^{(m)}$ and the computation of the relevant indices that determine the chiral spectrum on each matter curve.  We specify the bundles as in \cite{Donagi:2009ra}, by choosing divisor classes $\gamma_4$ and $\gamma_1$ on $\cloc^{(4)}$.  To ensure that $\CL_{\gamma}^{(m)}$ are honest line bundles these classes must satisfy the quantization rule
\begin{equation}\gamma_m + \frac{r_m}{2}\in H^2(\cloc^{(4)},\mathbb{Z})\label{C4quantizationrule}\end{equation}
while the condition $c_1(V) = c_1(\ploc_{4*}\CL_{\gamma}^{(4)})+c_1(\ploc_{1*}\CL_{\gamma}^{(1)})=0$ requires us to ensure that
\begin{equation}p_{4*}\gamma_4+p_{1*}\gamma_1=0\end{equation}
In models with a 4+1 split spectral cover there are three interesting kinds of $\gamma$'s that can be built from classes inherited from the ambient space $\mathbb{P}(\CO\oplus K_{S_2})$
\begin{equation}\begin{split}\gamma^{(u)} &= \left(4\sigma_{loc}-\eta_4\right)\cdot \cloc^{(4)} \\
\gamma^{(p)} &= \sigma_{loc}\cdot \cloc^{(4)}-\eta_4\cdot\cloc^{(1)} \\
\gamma^{(\rho)} &= \rho\cdot \left[\cloc^{(4)}-4\cloc^{(1)}\right]
\end{split}\label{gammabuildingblocks}\end{equation}
with $\rho$ a divisor class on $S_2$.  Note that this is an overcomplete set since
\begin{equation}\gamma^{(u)} = 4\gamma^{(p)} - \gamma^{(\rho=\eta_4)}\label{gammaurel}\end{equation}
We define $\gamma^{(u)}$ separately because it is a natural generalization of the inherited flux of the generic model \eqref{gammaugeneric} but will use it sparingly  when parametrizing the set of flux choices.

To have any hope of globally extending this data to $G$-fluxes on the full Calabi-Yau 4-fold, we should choose $\rho$ to be inherited from $B_3$.  We can build our net $\gamma$ from a linear combination of the classes in \eqref{gammabuildingblocks} provided we take into account the quantization rule \eqref{C4quantizationrule}.  The ramification divisor $r_1$ is trivial because the covering $p_1$ is smooth and 1-1 but this is not generally true for $r_4$
\begin{equation}r_4 = \left(2\sigma_{loc}+\eta_4\right)|_{\cloc^{(4)}}\end{equation}
A generic combination of the $\gamma^{(a)}$ in \eqref{gammabuildingblocks} that satisfies \eqref{C4quantizationrule}, then, takes the form
\begin{equation}\gamma_{\text{inherited}} = -\frac{1}{2}\gamma^{(u)}+p\gamma^{(p)}+\gamma^{(\rho)}\label{gammainherited}\end{equation}
for $p$ an integer and $\rho$ an integral class on $S_2$.  The relation \eqref{gammaurel} means that it would be redundant to make the coefficient of $\gamma^{(u)}$ arbitrary in \eqref{gammainherited}.

It is useful to compute the intersections of $\gamma^{(u)}$, $\gamma^{(p)}$, and $\gamma^{(\rho)}$ with the `upstairs' matter curves \eqref{upstairsmattercurves} since this determines the chiral spectrum induced by each.  Expressing the results as intersections in $S_2$ we find
\begin{equation}\begin{array}{c|ccc}
\cdot_{\cloc} & \gamma^{(u)} & \gamma^{(p)} & \gamma^{(\rho)} \\ \hline
\Sigma_{\mathbf{10},\uparrow}^{(4)} & -\eta_4(\eta_4-4\cs) & -\cs(\eta_4-4\cs) & \rho(\eta_4-4\cs) \\
\Sigma_{\mathbf{\overline{5}},\uparrow}^{(44)} & -2\cs(\eta_4-4\cs) & \eta_4^2 - 7\cs\eta_4 + 12\cs^2 & 2\rho(2\eta_4-5\cs) \\
\Sigma_{\mathbf{\overline{5}},\uparrow}^{(41)} & -\eta_4^2 + 6\cs\eta_4 - 8\cs^2 & -\eta_4^2 + 6\eta_4\cs - 8\cs^2 & -3\rho(\eta_4-3\cs)
\end{array}\label{gammachiralities}\end{equation}

For the models that we build in section \ref{sec:FEnriquesLocal}, this collection of fluxes will not be enough to engineer 3 chiral generations.  We therefore consider a tuning of $\cloc^{(4)}$ so that it contains additional non-inherited divisors.  More specifically, we set
\begin{equation}\begin{split}
a_4 &= c_4 h_0 - d_3 h_2 \\
a_3 &= c_3 h_0 + d_3 h_1 - d_2 h_2 \\
a_2 &= c_4h_0 + d_2h_1 \\
\alpha &= c_1 h_0
\end{split}\label{tuningchoice}\end{equation}
so that $\cloc^{(4)}$ takes the form
\begin{equation}\cloc^{(4)}:\,\,\, h_0\left[c_4V^2(V^2+U^2)+c_3V^3U+c_1U^3(e_1V-e_0U)\right] + V^2(h_1U-h_2V)(d_2U+d_3V)=0\label{C4tuned}\end{equation}
The new objects that we have introduced are sections of the indicated bundles on $S_2$
\begin{equation}\begin{array}{c|c}
\text{Section} & \text{Bundle} \\ \hline
h_0 & \CO(\xi) \\
h_1 & \CO(\chi) \\
h_2 & \CO(\chi-\cs) \\
d_m & \CO(\eta_4-m\cs - \chi) \\
c_m & \CO(\eta_4-m\cs - \xi)
\end{array}\label{tuningparams}\end{equation}
where $\chi$ and $\xi$ are two divisor classes on $S_2$ that we choose.  When $\cloc^{(4)}$ takes the form \eqref{C4tuned}, we introduce some new divisors including two particularly interesting ones
\begin{equation}\begin{split}D_1\,\,&:\,\, h_0 = h_1U-h_2V=0 \\
D_2\,\,&:\,\, h_0 = d_2U+d_3V=0
\end{split}\label{interestingdivisors}\end{equation}
The intersections of these divisors with our `upstairs' matter curves are trickier to compute since they are not complete intersections with $\cloc^{(4)}$ in $\mathbb{P}(\CO\oplus K_{S_2})$.  To proceed, we write the equations for the matter curves explicitly in the parametrization \eqref{tuningparams}
\begin{equation}\begin{array}{c|c}
\text{Matter Curve} & \text{Equations} \\ \hline
\Sigma_{\mathbf{10},\uparrow}^{(4)} & U=0 \\
& c_4h_0 - d_3h_2=0 \\ \hline
\Sigma_{\mathbf{\overline{5}},\uparrow}^{(44)} & \left[h_0(c_4+c_3e_0)+h_1(d_2+d_3e_0)-d_2e_0h_2\right]U^2 + \left(c_4h_0-d_3h_2\right)V^2 = 0 \\
& \left(c_3h_0 + d_3h_1 - d_2h_2\right)V^2 + c_1h_0e_1 U^2=0 \\ \hline
\Sigma_{\mathbf{\overline{5}},\uparrow}^{(\overline{4}1)} & e_1V - e_0U = 0 \\
& h_0\left[c_3UV + c_4(U^2+V^2)\right] + (d_2U + d_3V)(h_1U - h_2V)=0
\end{array}\end{equation}
where we only write the equation for the component $\Sigma_{\mathbf{\overline{5}},\uparrow}^{(\overline{4}1)}$ of $\Sigma_{\mathbf{\overline{5}},\uparrow}^{(41)}$ that sits inside $\cloc^{(4)}$.  We can now compute the intersections with $D_1$ and $D_2$ and express them as complete intersections inside the ambient space $\mathbb{P}(\CO\oplus K_{S_2})$
\begin{equation}\begin{array}{c|cc}
\cdot_{\cloc^{(4)}} & D_1 & D_2 \\ \hline
\Sigma_{\mathbf{10},\uparrow}^{(4)} & [U]\cdot [h_0]\cdot [h_2] & [U]\cdot [h_0]\cdot [d_3] \\
\Sigma_{\mathbf{\overline{5}},\uparrow}^{(44)} & [h_0]\cdot [h_1U - h_2V]\cdot [d_2h_2-d_3h_1] & [h_0]\cdot [d_2U+d_3V]\cdot [d_2h_2-d_3h_1] \\
\Sigma_{\mathbf{\overline{5}},\uparrow}^{(41)} & [h_0]\cdot [V-e_0U]\cdot [h_1U-h_2V] & [h_0]\cdot [V-e_0U]\cdot [d_2U+d_3V]
\end{array}\label{noninherspec}\end{equation}
The pair $D_1$ and $D_2$ is particularly nice because the difference
\begin{equation}\gamma_{\text{noninherited}} = D_1 - D_2\label{gammanoninherited}\end{equation}
satisfies
\begin{equation}\ploc_{4*}\gamma_{\text{noninherited}}=0\end{equation}
so that it represents a very simple non-inherited flux.  Its contribution to the chiral spectrum is easily computed
\begin{equation}\begin{array}{c|c}
\cdot_{\cloc} & \gamma_{\text{noninherited}} \\ \hline
\Sigma_{\mathbf{10},\uparrow}^{(4)} & \xi(2\chi+2\cs-\eta_4) \\
\Sigma_{\mathbf{\overline{5}},\uparrow}^{(44)} & 0 \\
\Sigma_{\mathbf{\overline{5}},\uparrow}^{(41)} & \xi(2\chi+2\cs - \eta_4)
\end{array}\label{noninherspectrum}\end{equation}
In section \ref{subsubsec:extendingnoninherited} we will explicitly construct a global $G$-flux that extends $\gamma_{\text{noninherited}}$ whenever the classes $\xi$ and $\chi$ are inherited from $B_3$.

\section{F-Enriques: The Local Model}
\label{sec:FEnriquesLocal}

We now proceed to construct a local model for an F-theory $SU(5)_{\rm GUT}$ GUT where the GUT divisor $S_{\rm GUT}$ is an Enriques surface and GUT-breaking is achieved through a discrete Wilson line.

\subsection{Generalities}

Our construction strategy is to realize the model as a quotient of one where the GUT divisor is a K3 surface.  Models of this type are special because they correspond to choosing Higgs bundles on the Enriques that can be lifted to its double cover.  The global completions of these models will be Calabi-Yau 4-folds that can be obtained as global guotients and $G$-fluxes that lift to the covering space.  While we found it difficult to find other suitable examples of elliptically fibered Calabi-Yau 4-folds with $A_4$ singularities along an Enriques surface, it may be possible to alleviate some of the problems we encounter in the Higgs spectrum by relaxing this condition and considering more general Higgs bundles and their completions{\footnote{The conclusions about adjoint matter on $S_{\rm GUT}$ in section \ref{subsec:bulkspectrum}, however, will remain unaffected.}}.

We start, then, with a parent theory where $S_2=K3$ and the Higgs bundle spectral cover sits inside $\mathbb{P}(\CO\oplus K_{S_2}) = \mathbb{P}^1\times S_2=\mathbb{P}^1\times K3$ and is given by
\begin{equation}\cloc:\,\, b_0 U^5 + b_2 U^3 V^2 + b_3 U^2 V^3 + b_4 UV^4 + b_5 V^5\label{clocK3}\end{equation}
Because $S_2=K3$, $\cs=0$ so that the objects appearing here are sections of the bundles
\begin{equation}\begin{array}{c|c}
\text{Section} & \text{Bundle} \\ \hline
U,V & \CO(\sigma_{loc}) \\
b_m & \eta
\end{array}\end{equation}
where $\sigma_{loc}$ is a hyperplane of the $\mathbb{P}^2$ and $\eta$ is related to the normal bundle of $S_2=K3$ inside $B_3$ by
\begin{equation}N_{S_2/B_3}=\CO(\eta)\end{equation}

Once we have a K3-based model, we specify a freely acting $\zenriques$ involution on the K3 and construct a spectral cover $\cloc$ that is $\zenriques$-invariant.  This allows us to take a quotient by the combination of the geometric $\zenriques$ and the $\mathbb{Z}_2$ center of $U(1)_Y$ to produce a model whose GUT divisor is an Enriques surface $S_{\rm GUT}=K3/\zenriques=$ where $SU(5)_{\rm GUT}$ can be broken to $U(1)_Y$ by a discrete Wilson line on the torsion cycle.

How do we ensure that $\cloc$ is invariant?  First, we need to specify how the $\zenriques$ acts on the $\mathbb{P}^1$ factor of the ambient $\mathbb{P}^1\times K3$ ins which $\cloc$ is defined.
The spectral cover of the daughter theory, where $S_{\rm GUT}$ is an Enriques surface, will be a divisor in the space $\mathbb{P}(\CO\oplus K_{\text{Enriques}})$ that is also given by an equation of the form \eqref{clocK3}.  The $\mathbb{P}^1$ fiber coordinate $V_{\text{daughter}}$ of the daughter, though, is not a section of $\sigma_{loc}$ but rather gets twisted by the anti-canonical bundle of the Enriques.  This tells us that the coordinate $V$ in the parent must be $\zenriques$-odd.  The $U$ coordinate is $\zenriques$-even so the $\zenriques$ acts on $\mathbb{P}^1$ as a simple reflection with two fixed points.

Given this description of how $\zenriques$ acts on $\mathbb{P}^1\times S_2$, we can ensure that $\cloc$ is $\zenriques$-invariant by choosing the $b_m$ to transform as
\begin{equation}\zenriques:\,\, b_m\rightarrow (-1)^m b_m\end{equation}
in order to ensure that $\cloc$ is truly $\zenriques$-invariant.
When $\cloc$ is built in this way, we are free to quotient the K3 model to build our Enriques model with a discrete Wilson line.
The easiest way to determine the spectrum of the quotient theory is to compute $\zenriques$-graded cohomologies in the parent.  Fields that carry even/odd hypercharge are counted by the corresponding even/odd parent cohomology groups.  Because the geometric $\zenriques$ acts freely on $S_2=K3$, we don't need to worry about any subtleties related to fixed points.  We usually work in the parent theory so we will use $S_2$ to denote the GUT divisor of that theory, $S_2=K3$, and $B_3$ to denote the base 3-fold in which it sits.  We reserve $S_{\rm GUT}$ for the Enriques surface.

\subsection{Bulk Spectrum}
\label{subsec:bulkspectrum}

Throughout this paper, our attention is mainly devoted to the spectrum of $\mathbf{10}$'s and $\mathbf{\overline{5}}$'s that localize along matter curves, giving rise to the three chiral generations of the MSSM and the Higgs doublets.  Before getting to those, however, we should first address the spectrum of fields that extend along the entire GUT divisor $S_{\rm GUT}${\footnote{While this paper was being completed, bulk matter in models with Wilson lines was addressed in a revision of \cite{Donagi:2008ca}.  We thank M.~Wijnholt for bringing this to our attention and for a helpful discussion about this topic.}}.
These descend from the $SU(5)_{\rm GUT}$ adjoint and, in a model with generic GUT divisor $S_{\rm GUT}$ and hypercharge bundle $L_Y$, are counted by the following cohomology groups \cite{Donagi:2008ca,Beasley:2008dc,Beasley:2008kw,Donagi:2008kj}
\begin{equation}\begin{array}{cc|c}
\text{Representation} & \text{Type of multiplet} & \text{Cohomology group} \\ \hline
(\mathbf{8},\mathbf{1})_0 & \text{Vector} & H^2(S_{\rm GUT},K_{S_{\rm GUT}}) \\
(\mathbf{1},\mathbf{3})_0 & \text{Vector} & H^2(S_{\rm GUT},K_{S_{\rm GUT}}) \\
(\mathbf{1},\mathbf{1})_0 & \text{Vector} & H^2(S_{\rm GUT},K_{S_{\rm GUT}}) \\ \hline
(\mathbf{8},\mathbf{1})_0 & \text{Chiral} & H^0(S_{\rm GUT},K_{S_{\rm GUT}}) \oplus H^1(S_{\rm GUT},K_{S_{\rm GUT}}) \\
(\mathbf{1},\mathbf{3})_0 & \text{Chiral} & H^0(S_{\rm GUT},K_{S_{\rm GUT}}) \oplus H^1(S_{\rm GUT},K_{S_{\rm GUT}}) \\
(\mathbf{1},\mathbf{1})_0 & \text{Chiral} & H^0(S_{\rm GUT},K_{S_{\rm GUT}}) \oplus H^1(S_{\rm GUT},K_{S_{\rm GUT}}) \\ \hline
(\mathbf{3},\mathbf{2})_{-5/6} & \text{Vector} & H^0(S_{\rm GUT},L_Y^{-1}) \\
(\mathbf{\overline{3}},\mathbf{2})_{+5/6} & \text{Vector} & H^0(S_{\rm GUT},L_Y) \\ \hline
(\mathbf{3},\mathbf{2})_{-5/6} & \text{Chiral} & H^1(S_{\rm GUT},L_Y) + H^2(S_{\rm GUT},L_H^{-1}) \\
(\mathbf{\overline{3}},\mathbf{2})_{+5/6} & \text{Chiral} & H^1(S_{\rm GUT},L_Y^{-1}) + H^2(S_{\rm GUT},L_Y)
\end{array}\end{equation}
An optimal spectrum would be to keep the $SU(3)\times SU(2)\times U(1)_Y$ adjoint vectors, which is automatic since $h^2(S_{\rm GUT},K_{S_{\rm GUT}}) = h^0(S_{\rm GUT},\CO_{S_{\rm GUT}})=1$, and remove everything else.  If $L_Y$ is a flat  bundle, though, it is easy to see that this can never happen.  Requiring the chiral adjoints of $SU(3)\times SU(2)\times U(1)_Y$ to be absent means that we must have
\begin{equation}h^1(S_{\rm GUT},\CO_{S_{\rm GUT}})=h^2(S_{\rm GUT},\CO_{S_{\rm GUT}})=0\end{equation}
which means that
\begin{equation}\int_{S_{\rm GUT}}\text{Td}(TS_{\rm GUT}) = 1\end{equation}
The holomorphic Euler character of any flat bundle on $S_{\rm GUT}$ is equivalent to its Todd genus, though
\begin{equation}c_1(L_Y)=0\quad\implies\quad \chi(S_2,L_Y) = \int_{S_{\rm GUT}}\text{Td}(TS_{\rm GUT})\end{equation}
so we have that
\begin{equation}\chi(S_{\rm GUT},L_Y) = h^0(S_{\rm GUT},L_Y)-h^1(S_{\rm GUT},L_Y)+h^2(S_{\rm GUT},L_Y) = 1\end{equation}
Since all $h^m(S_{\rm GUT},L_Y)$ cannot be vanishing, we are guaranteed to get some vector or chiral $(\mathbf{3},\mathbf{2})_{-5/6}$'s and $(\mathbf{\overline{3}},\mathbf{2})_{+5/6}$'s.  For the Enriques model with discrete Wilson line, $S_{\rm GUT}=\text{Enriques} = S_2/\zenriques$, $L_Y$ is the unique non-trivial flat line bundle on $S_{\rm GUT}$, and we have that
\begin{equation}\begin{split}
h^0(S_{\rm GUT},\CO_{S_{\rm GUT}}) &= h^{0,+}(K3,\CO_{K3}) = 1 \\
h^1(S_{\rm GUT},\CO_{S_{\rm GUT}}) &= h^{1,+}(K3,\CO_{K3}) = 0 \\
h^2(S_{\rm GUT},\CO_{S_{\rm GUT}}) &= h^{2,+}(K3,\CO_{K3}) = 0 \\
h^0(S_{\rm GUT},L_Y^{\pm 1}) &= h^{0,-}(K3,\CO_{K3}) = 0 \\
h^1(S_{\rm GUT},L_Y^{\pm 1}) &= h^{1,-}(K3,\CO_{K3}) = 0 \\
h^2(S_{\rm GUT},L_Y^{\pm 1}) &= h^{2,-}(K3,\CO_{K3}) = 1
\end{split}\end{equation}
where the $\pm$ refers to the $\zenriques$ grading specified by the Enriques involution.  We see that, in addition to vector multiplets for the MSSM gauge group, we get a vector-like pair of chiral multiplets in the representation $(\mathbf{3},\mathbf{2})_{-5/6}\oplus (\mathbf{\overline{3}},\mathbf{2})_{+5/6}$.  We emphasize that the presence of some kind of vector-like exotic matter is not a specific issue with this Enriques model but rather a general property of any model that breaks $SU(5)_{\rm GUT}\rightarrow SU(3)\times SU(2)\times U(1)_Y$ with a flat $U(1)_Y$ bundle on a holomorphic surface $S_{\rm GUT}$.

\subsection{Building the K3: some global issues}

Before delving into the details of local model building, we need to be more specific about what precisely the 3-fold $B_3$ is and how $S_2=K3$ sits inside it.  This is because a necessary condition for the local model to admit a global completion is that the objects $b_m$ in \eqref{clocK3} are inherited from holomorphic sections on $B_3$.  Further, if we need to impose extra conditions like factorization of $\cloc$, these too must be expressible in terms of sections that inherit from $B_3$.  In this way, the nature of $B_3$ determines the toolbox of holomorphic sections that are available to us for building our local model.

Finding a suitable $B_3$ is somewhat tricky because we need the $\zenriques$ involution on $S_2=K3$ to globally extend.  More specifically, what we need is a 3-fold $B_3$ with the following properties
\begin{itemize}
\item $B_3$ must be able to serve as the base of an elliptically fibered Calabi-Yau 4-fold with section that exhibits an $A_4$ singularity along an effective anti-canonical divisor $S_2=K3$

\item $B_3$ must admit a $\zenriques$ involution that acts freely on $S_2=K3$
\end{itemize}
From \eqref{bmbundles} we see that the first of these amounts to the requirement that
\begin{equation}h^0(B_3,\CO_{B_3}([6-m]\cb - [5-m]S_2))> 0\text{  for }m=0,2,3,4,5\end{equation}
Since $\cb=S_2$ for any anti-canonical divisor $S_2$ these bundles are the same for all $m$.  Requiring enough freedom to make distinct choices for all of the $b_m$'s which are independent from the defining equation for $S_2$ itself, we actually need (recall that we must specify 5 $b_m$'s as well as a holomorphic section $z$ whose vanishing defines the GUT divisor)
\begin{equation}h^{0,+}(B_3,\CO_{B_3}(S_2))\ge 4\qquad h^{0,-}(B_3,\CO_{B_3}(S_2))\ge 2\end{equation}

One simple way to construct a $B_3$ of this type is as follows.  We start with $\mathbb{P}^5$ and denote its homogeneous coordinates by $[u_1,u_2,u_3,v_1,v_2,v_3]$.  We then define the $\zenriques$ action by
\begin{equation}\zenriques:\,\,\, u_i\rightarrow u_i\qquad v_i\rightarrow -v_i\qquad i=1,2,3\label{Z2P5}\end{equation}
The fixed point locus of $\zenriques$ is a disjoint union of two $\mathbb{P}^2$'s, which we denote $\mathbb{P}^2_u$ and $\mathbb{P}^2_v$
\begin{equation}\mathbb{P}^2_u:\,\, [u_1,u_2,u_3,0,0,0]\qquad \mathbb{P}^2_v:\,\,[0,0,0,v_1,v_2,v_3]\end{equation}
We now realize a K3 surface in $\mathbb{P}^5$ as the intersection of 3 $\zenriques$-invariant quadrics.  To be specific, we can choose these as
\begin{equation}Q_2^{(a)} = f_2^{(a)}(u_1,u_2,u_3)- g_2^{(a)}(v_1,v_2,v_3)\qquad a=1,2,3\label{firstquadrics}\end{equation}
for $f_2^{(a)}$ and $g_2^{(a)}$ quadrics in the $u_i$ and $v_i$, respectively.  As a surface, we expect our K3 to generically miss the locus of fixed points, which itself is a disjoint union of surfaces in $\mathbb{P}^5$.  The choice \eqref{firstquadrics} manifestly has this property provided $f_2^{(1)}=f_2^{(2)}=f_2^{(3)}=0$ have no solutions in $\mathbb{P}^2$ and similar for $g_2^{(1)}=g_2^{(2)}=g_2^{(3)}=0$.  Provided the $f_m^{(i)}$ and $g_n^{(j)}$ are chosen suitably generically, then, the $\zenriques$ action on $\mathbb{P}^5$ descends to the desired freely acting $\zenriques$ on our K3.  Putting the K3 inside a 3-fold $B_3$ is also quite easy: we simply define $B_3$ to be the intersection of 2 of the quadrics.  It is easy to check that
\begin{equation}h^0_+(B_3,\CO_{B_3}(S_2)) = 10\qquad h^0_-(B_3,\CO_{B_3}(S_2)) = 9\end{equation}
so that there are more than enough available sections to construct a Tate model \eqref{genericTate}.  Note that this $B_3$ will meet the fixed point locus in 8 isolated points but these will be away from $S_2=K3$.  As we discuss in more detail in section \ref{subsubsec:geometrictadpole}, these fixed points will correspond to the locations of O3 planes \cite{Denef:2005mm,Collinucci:2008zs,Collinucci:2009uh,Blumenhagen:2009up} that we do not need to worry about apart from their contribution to the 3-brane tadpole.

\subsubsection{Increasing $\text{Pic}(B_3)$}

This simple setup, with $B_3$ as an intersection of 2 quadrics and $S_2$ the restriction of a third quadric, has an important problem.  The Picard number of $B_3$ is 1; the only line bundles on $B_3$ descend from multiples of the hyperplane bundle $\CO_{\mathbb{P}^5}(H)$ of the $\mathbb{P}^5$.  Consequently, the only holomorphic sections on $S_2$ that are inherited from $B_3$ are sections of line bundles that also descend from $\CO_{\mathbb{P}^5}(H)$.  Since
\begin{equation}B_3 = (2H)^2\qquad S_2 = (2H)^3\end{equation}
and
\begin{equation}H\cdot_{S_2}H = 8\end{equation}
we see that any pair of holomorphic sections on $S_2$ that are inherited from $B_3$ will simultaneously vanish at $8n$ points for some integer $n$.  This is problematic because chirality computations reduce to complete intersections in $S_2$ of classes that descend from $B_3$.  In the daughter Enriques theory we will need to get multiples of 3 in order to generate 3 chiral generations.  This corresponds to multiples of 6 in the parent K3-based theory.  Multiples of 6 are unattainable with a realization of K3 where intersections of inherited classes are all multiples of 8.

To deal with this, we need to increase the Picard number of $B_3$ to give us more bundles and, correspondingly, a greater variety of holomorphic sections to choose from.  We do this by introducing nodal singularities into some of the quadrics of \eqref{firstquadrics}.  Upon resolving the resulting singularities of $S_2$ and $B_3$, we obtain a new 3-fold with a larger Picard group.

To make this explicit, we choose the quadrics that define $S_2$ to take the form
\begin{equation}\begin{split}Q_2^{(1)} &= u_1 f_1^{(1)}(u_1,u_2,u_3) - v_1 g_1^{(1)}(v_1,v_2,v_3) \\
Q_2^{(2)} &= u_2 f_1^{(2)}(u_1,u_2,u_3) - v_2 g_1^{(2)}(v_1,v_2,v_3) \\
Q_2^{(3)} &= f_2^{(3)}(u_1,u_2,u_3) - g_2^{(3)}(v_1,v_2,v_3)
\end{split}\end{equation}
where each $f_i^{(m)}$ and $g_j^{(m)}$ is of degree $i,j$ in the indicated variables.  We then blow up $\mathbb{P}^5$ twice along
\begin{itemize}
\item $u_1=v_1=0$
\item $u_2=v_2=0$
\end{itemize}
The result is a simple toric space $X_5$ that we describe by presenting the $P|Q$ matrix with $P$ listing the vertices of the toric fan and $Q$ the GLSM charges.  We illustrate the coordinate names that are associated to each vertex as well as the divisor class
\begin{equation}\begin{pmatrix}\tilde{u}_1 \\ \tilde{v}_1 \\ \delta_1 \\ u_2 \\ v_2 \\ \delta_2 \\ u_3 \\ v_3\end{pmatrix}
\leftrightarrow
\left(\begin{array}{ccccc|ccc}
1 & 0 & 0 & 0 & 0 & 1 & -1 & 0 \\
0 & 1 & 0 & 0 & 0 & 1 & -1 & 0 \\
1 & 1 & 0 & 0 & 0 & 0 & 1 & 0  \\
0 & 0 & 1 & 0 & 0 & 1 & 0 & -1 \\
0 & 0 & 0 & 1 & 0 & 1 & 0 & -1 \\
0 & 0 & 1 & 1 & 0 & 0 & 0 & 1 \\
0 & 0 & 0 & 0 & 1 & 1 & 0 & 0 \\
-1&-1&-1&-1&-1 &1  & 0 & 0
\end{array}\right)
\leftrightarrow
\begin{pmatrix}
H-E_1 \\ H-E_1 \\ E_1 \\ H-E_2 \\ H-E_2 \\ E_2 \\ H \\ H\end{pmatrix}
\label{W5data}\end{equation}
The divisor class $H$ is inherited from the hyperplane of $\mathbb{P}^5$ while $E_1$ and $E_2$ are the exceptional divisors of our two blow-ups.  The relation of the new holomorphic sections introduced above to the old homogeneous coordinates of $\mathbb{P}^5$ are
\begin{equation}\begin{split}
u_1 &= \tilde{u}_1\delta_1 \\
v_1 &= \tilde{v}_1\delta_1 \\
u_2 &= \tilde{u}_2\delta_2 \\
v_2 &= \tilde{v}_2\delta_2
\end{split}\end{equation}
and the generators of the Stanley-Reisner ideal are
\begin{equation}SR = (\tilde{u}_1\tilde{v}_1,\tilde{u}_2\tilde{v}_2,\delta_1\delta_2 u_3v_3)\end{equation}
The $\zenriques$ action on $\mathbb{P}^5$ lifts to an action on $X_5$ where the new holomorphic sections transform as
\begin{equation}\tilde{u}_j\rightarrow \tilde{u}_j\quad \tilde{v}_j\rightarrow -\tilde{v}_j\quad \delta_j\rightarrow\delta_j\label{Z2W5}\end{equation}
After the blow-ups, we replace the nodal quadrics $Q_2^{(1)}$ and $Q_2^{(2)}$ with their proper transforms,  which we do not notationally distinguish from the original ones
\begin{equation}\begin{split}
Q_2^{(1)} &= \tilde{u}_1 f_1^{(1)}(\tilde{u}_1\delta_1,\tilde{u}_2\delta_2,u_3) - \tilde{v}_1 g_1^{(1)}(\tilde{v}_1\delta_1,\tilde{v}_2\delta_2,v_3) \\
Q_2^{(2)} &= \tilde{u}_2 f_1^{(2)}(\tilde{u}_1\delta_1,\tilde{u}_2\delta_2 ,u_3) - \tilde{v}_1 g_1^{(2)}(\tilde{v}_1\delta_1,\tilde{v}_2\delta_2,v_3)
\end{split}\label{firsttwoquadrics}\end{equation}
These are divisors in the classes $2H-E_1$ and $2H-E_2$, respectively.  Our K3 surface is the intersection of these with $Q_2^{(3)}$ and hence will be in the class
\begin{equation}S_2 = (2H-E_1)(2H-E_2)(2H)\end{equation}
We obtain a 3-fold $B_3$ containing $S_2=K3$ by intersecting two of these three divisors in $X_5$.  In the rest of this paper we make the explicit choice
\begin{equation}B_3=(2H-E_2)(2H)\end{equation}
so that, as a divisor in $B_3$, $S_2$ is in the class $2H-E_1$.  We can count the even and odd holomorphic sections of $\CO_{B_3}(S_2)=\CO_{B_3}(2H-E_1)$
\begin{equation}h^0_+(B_3,\CO_{B_3}(S_2))=6\qquad h^0_-(B_3,\CO_{B_3}(S_2))=5
\end{equation}
to see explicitly that there is still enough freedom to distinctly choose all of the $b_m$'s that we need.  It is also easy to determine the intersection table in $S_2$ of the divisors that are inherited from $B_3$
\begin{equation}\begin{array}{c|ccc}
\cdot_{S_2} & H & E_1 & E_2 \\ \hline
H & 8 & 4 & 4 \\
E_1 & 4 & 0 & 2 \\
E_2 & 4 & 2 & 0
\end{array}\label{S2intersectiontable}\end{equation}
The result $E_1\cdot_{S_2}E_2=2$ provides us with a glimmer of hope.  If we can craft the spectral data of the local model carefully enough, we can use this fact to help us achieve the equivalent of 6 chiral generations in the parent theory and, correspondingly, 3 chiral generations in the daughter.

We close this subsection by noting that one could easily add another bundle to $B_3$ by blowing up along $u_3=v_3=0$ and choosing the third quadric $Q_2^{(3)}$ to have the form $u_3 f_1^{(3)}(u_1,u_2,u_3)=v_3 g_1^{(3)}(v_1,v_2,v_3)$.  In the end, this freedom buys us very little in our model building efforts so we do not introduce this extra complexity.  The restriction of the exceptional divisor under this blow-up, $E_3$, would introduce another degree of freedom for specifying $\rho$ \eqref{therhodef} but it enters in the same way as $E_2$ in essentially all computations and we find, at the end of the day, that the $E_2$ part of $\rho$ cannot be introduced without overshooting the D3-brane tadpole{\footnote{More specifically, if we performed this additional blow-up and replaced $\rho$ in \eqref{therhodef} with $\rho= aH + bE_1 + c E_2 + d E_3$ then we would simply modify \eqref{theabcpconst} according to $c\rightarrow c+d$.  Further, it turns out that the D3-brane tadpole in \eqref{fluxD3} is modified according to $c^2\rightarrow c^2+d^2$ so avoiding overshoot, which requires $c=0$, also requires $d=0$.}}.

\subsection{Local Model: Spectral Cover, Flux, and Chiral Spectrum}
\label{subsec:EnriquesLocalSpec}

We now turn to the construction of the spectral cover $\cloc$ and flux $\gamma$ of our local model.

\subsubsection{Spectral Cover and Matter Curves}

Our starting point is a 4+1 split spectral cover of essentially the form \eqref{Clocsplit}
\begin{equation}\cloc^{(4)}:\,\,\, \left(a_4V^4 + a_3 V^3U + a_2V^2U^2 + \alpha [e_1 V-e_0U]\right)\left(e_1 V+e_0U\right)\end{equation}
For our particular choices of $S_2=K3$ and $B_3$, the objects appearing here are sections of the indicated bundles on $\mathbb{P}(\CO\oplus K_{S_2})=\mathbb{P}^1\times S_2$
\begin{equation}\begin{array}{c|c}
\text{Section} & \text{Bundle} \\ \hline
U & \CO(\sigma_{loc}) \\
V & \CO(\sigma_{loc}) \\
a_m & \CO(2H-E_1)\ \\
\alpha & \CO(2H-E_1) \\
e_m & \CO
\end{array}\end{equation}
For this spectral data to descend to the $\zenriques$ quotient, we will require the $a_m$ to be odd for $m$ odd
\begin{equation}\zenriques :a_m\mapsto (-1)^ma_m\label{amcharges}\end{equation}
and define the action on $e_1$ to be $e_1\rightarrow -e_1$ corresponding to choosing $e_1$ in the daughter theory to be a local section of a twist two line bundle that lifts to the trivial bundle upstairs.

What must we require in the parent K3-based theory to produce a model with 3 chiral generations in the quotient?  First, recall the $U(1)$ charge assignments for our matter curves
\begin{equation}\begin{array}{c|c}\text{Matter Curve} & U(1)\text{ charge} \\ \hline
\Sigma_{\mathbf{10},\uparrow}^{(4)} & +1 \\
\Sigma_{\mathbf{\overline{5}},\uparrow}^{(44)} & +2 \\
\Sigma_{\mathbf{\overline{5}},\uparrow}^{(41)} & -3
\end{array}\end{equation}
where we indicate the charges for $\mathbf{10}$'s and $\mathbf{\overline{5}}$'s which are related to the charges for $\mathbf{\overline{10}}$'s and $\mathbf{5}$'s through multiplication by $-1$.  The 3 chiral generations of $\mathbf{10}_M$'s must sit on $\Sigma_{\mathbf{10},\uparrow}^{(4)}$.  For the 3 chiral generations of $\mathbf{\overline{5}}_M$'s and the Higgs doublets from $\mathbf{5}_H\oplus\mathbf{\overline{5}}_H$, however, we in principle have a choice of two matter curves where each could live.  This choice is unique, though, if  we want to allow top and bottom Yukawa couplings
\begin{equation}\mathbf{10}_M\times\mathbf{10}_M\times\mathbf{5}_H\quad+\quad\mathbf{10}_M\times\mathbf{\overline{5}}_M\times\mathbf{\overline{5}}_H\end{equation}
We require 3 chiral generations of $\mathbf{\overline{5}}_M$'s on $\Sigma_{\mathbf{\overline{5}},\uparrow}^{(41)}$ and the pair of Higgs doublets on $\Sigma_{\mathbf{\overline{5}},\uparrow}^{(44)}$.  We summarize this as
\begin{equation}\begin{array}{c|c}
SU(5)_{\rm GUT}\text{ Multiplet} & \text{Matter Curve} \\ \hline
\mathbf{10}_M & \Sigma_{\mathbf{10},\uparrow}^{(44)} \\
\mathbf{\overline{5}}_M & \Sigma_{\mathbf{\overline{5}},\uparrow}^{(41)} \\
\mathbf{5}_H\oplus\mathbf{\overline{5}}_H & \Sigma_{\mathbf{\overline{5}},\uparrow}^{(44)}
\end{array}\end{equation}

If we are interested in obtaining the right chiral spectrum on each matter curve, then, we require the requisite indices to be $+3$ on the $\Sigma_{\mathbf{10},\uparrow}^{(4)}$ and $\Sigma_{\mathbf{\overline{5}},\uparrow}^{(41)}$ matter curves in the quotient theory and $0$ on $\Sigma_{\mathbf{\overline{5}},\uparrow}^{(44)}$.  The $\zenriques$ acts freely on our $S_2=K3$, though, so this corresponds to the following requirement in the parent
\begin{equation}\begin{array}{c|c}
\text{Matter Curve, }\Sigma & \gamma\cdot_{\cloc}\Sigma \\ \hline
\Sigma_{\mathbf{10}}^{(4)} & 6 \\
\Sigma_{\mathbf{\overline{5}},\uparrow}^{(44)} & 0 \\
\Sigma_{\mathbf{\overline{5}},\downarrow}^{(41)} & 6
\end{array}\label{desiredparentspectrum}\end{equation}

\subsubsection{Fluxes and Chiral Spectrum}

To see how we can achieve this, let us recall that the inherited fluxes can be written as \eqref{gammainherited}
\begin{equation}\gamma_{\text{inherited}} = -\frac{1}{2}\gamma^{(u)}+p\gamma^{(p)}+\gamma^{(\rho)}\end{equation}
where the intersections of $\gamma^{(a)}$ with `upstairs' matter curves are determined by \eqref{gammachiralities} and \eqref{S2intersectiontable} for our choices of $S_2=K3$ and $B_3$
\begin{equation}\begin{array}{c|ccc}
\cdot_{\cloc} & \gamma^{(u)} & \gamma^{(p)} & \gamma^{(\rho)} \\ \hline
\Sigma_{\mathbf{10},\uparrow}^{(4)} & -(2H-E_1)^2 =-16 & 0 & \rho\cdot_{S_2}(2H-E_1) \\
\Sigma_{\mathbf{\overline{5}},\uparrow}^{(44)} & 0 & (2H-E_1)^2=16 & 4\rho\cdot_{S_2}(2H-E_1) \\
\Sigma_{\mathbf{\overline{5}},\uparrow}^{(41)} & -(2H-E_1)^2=-16  & -(2H-E_1)^2 = -16 & -3\rho(2H-E_1)
\end{array}\end{equation}
For $\rho$ an integer linear combination of $H$, $E_1$, and $E_2$ we have
\begin{equation}\rho = a H + b E_1 + c E_2\qquad\implies\qquad\rho\cdot_{S_2}(2H-E_1) = 12a + 8b + 6c\end{equation}
so that the spectrum induced by our $\gamma_{\text{inherited}}$ is
\begin{equation}\begin{array}{c|ccc}
\cdot_{\cloc} & \gamma_{\text{inherited}} \\ \hline
\Sigma_{\mathbf{10},\uparrow}^{(4)} & 2\left(6a+4b+3c+4\right) \\
\Sigma_{\mathbf{\overline{5}},\uparrow}^{(44)} & 8\left(6a+4b+3c+2p\right) \\
\Sigma_{\mathbf{\overline{5}},\uparrow}^{(41)} & -2\left(-4+18a+12b+9c+8p\right)
\end{array}\end{equation}
To get the desired spectrum \eqref{desiredparentspectrum} we would need to find integer solutions to
\begin{equation}\begin{split}0 &= 6a+4b+3c+2p \\
&= 6+4(p-2)
\end{split}\end{equation}
Clearly no such solution exists.  It is for this reason that we make the choices \eqref{tuningchoice} so that $\cloc$ takes the specialized form
\begin{equation}\cloc^{(4)}:\,\,\, h_0\left[c_4V^2(V^2+U^2)+c_3V^3U+c_1U^3(e_1V-e_0U)\right] + V^2(h_1U-h_2V)(d_2U+d_3V)=0\label{cloctuned}\end{equation}
We further choose the classes $\xi$ and $\chi$ in \eqref{tuningparams} to be
\begin{equation}\xi=E_2\qquad \chi = H-E_1\end{equation}
so that the ingredients in \eqref{cloctuned} are sections of the indicated bundles
\begin{equation}\begin{array}{c|cc}
\text{Section} & \text{Bundle} & \text{Required }\zenriques\text{ Parity}\\ \hline
h_0 & \CO_{B_3}(E_2)  & + \\
h_1 & \CO_{B_3}(H-E_1) & +\\
h_2 & \CO_{B_3}(H-E_1) & - \\
c_3 & \CO_{B_3}(2H-E_1-E_2) & - \\
c_1, c_4 & \CO_{B_3}(2H-E_1-E_2) & + \\
d_2 & \CO_{B_3}(H) & + \\
d_3 & \CO_{B_3}(H) & -
\end{array}\label{tunedbundles}\end{equation}
We should do a quick check that there are enough holomorphic sections available to specify $h_m$, $c_n$, and $d_p$ while keeping the appropriate even and odd conditions in tact.  One can verify by direct computation using \texttt{cohomcalg}, for instance, that all global sections of $\CO_{B_3}(H-E_1)$, $\CO_{B_3}(2H-E_1-E_2)$, and $\CO_{B_3}(H)$ are just the restrictions of the corresponding bundles from $X_5$.  It turns out that there are two global sections of $\CO_{B_3}(E_2)$ of which one is the restriction of the unique section $\delta_2$ of $\CO_{X_5}(E_2)$.  The other descends from $H^1(X_5,\CO_{X_5}(-2H+2E_2))$ and is odd under the $\zenriques$ involution \eqref{Z2W5}.  In total, we list the cohomology group to which each object $h_m,c_n,d_p$ must belong as well as the rank of that group
\begin{equation}\begin{array}{c|cc}
\text{Section} & \text{Cohomology Group} & \text{Dimension} \\ \hline
h_0 & H^{0,+}(B_3,\CO_{B_3}(E_2)) & 1 \\
h_1 & H^{0,+}(B_3,\CO_{B_3}(H-E_1)) & 1 \\
h_2 & H^{0,-}(B_3,\CO_{B_3})(H-E_1)) & 1 \\
c_3 & H^{0,-}(B_3,\CO_{B_3}(2H-E_1-E_2)) & 2 \\
c_1, c_4 & H^{0,+}(B_3,\CO_{B_3}(2H-E_1-E_2)) & 2 \\
d_2 & H^{0,+}(B_3,\CO_{B_3}(H)) & 3 \\
d_3 & H^{0,-}(B_3,\CO_{B_3}(H)) & 3
\end{array}\end{equation}
where the upper index denotes the grading with respect to $\zenriques$ \eqref{Z2W5}.  In each case there are enough sections available to build the model.

We can now consider the non-inherited $\gamma$ \eqref{gammanoninherited}
\begin{equation}\gamma_{\text{noninherited}} = D_1-D_2\label{gammanoninher}\end{equation}
where
\begin{equation}\begin{split}D_1:\,\,\, h_0 = h_1U-h_2V=0 \\
D_2:\,\,\, h_0 = d_2U+d_3V=0
\end{split}\end{equation}
From \eqref{noninherspectrum}, the contributions of $\gamma_{\text{noninherited}}$ to the chiral spectrum are given by
\begin{equation}\begin{array}{c|ccc}
\cdot_{\cloc} & \gamma_{\text{noninherited}} \\ \hline
\Sigma_{\mathbf{10},\uparrow}^{(4)} & -E_1\cdot_{S_2}E_2 = -2\\
\Sigma_{\mathbf{\overline{5}},\uparrow}^{(44)} & 0 \\
\Sigma_{\mathbf{\overline{5}},\uparrow}^{(41)} & -E_1\cdot_{S_2}E_2 = -2
\end{array}\label{noninherspecK3}\end{equation}
What we have achieved with $\gamma_{\text{noninherited}}$ is the ability to adjust the chiral spectrum by 2's on the two matter curves where the MSSM matter fields should live without doing anything to the Higgs curve.  This makes it very easy to write a flux that yields a 3 generation model.  A `minimal' choice combines a $\gamma_{\text{inherited}}$ with $p=0$ and $\rho=0$ with one unit of $\gamma_{\text{noninherited}}$
\begin{equation}\gamma_{\text{minimal}} = \gamma_{\text{noninherited}} - \frac{1}{2}\gamma^{(u)}\label{gammaminimal}\end{equation}
and gives exactly the desired chiral spectrum \eqref{desiredparentspectrum}.  More generally, we can consider
\begin{equation}\gamma = \gamma_{\text{minimal}}+\gamma_0\label{moregeneralgamma}\end{equation}
where $\gamma_0$ has vanishing intersection with all matter curves.  In general, we can write such a $\gamma_0$ as
\begin{equation}\gamma_0 = p\left(\gamma^{(p)}-2\gamma_{\text{noninherited}}\right)+\gamma^{(\rho)}
\label{gamma0def}\end{equation}
where
\begin{equation}\rho = a H + bE_1 + cE_2\label{therhodef}\end{equation}
and
\begin{equation}6a+4b+3c+2p=0\label{theabcpconst}\end{equation}
This yields a 3-parameter family of choices for $\gamma$ that produce the desired chiral spectrum \eqref{desiredparentspectrum}.

\subsubsection{Need for Global Completion}

One might hope that this abundance of possibilities will help deal with some of the global problems that flux can present.  More specifically, the bundle data $\gamma$ must descend from a globally well-defined $G$-flux on the Calabi-Yau 4-fold.
This flux will induce a 3-brane charge and may also introduce a $U(1)$ $D$-term as well.  The $D$-term is computed by $\int_{Y_4}\omega\wedge J\wedge G_4$ where $\omega$ is a $(1,1)$-form associated to the $U(1)$ in the Calabi-Yau 4-fold and $J$ is the K\"ahler form.  This is obviously a global computation for which we cannot obtain any insight from the local model perspective.

The same is true for the flux-induced 3-brane charge, which combines with a geometrically-induced 3-brane charge on the full Calabi-Yau 4-fold to yield the net 3-brane tadpole
\begin{equation}n_{D3,\text{ induced}} = \frac{1}{2}\int_{Y_4}G\wedge G - \frac{\chi(Y_4)}{24}\end{equation}
To cancel this with supersymmetry-preserving D3-branes we need $n_{D3,\text{ induced}}\le 0$ which can be problematic because $G$ is self-dual and hence
\begin{equation}\int_{Y_4}G\wedge G\ge 0\end{equation}
The potential violation of $n_{D3,\text{induced}}\le 0$ is referred to as the `overshoot problem' and was emphasized strongly in \cite{Blumenhagen:2009yv,Grimm:2009yu}.  We are unable to address this in the local model when $\cloc$ is nongeneric because we do not know $\chi(Y_4)$ and spectral cover methods are unable to reliably compute the global quantity $\int_{Y_4}G\wedge G${\footnote{In generic situations when the spectral cover is not split it turns out that we can actually compute both of these quantities \cite{Blumenhagen:2009yv,Grimm:2009yu} \cite{Marsano:2011hv}.  The basic reason for this is that all of the nontriviality is associated to singularities of $Y_4$ and these, in turn, are all found along $S_2$ in this case.  The local model is able to capture all essential properties of singularities on $S_2$ as it is really just a repackaging of the local geometric and flux data.  When $U(1)$'s are introduced, however, new singularities appear that extend beyond the local geometry near $S_2$ \cite{Grimm:2010ez}.  In these cases, one does not expect local model computations to be useful and indeed there are now examples where attempts to compute the 3-brane tadpole in the local model explicitly fail \cite{Krause:2011xj}.}}.

To address $D$-terms and the 3-brane tadpole it is necessary to embed the local model into a global one and construct explicit $G$-fluxes that globally extend the local model bundle data $\gamma$.  We can hope that the existence of a 3-parameter family of choices that yield the right spectrum will allow the 3-brane tadpole problem to be addressed in a satisfactory way.  While many choices of flux can in principle lead to a vanishing $D$-term, we are only able to find one that avoids the overshoot problem, namely the minimal choice \eqref{gammaminimal}.  In addition to this, we will also consider a second choice with $a=-1$, $b=1$, and $c=0$ in \eqref{gamma0def} that only overshoots by one.
We summarize these two choices for convenience below
\begin{equation}\gamma = -\frac{1}{2}\gamma^{(u)}+p\left(\gamma^{(p)}-2\gamma_{\text{noninherited}}\right)+\gamma^{(\rho)}\qquad\left\{\begin{array}{ll}p=0 & \rho=0 \\ p=1 & \rho = -H+E_1\end{array}\right.\label{goodgammas}\end{equation}
For the rest of this section, we take the preference for these two choices of $\gamma$ as a given.  We will later justify this in section \ref{subsec:tadpoleandD} through an honest D3-brane tadpole computation in the global completion.

\subsection{Precise Spectrum}
\label{subsec:precisespectrum}

We have managed to construct fluxes (i.e. $\gamma$'s) in the local model that yield the desired chiral spectrum \eqref{desiredparentspectrum}.  In this subsection, we aim to do a bit better by turning to the precise spectrum of all fields on the matter curves.  This is important not only to check whether vector-like exotics are hiding on the curves $\Sigma_{\mathbf{10},\uparrow}^{(4)}$ and $\Sigma_{\mathbf{\overline{5}},\uparrow}^{(41)}$ where our quarks and leptons are sitting but also to see whether there is any possibility of realizing a vector-like pair of Higgs doublets in the curve $\Sigma_{\mathbf{\overline{5}},\uparrow}^{(44)}$.  Naively speaking, we might think that there is no hope of finding Higgs doublets because we have not introduced any kind of symmetry to keep them massless.  Quite to the contrary, we will encounter the opposite problem: for the fluxes of interest \eqref{goodgammas} there will be \emph{too many} Higgs doublets!  We find the appearance of these doublets quite surprising and do not see an obvious symmetry reason behind it.  A more intuitive understanding for why they emerge would be very interesting and could provide new ideas for solving the $\mu$ problem in F-theory models.

In the rest of this subsection, we determine the relevant cohomology groups \eqref{cohomsforspectrum} for computing the spectrum on our three matter curves for each of the choices of $\gamma$ in \eqref{goodgammas}.  We work in the upstairs picture of \eqref{cohomsforspectrum} to avoid subtleties associated with the choice of spin structure $K_{\Sigma,\downarrow}^{1/2}$ on the downstairs curve.  Before moving to the computations, we summarize the basic strategy.

\subsubsection{Strategy}
\label{subsubsec:strategy}

Let us begin by introducing some convenient notation for the bundle on each matter curve whose cohomologies we need from \eqref{cohomsforspectrum}
\begin{equation}\begin{split}\CL_{\Sigma_{\mathbf{10},\uparrow}^{(4)}} &= \CO\left(\frac{r_4}{2}+\gamma_4\right)|_{\Sigma_{\mathbf{10}}^{(4)}} \\
\CL_{\Sigma_{\mathbf{\overline{5}},\uparrow}^{(44)}} &= \left[\CO\left(\frac{r_4}{2}+\gamma_4\right)|_{\Sigma_{\mathbf{\overline{5}}}^{(44)}}\right]\otimes \tau\left[\CO\left(\frac{r_4}{2}+\gamma_4\right)|_{\Sigma_{\mathbf{\overline{5}}}^{(44)}}\right] \\
\CL_{\Sigma_{\mathbf{\overline{5}},\uparrow}^{(\bar{4}1)}} &= \left[\CO\left(\frac{r_4}{2}+\gamma_4\right)|_{\Sigma_{\mathbf{\overline{5}}}^{(\bar{4}1)}}\right]\otimes \tau\left[\CO\left(\gamma_1\right)|_{\Sigma_{\mathbf{\overline{5}}}^{(4\bar{1})}}\right]
\end{split}\label{neededbundles}\end{equation}
Here $\tau$ maps a bundle to its image under the involution
\begin{equation}\ztau:\,\, V\rightarrow -V\end{equation}
We recall that $\Sigma_{\mathbf{\overline{5}}}^{(\bar{4}1)}$ is the component of $\Sigma_{\mathbf{\overline{5}}}^{(41)}$ that sits inside $\cloc^{(4)}$ and $\Sigma_{\mathbf{\overline{5}}}^{(4\bar{1})}$ is the component that sits inside $\cloc^{(1)}$.  We also recall that $\gamma_4$ ($\gamma_1$) denotes the piece of $\gamma$ that sits inside $\cloc^{(4)}$ ($\cloc^{(1)}$).  We compute the cohomologies of these bundles in the parent K3-based theory which determine the spectrum of the quotient according to the action of the $\zenriques$ involution.  More specifically, the K3 parent is quotiented by a combination of $\zenriques$ and the $\mathbb{Z}_2$ center of $U(1)_Y$.  To be completely clear, we list the cohomology groups that determine the daughter theory spectrum below
\begin{equation}
\begin{array}{c|c|c}
\text{Matter Curve} & SU(3)\times SU(2)\times U(1)_Y & \text{Cohomology} \\ \hline
\Sigma_{\mathbf{10},\uparrow}^{(4)}
& (\mathbf{1},\mathbf{1})_{+1} & h^{0,+}(\tencurve,\CL_{\tencurve})\\
& (\mathbf{3},\mathbf{2})_{+1/6} & h^{0,-}(\tencurve,\CL_{\tencurve}) \\
& (\mathbf{\overline{3}},\mathbf{1})_{-2/3} &  h^{0,+}(\tencurve,\CL_{\tencurve}) \\
& (\mathbf{1},\mathbf{1})_{-1} &  h^{1,+}(\tencurve,\CL_{\tencurve}) \\
& (\mathbf{\overline{3}},\mathbf{2})_{-1/6} &  h^{1,-}(\tencurve,\CL_{\tencurve}) \\
& (\mathbf{3},\mathbf{1})_{+2/3} &  h^{1,+}(\tencurve,\CL_{\tencurve}) \\ \hline
\fivefourcurve & (\mathbf{\overline{3}},\mathbf{1})_{+1/3} & h^{0,+}_-(\fivefourcurve,\CL_{\fivefourcurve}) \\
& (\mathbf{1},\mathbf{2})_{-1/2} & h^{0,-}_-(\fivefourcurve,\CL_{\fivefourcurve}) \\
& (\mathbf{3},\mathbf{1})_{-1/3} & h^{1,+}_-(\fivefourcurve,\CL_{\fivefourcurve}) \\
& (\mathbf{1},\mathbf{2})_{+1/2} & h^{1,-}_-(\fivefourcurve,\CL_{\fivefourcurve}) \\ \hline
\fiveonecurve & (\mathbf{\overline{3}},\mathbf{1})_{+1/3} & h^{0,+}(\fiveonecurve,\CL_{\fiveonecurve}) \\
& (\mathbf{1},\mathbf{2})_{-1/2} & h^{0,-}(\fiveonecurve,\CL_{\fiveonecurve}) \\
& (\mathbf{3},\mathbf{1})_{-1/3} & h^{1,+}(\fiveonecurve,\CL_{\fiveonecurve}) \\
& (\mathbf{1},\mathbf{2})_{+1/2} & h^{1,-}(\fiveonecurve,\CL_{\fiveonecurve})
\end{array}\label{neededcohoms}\end{equation}
Here the upper $\pm$ denotes even or odd under $\zenriques$ while the subscript $-$ for the $\fivefourcurve$ cohomologies denotes that we take the odd cohomology with respect $\ztau$, which interchanges the two sheets of $\fivefourcurve$.

If we had not introduced $\gamma_{\text{noninherited}}$ then the computation would be a completely straightforward application of the techniques in \cite{Blumenhagen:2010pv,Jow,Blumenhagen:2010ed}, which are nicely automated in the \texttt{cohomCalg} package \cite{cohomCalg:Implementation}.  The reason for this is twofold.  First, each matter curve $\Sigma_{\uparrow}$ can be written as a complete intersection of 5 divisors in the toric 6-fold $X_5\times \mathbb{P}^1$
\begin{equation}\Sigma_{\uparrow} = \prod_{i=1}^5 \CD_i\end{equation}
Written explicitly, these intersections are
\begin{equation}
\begin{array}{c|cc}
\text{Curve} & \text{Equations in }S_2\times\mathbb{P}^1 & \text{Class in }X_5\times\mathbb{P}^1 \\ \hline
\Sigma_{\mathbf{10},\uparrow}^{(4)} & U=c_4h_0-d_3h_2=0 & (2H-E_1)(2H-E_2)(2H)(\sigma_{loc})(2H-E_1) \\ \hline
\Sigma_{\mathbf{\overline{5}},\uparrow}^{(44)} & a_3V^2+\alpha e_1 U^2 = 0& (2H-E_1)(2H-E_2)(2H)(2\sigma_{loc}+2H-E_1)(2\sigma_{loc}+2H-E_1) \\
&  a_4V^2 + (a_2+a_3e_0)U^2=0  \\ \hline
\Sigma_{\mathbf{\overline{5}},\uparrow}^{(\bar{4}1)} & e_1V-e_0U=0 & (2H-E_1)(2H-E_2)(2H)(\sigma_{loc})(2\sigma_{loc}+2H-E_1) \\
& h_0(c_3UV+c_4(U^2+V^2))\\
& +(d_2U+d_3V)(h_1U-h_2V)=0
\end{array}\label{mattcurvecompleteints}\end{equation}
where we used the fact that $S_2\times\mathbb{P}^1=(2H-E_1)(2H-E_2)(2H)$.
Second, the line bundles $\CO(\frac{r_4}{2}+\gamma_4)$ and $\CO(\gamma_1)$
that one obtains from `inherited' $\gamma$'s are inherited from $X_5\times\mathbb{P}^1$ in the sense that each is the restriction of a bundle $\CO(\CD)$ on that space.  Cohomologies of these bundles can be related to cohomologies of a variety of related bundles on the toric space $X_5\times\mathbb{P}^1$ using Koszul sequences following \cite{Blumenhagen:2010ed} which, in turn, can be explicitly computed from the algorithm of \cite{Blumenhagen:2010pv}.

In fact, we can follow precisely this procedure for the Higgs curve $\Sigma_{\mathbf{\overline{5}},\uparrow}^{(44)}$ because $\CL_{\Sigma_{\mathbf{\overline{5}},\uparrow}^{(44)}}$ is completely unaffected by the presence of $\gamma_{\text{noninherited}}$.  To see this, look back to the contribution of $\gamma_{\text{noninherited}}$ to the chiral index \eqref{noninherspec}.  The entries in that table actually indicate precisely how $\gamma_{\text{noninherited}}$ affects the restriction of the bundle $\CO(\frac{r_4}{2}+\gamma_4)$ to $\Sigma_{\mathbf{\overline{5}},\uparrow}^{(44)}$.  In particular, the effect of adding $\gamma_{\text{noninherited}}$ is to twist this bundle by $\CO_{\Sigma_{\mathbf{\overline{5}},\uparrow}^{(44)}}(\sum_i p_i - \sum_j q_j)$ where
\begin{equation}\begin{split}p_i &= \text{points in }\mathbb{P}^1\times S_2\text{ where }h_0=h_1U-h_2V=d_2h_2-d_3h_1=0 \\
q_j &= \text{points in }\mathbb{P}^1\times S_2\text{ where }h_0 = d_2U+d_3V=d_2h_2-d_3h_1=0
\end{split}\end{equation}
It is easy to see, however, that $\CO_{\Sigma_{\mathbf{\overline{5}},\uparrow}^{(44)}}(\sum_ip_i - \sum_j q_j)$ is antisymmetric under the involution $\ztau$ so it cancels in the tensor product that we must take to obtain $\CL_{\Sigma_{\mathbf{\overline{5}},\uparrow}^{(44)}}$ in \eqref{neededbundles}.  Effectively, this means we can neglect the non-inherited piece when studying the spectrum of fields on $\Sigma_{\mathbf{\overline{5}},\uparrow}^{(44)}$.

This simplification does not happen for the other matter curves so for them we must deal with $\gamma_{\text{noninherited}}$ head on.  In these cases our strategy is to realize that
\begin{equation}\CO(\gamma_{\text{noninherited}})|_{\Sigma_{\uparrow}} = \CO(D_1-D_2)|_{\Sigma_{\uparrow}} = \CO(E_2-2D_2)|_{\Sigma_{\uparrow}}\label{noninheritedprocessing}\end{equation}
where $D_1$ and $D_2$ are the two divisors in $\mathbb{P}^1\times S_2$ from \eqref{interestingdivisors} that are used to build $\gamma_{\text{noninherited}}$ and $E_2$ is the divisor class of $h_0=0$ \eqref{tunedbundles}.  Because of \eqref{noninheritedprocessing}, each of the bundles $\CL_{\Sigma_{\mathbf{10}}^{(4)}}$ and $\CL_{\Sigma_{\mathbf{\overline{5}},\uparrow}^{(\bar{4}1)}}$ can be written as the tensor product of an inherited bundle with a bundle of the form $\CO_{\Sigma_{\uparrow}}(-\sum_ap_a)$
\begin{equation}\begin{split}\CL_{\Sigma_{\mathbf{10},\uparrow}^{(4)}} &= \CO(\CD_{\mathbf{10}^{(4)}}|_{\Sigma_{\mathbf{10}}^{(4)}}-\sum_a P_a) \\
\CL_{\Sigma_{\mathbf{\overline{5}},\uparrow}^{(\bar{4}1)}} &= \CO(\CD_{\mathbf{\overline{5}}^{(41)}}|_{\Sigma_{\mathbf{\overline{5}},\uparrow}^{(\bar{4}1)}}-\sum_b Q_b)
\end{split}\end{equation}
In these cases, we can use the exact sequence
\begin{equation}0\rightarrow \CO_{\Sigma_{\uparrow}}\left(\CD_R|_{\Sigma_{R,\uparrow}}-\sum_aq_a\right)\rightarrow \CO_{\Sigma_{\uparrow}}\left(\CD_R|_{\Sigma_{R,\uparrow}}\right)\xrightarrow{f}\mathbb{C}^n\rightarrow 0\end{equation}
where $n$ is the number of points $q_a$ that we have removed, with multiplicities.  This leads to a long exact cohomology sequence
\begin{equation}\begin{split}0&\rightarrow H^0\left(\Sigma_{R,\uparrow},\CO_{\Sigma_{R,\uparrow}}\left(\CD_R|_{\Sigma_{R,\uparrow}}-\sum_a q_a\right)\right))\rightarrow H^0\left(\Sigma_{R,\uparrow},\CO_{\Sigma_{R,\uparrow}}\left(\CD_R|_{\Sigma_{R,\uparrow}}\right)\right)\xrightarrow{f}\mathbb{C}^n \\
&\rightarrow H^1\left(\Sigma_{R,\uparrow},\CO_{\Sigma_{R,\uparrow}}\left(\CD_R|_{\Sigma_{R,\uparrow}} -\sum_aq_a \right)\right))\rightarrow H^1\left(\Sigma_{R,\uparrow},\CO_{\Sigma_{R,\uparrow}}\left(\CD_R|_{\Sigma_{R,\uparrow}}\right)\right)\rightarrow 0
\end{split}\label{longexactpointsremoved}\end{equation}
The methods of \cite{Blumenhagen:2010pv,Jow,Blumenhagen:2010ed} can tell us about cohomologies of $\CO_{\Sigma_{R,\uparrow}}(\CD_R|_{\Sigma_{R,\uparrow}})$ along with enough information to deduce the rank of the map $f$, from which we can deduce the cohomologies of $\CO_{\Sigma_{R,\uparrow}}(\CD_R|_{\Sigma_{R,\uparrow}}-\sum_aq_a)$.

\subsubsection{Spectrum for Minimal Flux}
\label{subsubsec:minimalspectrum}

We now proceed to study the spectrum on each matter curve for the minimal choice $\gamma_{\text{minimal}}$ \eqref{gammaminimal}.  The relevant bundles \eqref{neededbundles} on each matter curve for this choice of $\gamma$ are
\begin{equation}\begin{split}
\CL_{\Sigma_{\mathbf{10},\uparrow}^{(4)}} &= \CO_{X_5\times\mathbb{P}^1}(2H-E_1-\sigma_{loc})|_{\Sigma_{\mathbf{10},\uparrow}^{(4)}}\otimes \CO_{\cloc^{(4)}}(\gamma_{\text{noninherited}})|_{\Sigma_{\mathbf{10},\uparrow}^{(4)}} \\
&= \CO_{X_5\times\mathbb{P}^1}(2H-E_1+E_2)|_{\Sigma_{\mathbf{10},\uparrow}^{(4)}}\otimes \CO_{\Sigma_{\mathbf{10},\uparrow}^{(4)}}(-2Q_{\mathbf{10}}) \\
\CL_{\Sigma_{\mathbf{\overline{5}},\uparrow}^{(44)}} &= \CO_{X_5\times\mathbb{P}^1}(2[2H-E_1-\sigma_{loc}])|_{\Sigma_{\mathbf{\overline{5}},\uparrow}^{(44)}} \\
\CL_{\Sigma_{\mathbf{\overline{5}},\uparrow}^{(\bar{4}1)}} &= \CO_{X_5\times\mathbb{P}^1}(2H-E_1-\sigma_{loc})|_{\Sigma_{\mathbf{\overline{5}},\uparrow}^{(\bar{4}1)}}\otimes \CO_{\cloc^{(4)}}(\gamma_{\text{noninherited}})|_{\Sigma_{\mathbf{\overline{5}},\uparrow}^{(\bar{4}1)}} \\
&= \CO_{X_5\times\mathbb{P}^1}(2H-E_1+E_2)|_{\Sigma_{\mathbf{\overline{5}},\uparrow}^{(\bar{4}1)}}\otimes \CO_{\Sigma_{\mathbf{\overline{5}},\uparrow}^{(\bar{4}1)}}(-2Q_{\mathbf{\overline{5}}})
\end{split}\end{equation}
We have explicitly used the fact that $\CO_{\cloc^{(4)}}(\gamma_{\text{noninherited}})|_{\Sigma_{\mathbf{\overline{5}},\uparrow}^{(44)}} = \CO_{\Sigma_{\mathbf{\overline{5}},\uparrow}^{(44)}}$ and have rewritten the bundles $\CL_{\Sigma_{\mathbf{10},\uparrow}^{(4)}}$ and $\CL_{\Sigma_{\mathbf{\overline{5}},\uparrow}^{(\bar{4}1)}}$ according to the strategy outlined in section \ref{subsubsec:strategy}.  The divisors $Q_{\mathbf{10}}$ and $Q_{\mathbf{\overline{5}}}$ on these curves are defined as
\begin{equation}\begin{split}Q_{\mathbf{10}} &= \text{set of 4 points on }\Sigma_{\mathbf{10},\uparrow}^{(4)}\text{ where }h_0=d_3=0 \\
Q_{\mathbf{\overline{5}}} &= \text{set of 4 points on }\Sigma_{\mathbf{\overline{5}},\uparrow}^{(\bar{4}1)}\text{ where }h_0=d_2+d_3=0
\end{split}\label{Qpointsdefined}\end{equation}
Let us turn first to the spectrum on $\Sigma_{\mathbf{10},\uparrow}^{(4)}$ which is determined by $h^m(\Sigma_{\mathbf{10},\uparrow}^{(4)},\CL_{\Sigma_{\mathbf{10},\uparrow}^{(4)}})$.  Using \texttt{cohomcalg} \cite{Blumenhagen:2010pv,Jow,Blumenhagen:2010ed,cohomCalg:Implementation}, we immediately see that
\begin{equation}\begin{split}h^0(\Sigma_{\mathbf{10},\uparrow}^{(4)},\CO_{X_5\times\mathbb{P}^1}(2H-E_1+E_2)|_{\Sigma_{\mathbf{10},\uparrow}^{(4)}}) &= 14 \\
h^1(\Sigma_{\mathbf{10},\uparrow}^{(4)},\CO_{X_5\times\mathbb{P}^1}(2H-E_1+E_2)|_{\Sigma_{\mathbf{10},\uparrow}^{(4)}}) &= 0
\end{split}\end{equation}
From the sequence \eqref{longexactpointsremoved} this means that
\begin{equation}\begin{split}h^0(\Sigma_{\mathbf{10},\uparrow}^{(4)},\CL_{\Sigma_{\mathbf{10},\uparrow}^{(4)}}) &= \text{ker }f_{\mathbf{10}} \\
h^1(\Sigma_{\mathbf{10},\uparrow}^{(4)},\CL_{\Sigma_{\mathbf{10},\uparrow}^{(4)}}) &= \text{coker }f_{\mathbf{10}}
\end{split}\end{equation}
where $f_{\mathbf{10}}$ is the map
\begin{equation}H^0(\Sigma_{\mathbf{10},\uparrow}^{(4)},\CO_{X_5\times\mathbb{P}^1}(2H-E_1+E_2)|_{\Sigma_{\mathbf{10},\uparrow}^{(4)}}) \xrightarrow{f_{\mathbf{10}}} \mathbb{C}^8\end{equation}
given by restriction of the sections in $H^0(\Sigma_{\mathbf{10},\uparrow}^{(4)},\CO_{X_5\times\mathbb{P}^1}(2H-E_1+E_2)|_{\Sigma_{\mathbf{10},\uparrow}^{(4)}}$ to their values and their first derivatives at the 4 points of $Q_{\mathbf{10}}$ \eqref{Qpointsdefined}.  This is a map from a 14-dimensional space to an 8-dimensional one so the kernel will generically have dimension 6 and the cokernel dimension 0.  In Appendix
\ref{appsubsubsec:minimalten} we check explicitly that this is generically the case in our setup so that
\begin{equation}\begin{split}h^0(\Sigma_{\mathbf{10},\uparrow}^{(4)},\CL_{\Sigma_{\mathbf{10},\uparrow}^{(4)}}) &= 6 \\
 h^1(\Sigma_{\mathbf{10},\uparrow}^{(4)},\CL_{\Sigma_{\mathbf{10},\uparrow}^{(4)}}) &= 0
 \end{split}\end{equation}
 We get exactly 6 $\mathbf{10}$'s and 0 $\mathbf{\overline{10}}$'s on $\Sigma_{\mathbf{10},\uparrow}^{(4)}$ in the parent K3-based theory.  Because $\zenriques$ acts freely on $\tencurve$ the index over even cohomologies is exactly 3 which tells us that
 \begin{equation}h^{0,+}(\tencurve,\CL_{\tencurve})=h^{0,-}(\tencurve,\CL_{\tencurve}) = 3\end{equation}
and  we get exactly 3 complete generations of $\mathbf{10}$'s with no $\mathbf{\overline{10}}$'s in the quotient.

The story for $\Sigma_{\mathbf{\overline{5}},\uparrow}^{(\bar{4}1)}$ is almost identical because it is equivalent to $\Sigma_{\mathbf{10},\uparrow}^{(4)}$ as a divisor in $\cloc^{(4)}$ and the bundle of interest differs only in the precise location of the 4 points in $Q_{\mathbf{\overline{5}}}$.  Because of this we automatically have
\begin{equation}\begin{split}h^0(\Sigma_{\mathbf{\overline{5}},\uparrow}^{(\bar{4}1)},\CO_{X_5\times\mathbb{P}^1}(2H-E_1+E_2)|_{\Sigma_{\mathbf{\overline{5}},\uparrow}^{(\bar{4}1)}}) &= 14 \\
 h^1(\Sigma_{\mathbf{\overline{5}},\uparrow}^{(\bar{4}1)},\CO_{X_5\times\mathbb{P}^1}(2H-E_1+E_2)|_{\Sigma_{\mathbf{\overline{5}},\uparrow}^{(\bar{4}1)}}) &= 0
 \end{split}\end{equation}
 so that
 \begin{equation}\begin{split}h^0(\Sigma_{\mathbf{\overline{5}},\uparrow}^{(\bar{4}1)},\CL_{\Sigma_{\mathbf{\overline{5}},\uparrow}^{(\bar{4}1)}}) &= \text{ker }f_{\mathbf{\overline{5}}} \\
 h^1(\Sigma_{\mathbf{\overline{5}},\uparrow}^{(\bar{4}1)},\CL_{\Sigma_{\mathbf{\overline{5}},\uparrow}^{(\bar{4}1)}}) &= \text{coker }f_{\mathbf{\overline{5}}}
 \end{split}\end{equation}
where $f_{\mathbf{\overline{5}}}$ is the map
\begin{equation}H^0(\Sigma_{\mathbf{\overline{5}},\uparrow}^{(\bar{4}1)},\CO_{X_5\times\mathbb{P}^1}(2H-E_1+E_2)|_{\Sigma_{\mathbf{\overline{5}},\uparrow}^{(\bar{4}1)}})\xrightarrow{f_{\mathbf{\overline{5}}}}\mathbb{C}^8\end{equation}
given by restriction of the sections of $H^0(\Sigma_{\mathbf{\overline{5}},\uparrow}^{(\bar{4}1)},\CO_{X_5\times\mathbb{P}^1|}(2H-E_1+E_2)|_{\Sigma_{\mathbf{\overline{5}},\uparrow}^{(\bar{4}1)}}$ to their values and their first derivatives at the 4 points of $Q_{\mathbf{\overline{5}}}$ \eqref{Qpointsdefined}.  In Appendix \ref{appsubsubsec:minimalfiveone} we argue that the map $f_{\mathbf{\overline{5}}}$ is sufficiently generic that its kernel is 6-dimensional and cokernel is empty and hence
\begin{equation}\begin{split}h^0(\Sigma_{\mathbf{\overline{5}},\uparrow}^{(\bar{4}1)},\CL_{\Sigma_{\mathbf{\overline{5}},\uparrow}^{(\bar{4}1)}}) &= 6 \\
h^1(\Sigma_{\mathbf{\overline{5}},\uparrow}^{(\bar{4}1)},\CL_{\Sigma_{\mathbf{\overline{5}},\uparrow}^{(\bar{4}1)}}) &= 0
\end{split}\end{equation}
The involution $\zenriques$ acts freely on $\fiveonecurve$ so, just as we had for $\tencurve$, the cohomologies of interest are
\begin{equation}h^{0,+}(\fiveonecurve,\CL_{\fiveonecurve}) = h^{0,-}(\fiveonecurve,\CL_{\fiveonecurve}) = 3\end{equation}
and we get exactly 3 full generations of $\mathbf{\overline{5}}$'s in the quotient theory.

Finally we should deal with the Higgs curve $\Sigma_{\mathbf{\overline{5}},\uparrow}^{(44)}$.  On one hand this is a little easier because we don't have to worry about the noninherited part of the bundle.  It is a little trickier, though, because the spectrum of the K3-based parent theory is given by odd cohomologies with respect to $\ztau$
\begin{equation}h^m_-(\Sigma_{\mathbf{\overline{5}},\uparrow}^{(44)},\CL_{\Sigma_{\mathbf{\overline{5}},\uparrow}^{(44)}})\end{equation}
We can use \texttt{cohomcalg} \cite{Blumenhagen:2010pv,Blumenhagen:2010ed,cohomCalg:Implementation} to help with this computation.  As we describe in Appendix \ref{appsubsubsec:minimalfivefour}, the cohomologies of interest are ultimately determined from an exact sequence of the form
\begin{equation}\begin{split}0&\rightarrow H^0_-(\Sigma_{\mathbf{\overline{5}},\uparrow}^{(44)},\CL_{\Sigma_{\mathbf{\overline{5}},\uparrow}^{(44)}}) \rightarrow H^1_-(X_5\times\mathbb{P}^1,\CI_4)\xrightarrow{g} H^1_-(X_5\times\mathbb{P}^1,\CO(\CD))\\
&\rightarrow H^1_-(\Sigma_{\mathbf{\overline{5}},\uparrow}^{(44)},\CL_{\Sigma_{\mathbf{\overline{5}},\uparrow}^{(44)}}) \rightarrow H^2_-(X_5\times\mathbb{P}^1,\CI_4)\rightarrow 0
\end{split}\label{longseqwithg}\end{equation}
We define $\CO(\CD)$ and $\CI_4$ in Appendix \ref{appsubsubsec:minimalfivefour}.  Here the important properties of these sheaves are that
\begin{equation}\begin{split} h^0_-(X_5\times\mathbb{P}^1,\CO(\CD)) &= 51 \\
h^0_-(X_5\times\mathbb{P}^1,\CI_4) &= 54 \\
h^1_-(X_5\times\mathbb{P}^1,\CI_4) &= 3
\end{split}\end{equation}
This means that
\begin{equation}h^0_-(\Sigma_{\mathbf{\overline{5}},\uparrow}^{(44)},\CL_{\Sigma_{\mathbf{\overline{5}},\uparrow}^{(44)}}) = h^1_-(\Sigma_{\mathbf{\overline{5}},\uparrow}^{(44)},\CL_{\Sigma_{\mathbf{\overline{5}},\uparrow}^{(44)}}) = \text{ker }g\end{equation}
where $g$ is the map indicated in \eqref{longseqwithg}.  As a map from a 54-dimensional space to a 51-dimensional one, this kernel has dimension at least 3.  In Appendix \ref{appsubsubsec:minimalfivefour} we show that, in our setup, the dimension is in fact  4 so that $h^m_-(\fivefourcurve,\CL_{\fivefourcurve})=4$ for $m=0,1$.  The parent K3-based theory thus exhibits 4 vector-like pairs of $\mathbf{\overline{5}}/\mathbf{5}$'s.  We also show in Appendix \ref{appsubsubsec:minimalfivefour} that all 4 of these are $\zenriques$-odd
\begin{equation}\begin{split}0 &= h^{0,+}_-(\fivefourcurve,\CL_{\fivefourcurve}) = h^{1,+}_-(\fivefourcurve,\CL_{\fivefourcurve}) \\
4 &= h^{0,-}_-(\fivefourcurve,\CL_{\fivefourcurve}) = h^{1,-}_-(\fivefourcurve,\CL_{\fivefourcurve})
\end{split}\end{equation}
so that, from \eqref{neededcohoms}, we get 4 pairs of Higgs doublets and no triplets when we pass to the quotient.

It is interesting to note that the Higgs curve $\Sigma_{\mathbf{\overline{5}},\uparrow}^{(44)}$, which has genus 33, always gives us vector-like pairs of massless fields even though we didn't introduce any symmetry to protect them.  Of course, generic curves of this genus will not yield such pairs so the subspace of the moduli space of genus 33 curves that our explicit embedding of $\Sigma_{\mathbf{\overline{5}},\uparrow}^{(44)}$ allows us to explore is very special.  A better understanding of why this happened may prove useful for solving the $\mu$ problem.

We now summarize the spectrum that we found with the `minimal' flux as
\begin{equation}
\begin{array}{c|c|c}
\text{Matter Curve} & SU(3)\times SU(2)\times U(1)_Y & \text{Cohomology} \\ \hline
\Sigma_{\mathbf{10},\uparrow}^{(4)}
& (\mathbf{1},\mathbf{1})_{+1} & h^{0,+}(\tencurve,\CL_{\tencurve})=3\\
\text{(MSSM matter)} & (\mathbf{3},\mathbf{2})_{+1/6} & h^{0,-}(\tencurve,\CL_{\tencurve})=3 \\
& (\mathbf{\overline{3}},\mathbf{1})_{-2/3} &  h^{0,+}(\tencurve,\CL_{\tencurve})=3 \\
& (\mathbf{1},\mathbf{1})_{-1} &  h^{1,+}(\tencurve,\CL_{\tencurve})=0 \\
& (\mathbf{\overline{3}},\mathbf{2})_{-1/6} &  h^{1,-}(\tencurve,\CL_{\tencurve})=0 \\
& (\mathbf{3},\mathbf{1})_{+2/3} &  h^{1,+}(\tencurve,\CL_{\tencurve})=0 \\ \hline
\fivefourcurve & (\mathbf{\overline{3}},\mathbf{1})_{+1/3} & h^{0,+}_-(\fivefourcurve,\CL_{\fivefourcurve})=0 \\
\text{(Higgs fields)}& (\mathbf{1},\mathbf{2})_{-1/2} & h^{0,-}_-(\fivefourcurve,\CL_{\fivefourcurve})=4 \\
& (\mathbf{3},\mathbf{1})_{-1/3} & h^{1,+}_-(\fivefourcurve,\CL_{\fivefourcurve})=0 \\
& (\mathbf{1},\mathbf{2})_{+1/2} & h^{1,-}_-(\fivefourcurve,\CL_{\fivefourcurve})=4 \\ \hline
\fiveonecurve & (\mathbf{\overline{3}},\mathbf{1})_{+1/3} & h^{0,+}(\fiveonecurve,\CL_{\fiveonecurve})=3 \\
\text{(MSSM matter)} & (\mathbf{1},\mathbf{2})_{-1/2} & h^{0,-}(\fiveonecurve,\CL_{\fiveonecurve})=3 \\
& (\mathbf{3},\mathbf{1})_{-1/3} & h^{1,+}(\fiveonecurve,\CL_{\fiveonecurve})=0 \\
& (\mathbf{1},\mathbf{2})_{+1/2} & h^{1,-}(\fiveonecurve,\CL_{\fiveonecurve})=0
\end{array}\end{equation}

\subsubsection{Spectrum for the `Non-minimal' Flux}
\label{subsubsec:nonminflux}

We now turn to the spectrum for the `non-minimal' flux choice in \eqref{goodgammas} with $p=1$ and $\rho=-H+E_1$.  The relevant bundles \eqref{neededbundles} on each matter curve for this $\gamma$ are
\begin{equation}\begin{split}
\CL_{\tencurve} &= \CO_{X_5\times\mathbb{P}^1}(H)|_{\tencurve}\otimes \CO_{\cloc^{(4)}}(-\gamma_{\text{noninherited}})|_{\tencurve} \\
&= \CO_{X_5\times\mathbb{P}^1}(H+E_2)|_{\tencurve}\otimes \CO_{\tencurve}(-2Q_{\mathbf{10}}') \\
\CL_{\fivefourcurve} &= \CO_{X_5\times\mathbb{P}^1}(2H)|_{\fivefourcurve} \\
\CL_{\fiveonecurve} &= \CO_{X_5\times\mathbb{P}^1}(3H-3E_1)|_{\fiveonecurve}\otimes \CO_{\cloc^{(4)}}(-\gamma_{\text{noninherited}}) \\
&= \CO_{X_5\times\mathbb{P}^1}(3H-3E_1+E_2)|_{\fiveonecurve}\otimes \CO_{\fiveonecurve}(-2Q_{\mathbf{\overline{5}}}')
\end{split}\end{equation}
where again we have explicitly used the fact that $\CO_{\cloc^{(4)}}(\gamma_{\text{noninherited}})|_{\fivefourcurve} = \CO_{\fivefourcurve}$ and have rewritten the bundles $\CL_{\tencurve}$ and $\CL_{\fiveonecurve}$ according to the strategy outlined in section \ref{subsubsec:strategy}.  The divisors $Q_{\mathbf{10}}'$ and $Q_{\mathbf{\overline{5}}}'$ on these curves are defined as
\begin{equation}\begin{split}Q_{\mathbf{10}}' &= \text{set of 2 points on }\tencurve\text{ where }h_0=h_2=0 \\
Q_{\mathbf{\overline{5}}}' &= \text{set of 2 points on }\fiveonecurve\text{ where }h_0 = h_1-h_2=0
\end{split}\end{equation}
We turn first to the spectrum on $\tencurve$, which is determined by $h^m(\tencurve,\CL_{\tencurve})$.  Using \texttt{cohomcalg} \cite{Blumenhagen:2010pv,Jow,cohomCalg:Implementation}, we immediately see that
\begin{equation}\begin{split}h^0(\tencurve,\CO_{X_5\times\mathbb{P}^1}(H+E_2)|_{\tencurve}) &= 10 \\
h^1(\tencurve,\CO_{X_5\times\mathbb{P}^1}(H+E_2)|_{\tencurve}) &= 0
\end{split}\end{equation}
From the sequence \eqref{longexactpointsremoved} this means that
\begin{equation}\begin{split}h^0(\tencurve,\CL_{\tencurve})&=\text{ker }f'_{\mathbf{10}} \\
h^1(\tencurve,\CL_{\tencurve}) &= \text{coker }f'_{\mathbf{10}}
\end{split}\end{equation}
where $f'_{\mathbf{10}}$ is the map
\begin{equation}H^0(\tencurve,\CO_{X_5\times\mathbb{P}^1}(H+E_2)|_{\tencurve})\xrightarrow{f'_{\mathbf{10}}}\mathbb{C}^4\end{equation}
given by restriction of the sections in $H^0(\tencurve,\CO_{X_5\times\mathbb{P}^1}(H)|_{\tencurve})$ to their values and first derivatives at the 2 points of $Q_{\mathbf{10}}'$.  This is a map from a 10-dimensional space to a 4-dimensional one so the kernel will generically have dimension 6 and the cokernel dimension 0.  In Appendix \ref{appsubsubsec:minimalten} we check explicitly that this is generically the case in our setup so that
\begin{equation}\begin{split}h^0(\tencurve,\CL_{\tencurve}) &= 6 \\
h^1(\tencurve,\CL_{\tencurve}) &= 0
\end{split}\end{equation}
As before, this gives exactly 6 $\mathbf{10}$'s and 0 $\mathbf{\overline{10}}$'s in the parent K3-based theory.  Since $\zenriques$, the index over even cohomologies is exactly three and we get
\begin{equation}h^{0,+}(\tencurve,\CL_{\tencurve}) = h^{0,-}(\tencurve,\CL_{\tencurve}) = 3\end{equation}
and hence exactly 3 complete generations of $\mathbf{10}$'s with no $\mathbf{\overline{10}}$'s in the quotient.

The story for $\fiveonecurve$ is similar since
\begin{equation}\begin{split}h^0(\fiveonecurve,\CO_{X_5\times\mathbb{P}^1}(3H-3E_1+E_2)|_{\fiveonecurve}) &= 10 \\
h^1(\fiveonecurve,\CO_{X_5\times\mathbb{P}^1}(3H-3E_1+E_2)|_{\fiveonecurve}) &= 0
\end{split}\end{equation}
We have that
\begin{equation}\begin{split}
h^0(\fiveonecurve,\CL_{\fiveonecurve}) &= \text{ker }f'_{\mathbf{\overline{5}}} \\
h^1(\fiveonecurve,\CL_{\fiveonecurve}) &= \text{coker }f'_{\mathbf{\overline{5}}}
\end{split}\end{equation}
where $f'_{\mathbf{\overline{5}}}$ is the map
\begin{equation}H^0(\fiveonecurve,\CO_{X_5\times\mathbf{P}^1}(3H-3E_1+E_2)|_{\fiveonecurve})\xrightarrow{f'_{\mathbf{\overline{5}}}}\mathbb{C}^4\end{equation}
given by restriction of the sections of $H^0(\fiveonecurve,\CO_{X_5\times\mathbb{P}^1}(3H-3E_1+E_2)|_{\fiveonecurve})$ to their values and first derivatives at the 2 points of $Q_{\mathbf{\overline{5}}}'$.  In Appendix \ref{app:nonminmatt} we argue that the map $f'_{\mathbf{\overline{5}}}$ is sufficiently generic in our setup that its kernel is 6-dimensional and its cokernel is empty and hence
\begin{equation}\begin{split}h^0(\fiveonecurve,\CL_{\fiveonecurve}) &= 6 \\
h^1(\fiveonecurve,\CL_{\fiveonecurve}) &= 0
\end{split}\end{equation}
The involution $\zenriques$ acts freely on $\fiveonecurve$ so just as we had for $\tencurve$ the cohomologies of interest are
\begin{equation}h^{0,+}(\fiveonecurve,\CL_{\fiveonecurve}) = h^{0,-}(\fiveonecurve,\CL_{\fiveonecurve}) = 3\end{equation}
and we get exactly 3 full generations of $\mathbf{\overline{5}}$'s in the quotient theory.

Finally we look to the Higgs curve $\fivefourcurve$.  We can use \texttt{cohomcalg} to help with this computation.  As we describe in Appendix \ref{app:nonminhiggs}, the dimensions of the cohomologies of interest are completely determined without looking at the behavior of any maps in detail.  In particular, we have
\begin{equation}\begin{split}h^{0,+}_-(\fivefourcurve,\CL_{\fivefourcurve}) &= 4 \\
h^{0,-}_-(\fivefourcurve,\CL_{\fivefourcurve}) &= 2 \\
h^{1,+}_-(\fivefourcurve,\CL_{\fivefourcurve}) &= 4 \\
h^{0,-}_-(\fivefourcurve,\CL_{\fivefourcurve}) &= 2
\end{split}\label{nonminhiggsresults}\end{equation}
where as usual the upper $\pm$ denotes parity with respect to $\zenriques$ while the lower $\pm$ denotes parity with respect to $\ztau$, which interchanges the two sheets of $\fivefourcurve$.  The results \eqref{nonminhiggsresults} correspond to a spectrum on $\fivefourcurve$ consisting of 4 vector-like pairs of triplets and 2 vector-like pairs of doublets.

We now summarize the spectrum that we found with the `non-minimal' flux as
\begin{equation}
\begin{array}{c|c|c}
\text{Matter Curve} & SU(3)\times SU(2)\times U(1)_Y & \text{Cohomology} \\ \hline
\Sigma_{\mathbf{10},\uparrow}^{(4)}
& (\mathbf{1},\mathbf{1})_{+1} & h^{0,+}(\tencurve,\CL_{\tencurve})=3\\
\text{(MSSM matter)} & (\mathbf{3},\mathbf{2})_{+1/6} & h^{0,-}(\tencurve,\CL_{\tencurve})=3 \\
& (\mathbf{\overline{3}},\mathbf{1})_{-2/3} &  h^{0,+}(\tencurve,\CL_{\tencurve})=3 \\
& (\mathbf{1},\mathbf{1})_{-1} &  h^{1,+}(\tencurve,\CL_{\tencurve})=0 \\
& (\mathbf{\overline{3}},\mathbf{2})_{-1/6} &  h^{1,-}(\tencurve,\CL_{\tencurve})=0 \\
& (\mathbf{3},\mathbf{1})_{+2/3} &  h^{1,+}(\tencurve,\CL_{\tencurve})=0 \\ \hline
\fivefourcurve & (\mathbf{\overline{3}},\mathbf{1})_{+1/3} & h^{0,+}_-(\fivefourcurve,\CL_{\fivefourcurve})=4 \\
\text{(Higgs fields)}& (\mathbf{1},\mathbf{2})_{-1/2} & h^{0,-}_-(\fivefourcurve,\CL_{\fivefourcurve})=2 \\
& (\mathbf{3},\mathbf{1})_{-1/3} & h^{1,+}_-(\fivefourcurve,\CL_{\fivefourcurve})=4 \\
& (\mathbf{1},\mathbf{2})_{+1/2} & h^{1,-}_-(\fivefourcurve,\CL_{\fivefourcurve})=2 \\ \hline
\fiveonecurve & (\mathbf{\overline{3}},\mathbf{1})_{+1/3} & h^{0,+}(\fiveonecurve,\CL_{\fiveonecurve})=3 \\
\text{(MSSM matter)} & (\mathbf{1},\mathbf{2})_{-1/2} & h^{0,-}(\fiveonecurve,\CL_{\fiveonecurve})=3 \\
& (\mathbf{3},\mathbf{1})_{-1/3} & h^{1,+}(\fiveonecurve,\CL_{\fiveonecurve})=0 \\
& (\mathbf{1},\mathbf{2})_{+1/2} & h^{1,-}(\fiveonecurve,\CL_{\fiveonecurve})=0
\end{array}\end{equation}

\section{Global Generalities--Resolved Calabi-Yau and $G$-flux}
\label{sec:globalgeneral}

We now proceed to discuss global F-theory compactifications with honest $(2,2)$ $G$-fluxes.  Ultimately, we are interested in a global completion of the local model described in the previous section.  Before getting to that, however, we discuss the global completion of generic local models of the type described in section \ref{subsec:localmodelsgen}.  The Tate divisor formalism has recently been used to construct explicit global extensions of local model fluxes for generic unsplit $\cloc$ \cite{Marsano:2011hv}.  The objective of this section is to extend those results in two ways.  Firstly, we describe how the Tate divisor formalism applies to global extensions of local models with a 4+1 split of $\cloc$.  In this way, we construct explicit extensions of the inherited fluxes of section \ref{subsubsec:chiralitygen} as well as the explicit divisor that yields an additional global $U(1)$.  This clarifies the nature of the Tate divisor in the special case of `$U(1)$-restricted Tate models' \cite{Grimm:2010ez}, of which the model in this paper is an example, and demonstrates the natural role played by the Tate divisor in their resolution.  We then construct explicit extensions of the noninherited fluxes of section \ref{subsubsec:chiralitygen} that arise only when the form of $\cloc$ is specially chosen \eqref{C4tuned} and the Tate divisor respects that choice.

Throughout this discussion, we will be working with an explicit resolution of the singular Calabi-Yau 4-fold.  In the work \cite{Marsano:2011hv}, the Esole-Yau procedure \cite{Esole:2011sm} was used for this purpose.  In the present work, we work instead with the procedure outlined by \cite{Krause:2011xj}, who provided an explicit resolution of precisely the type of 4-fold that we are considering.  We will therefore begin with a brief review of the resolution procedure of \cite{Krause:2011xj} in order to set some convenient notation.

\subsection{Resolved Geometry and Fluxes for the Generic Spectral Cover}

We begin with a review of the resolution procedure of \cite{Krause:2011xj} as applied to a generic elliptically fibered Calabi-Yau 4-fold $Y_4$ that exhibits $A_4$ singular fibers over a surface $S_2$ inside the base $B_3$.  As described in section \ref{sec:generalsetup}, we build such a $Y_4$ as a hypersurface inside the 5-fold
\begin{equation}W_5 = \mathbb{P}(\CO\oplus K_{B_3}^{-2}\oplus K_{B_3}^{-3})\end{equation}
specified by the defining equation
\begin{equation}vy^2 = x^3 + \bh_0z^5 v^3 + \bh_2 z^3 v^2 x + \bh_3 z^2 v^2 y + \bh_4 z vx^2 + \bh_5 v xy\label{defeqn}\end{equation}
Following \cite{Krause:2011xj}, we resolve this 4-fold by performing a series of 4 blow-ups in $W_5$ and passing from $Y_4$ to its proper transform.  The blow-ups are as follows:

\begin{enumerate}
\item Blow-up along $x=y=z=0$ to get the once blown up space $W_5^{(1)}$.  This gives an exceptional divisor $\CE_1$.  The pullbacks of the holomorphic sections $x$, $y$, and $z$ are products of holomorphic sections on $W_5^{(1)}$:
\begin{equation}\begin{split}x &= x_1\delta_1 \\
y &= y_1\delta_1 \\
z &= z_1\delta_1
\end{split}\end{equation}
where $\delta_1$ is the unique holomorphic section of $\CO(\CE_1)$ whose vanishing defines $\CE_1$.

\item Blow-up along $y_1=\delta_1=0$ to get the twice blown up space $W_5^{(2)}$.  This gives an exceptional divisor $\CE_2$.  The pullbacks of the holomorphic sections $y_1$ and $\delta_1$ are products of holomorphic sections on $W_5^{(2)}$:
\begin{equation}\begin{split}y_1 &= y_{12}\delta_2 \\
\delta_{1} &= \delta_{12}\delta_2 \end{split}\end{equation}
where $\delta_2$ is the unique holomorphic section of $\CO(\CE_2)$ whose vanishing defines $\CE_2$.

\item Blow-up along $x_1=\delta_2=0$ to get the thrice blown up space $W_5^{(3)}$.  This gives an exceptional divisor $\CE_3$.  The pullbacks of the holomorphic sections $x_1$ and $\delta_2$ are products of holomorphic sections on $W_5^{(3)}$:
\begin{equation}\begin{split}x_1 &= x_{13}\delta_3 \\
\delta_2 &= \delta_{23}\delta_3
\end{split}\end{equation}
where $\delta_3$ is the unique holomorphic section of $\CO(\CE_3)$ whose vanishing defines $\CE_3$.

\item Blow-up along $y_{12}=\delta_3=0$ to get the four-times blown up space $W_5^{(4)}$.  This gives an exceptional divisor $\CE_4$.  The pullbacks of the holomorphic sections $y_{12}$ and $\delta_3$ are products of holomorphic sections on $W_5^{(4)}$:
\begin{equation}\begin{split}y_{12} &= y_{124}\delta_4 \\
\delta_3 &= \delta_{34}\delta_4
\end{split}\end{equation}
where $\delta_4$ is the unique holomorphic section of $\CO(\CE_4)$ whose vanishing defines $\CE_4$.
\end{enumerate}
After all of these blow-ups, the original holomorphic sections $x$, $y$, and $z$ on $W_5$ split in the following way on $W_5^{(4)}$
\begin{equation}\begin{split}
x &= x_{13}\delta_{12}\delta_{23}\delta_{34}^2\delta_4^2 \\
y &= y_{124}\delta_{12}\delta_{23}^2\delta_{34}^2\delta_4^3 \\
z &= z_1\delta_{12}\delta_{23}\delta_{34}\delta_4
\end{split}\end{equation}
where the sections appearing here are associated to bundles as follows
\begin{equation}\begin{array}{c|c}
\text{Section} & \text{Bundle} \\ \hline
x_{13} & \CO(\sigma+2\cb - \CE_1-\CE_2) \\
y_{124} & \CO(\sigma + 3\cb - \CE_1-\CE_2-\CE_4) \\
z_1 & \CO(S_2-\CE_1) \\
\delta_{12} & \CO(E_1-E_2) \\
\delta_{23} & \CO(E_2-E_3) \\
\delta_{34} & \CO(E_3-E_4) \\
\delta_4 & \CO(E_4)
\end{array}\end{equation}
 Several sets of sections that arise in the blow-ups cannot simultaneously vanish.  For later use, we list those below
\begin{equation}\{x_{13},y_{124},z_1\}\,\,\, \{z_1,\delta_4\},\,\,\,\{ z_1,\delta_{34}\},\,\,\, \{y_{124},\delta_{12}\},\,\,\, \{y_{124},\delta_{34}\},\,\,\, \{x_{13},\delta_{23}\},\,\,\, \{\delta_4,\delta_{12}\}\end{equation}
The pullback of our defining equation \eqref{defeqn} to $W_5^{(4)}$ is
\begin{equation}\begin{split}0 &= \delta_{12}^2\delta_{23}^3\delta_{34}^4\delta_4^5 \\
&\times\left[\delta_{12}\delta_{34}\left(\bh_4x_{13}^2 vz_1 + \bh_2 x_{13}v^2z_1^3\delta_{12}\delta_{23} + \bh_0 v^3 z_1^5\delta_{12}^2\delta_{23}^2+x_{13}^3\delta_{34}\delta_4\right)\right.\\
&\qquad\qquad\left.+y_{124}\left(\bh_5x_{13}v+\bh_3v^2z_1^2\delta_{12}\delta_{23}-y_{124}v\delta_{23}\delta_4\right)\right]\end{split}\label{pullbackofY4toX5}\end{equation}
so that the proper transform $Y_4^{(4)}$ is a divisor of $W_5^{(4)}$ in the class
\begin{equation}Y_4^{(4)} = 3\sigma + 6\cb - 2\CE_1 - \CE_2-\CE_3-\CE_4\end{equation}
Because the canonical class of $W_5^{(4)}$ is
\begin{equation}K_{W_5^{(4)}} = -3\sigma=6\cb + 2\CE_1+\CE_2+\CE_3+\CE_4\end{equation}
the 4-fold $Y_4^{(4)}$ is an anti-canonical divisor and hence a Calabi-Yau.  When the $\bh_m$ are generic, this Calabi-Yau is smooth.  As noted in \cite{Esole:2011sm}, the resolution considered here is just one of six birationally equivalent resolutions that are also discussed in \cite{Krause:2011xj}.  In all that follows, we focus on the specific resolution described above.

\subsubsection{Cartan Divisors and Matter Surfaces}
\label{subsubsec:unsplitmattsurfaces}

We would now like to repeat the analysis of fluxes and the Tate divisor in \cite{Marsano:2011hv} for the resolution procedure of \cite{Krause:2011xj}.  Before doing that, it is necessary to review the structure of Cartan divisors and matter surfaces.  This can be obtained from \cite{Krause:2011xj} so we are very brief, presenting results in our current notation.

Above $S_2$, the (resolved) singular fiber splits into multiple components $C_i$.  Each component yields a curve $C_i$ as well as a Cartan divisor obtained by fibering $C_i$ over $S_2$.  Because of the identification of singular fibers over codimension 1 loci in $B_3$ with Dynkin diagrams, we will often refer to components of singular fibers like $C_i$ as nodes.

In $Y_4^{(4)}$, the Cartan divisors are obtained by restricting the irreducible components of the exceptional divisors of the blow-ups.  The nodes of the $A_4$ singular fiber can be obtained from these by restricting the Cartan divisors to points on $S_2$.  We introduce notation for the Cartan divisors and nodes as follows
\begin{equation}\begin{array}{c|c|c}
\text{Divisor Class} & \text{Cartan Divisor Notation} & \text{Node} \\ \hline
\CE_1-\CE_2 & \CD_{12} & C_{12} \\
\CE_2 - \CE_3 & \CD_{23} & C_{23} \\
\CE_3-\CE_4 & \CD_{34} & C_{34} \\
\CE_4 & \CD_4 & C_4 \\
S_2-\CE_1 & \CD_0 & C_0
\end{array}\end{equation}
The intersections of Cartan divisors with nodes should reproduce the $A_4$ Dynkin diagram and this is easy to demonstrate explicitly
\begin{equation}\begin{array}{c|ccccc}
\cdot_{Y_4^{(4)}} & C_{12} & C_{34} & C_4 & C_{23} & C_0 \\ \hline
\CD_{12} & -2 & 1 & 0 & 0 & 0 \\
\CD_{34} & 1 & -2 & 1 & 0 & 0 \\
\CD_4 & 0 & 1 & -2 & 1 & 0 \\
\CD_{23} & 0 & 0 & 1 & -2 & 1 \\
\CD_0 & 0 & 0 & 0 & 1 & -2
\end{array}\end{equation}
The node $C_0$ corresponds to the `extended' node of the $A_4$ Dynkin diagram and does not degenerate in the singular limit in which the resolutions are undone $Y_4^{(4)}\rightarrow Y_4$.

As described in section \ref{subsubsec:cod23}, some of the nodes become reducible above matter curves $\Sigma_{R,\downarrow}$ \eqref{mattercurvesgen} in $S_2$.  This yields new effective curves in the singular fiber above $\Sigma_{R,\downarrow}$ as well as `matter surfaces' $\CS_R$, which are irreducible components of $\pi^*\Sigma_{R,\downarrow}$ that contain one of the new effective curves.

We now describe the matter surfaces in more detail so that they can be used to study the chiral spectrum.  Consider first the $\mathbf{10}$ matter curve, $\Sigma_{\mathbf{10},\downarrow}$, which is given by the restriction of $\bh_5=0$ to $S_2$ \eqref{mattercurvesgen}
\begin{equation}\Sigma_{\mathbf{10},\downarrow}:\,\, \bh_5=z=0\end{equation}
If $\Sigma_{\mathbf{10},\downarrow}$ were a generic curve, this would split into 5 components of the form $\CD_a\cdot_{Y_4^{(4)}} [\bh_5]$ which contain the 5 nodes of the $A_4$ singular fiber respectively.  Because of the enhancement in singularity type above $\Sigma_{\mathbf{10},\downarrow}$, though, two of these components are further reducible and split in the following way:
\begin{equation}\begin{array}{cccccc}
\text{Surface split:} & [\bh_5]\cdot_{Y_4^{(4)}}\CD_{34} &\rightarrow & \CD_{23}\cdot_{Y_4^{(4)}}\CD_{34} &+& ([\bh_5]-\CD_{23})\cdot_{Y_4^{(4)}}\CD_{34} \\
\text{Node spit:} & (1,-2,1,0) &\rightarrow & (1,-1,1,-1) &+&(0,-1,0,1)
\end{array}\end{equation}

\begin{equation}\begin{array}{cccccccc}
\text{Surface split:} & [\bh_5]\cdot_{Y_4^{(4)}} \CD_{23} & \rightarrow & \CD_{12}\cdot_{Y_4^{(4)}}\CD_{23} &+& \CD_{34}\cdot_{Y_4^{(4)}}\CD_{23} &+& ([\bh_5]-\CD_{12}-\CD_{34})\cdot_{Y_4^{(4)}}\CD_{23} \\
\text{Node split:} & (0,0,1,-2) &\rightarrow &(-2,1,0,0) &+& (1,-1,1,-1) &+& (1,0,0,-1) \\
\end{array}\end{equation}

In each case, we have indicated the irreducible surface components as well as the Cartan charges of the corresponding node.  For our spectrum computations, we choose a distinguished $\mathbf{10}$ matter surface
\begin{equation}\CS_{\mathbf{10}} = \left([\bh_5]-\CD_{23}\right)\cdot_{Y_4^{(4)}}\CD_{34} = \left(\cb-\CE_2+\CE_3\right)\cdot_{Y_4^{(4)}}\left(\CE_3-\CE_4\right)\end{equation}
Studying the $G$-flux on $\CS_{\mathbf{10}}$ will tell us only about the spectrum of states that carry Cartan charges $(0,-1,0,1)$ but, provided we choose our fluxes to preserve $SU(5)_{\rm GUT}$, this is enough to determine the spectrum of all states in the $\mathbf{10}$ and $\mathbf{\overline{10}}$ representations.

Turning now to the $\mathbf{\overline{5}}$ matter curve, we recall that it is given by the restriction of $\hat{P}=0$ to $S_2$ \eqref{mattercurvesgen} where $\hat{P}$ is defined as below
\begin{equation}\Sigma_{\mathbf{\overline{5}},\downarrow}:\,\,\,\hat{P} = z=0\qquad \hat{P}\equiv \bh_0\bh_5^2 - \bh_2\bh_3\bh_5 + \bh_3^2\bh_4\end{equation}
In this case, only one of the surfaces $\CD_a\cdot_{Y_4^{(4)}}[\hat{P}]$ is reducible:
\begin{equation}\begin{array}{cccccc}
\text{Surface split:} & \CD_4\cdot_{Y_4^{(4)}} [\hat{P}] &\rightarrow& \CD_4\cdot_{Y_4^{(4)}}([\bh_5x_{13}]-\CD_{34}) &+& \CD_4\cdot_{Y_4^{(4)}} ([\hat{P}] - [\bh_5 x_{13}]+\CD_{34}) \\
\text{Node split:} & (0,1,-2,1) &\rightarrow & (0,1,-1,0) &+& (0,0,-1,1)
\end{array}\end{equation}
We make the following choice for our distinguished $\mathbf{\overline{5}}$ matter surface
\begin{equation}\CS_{\mathbf{\overline{5}}} = \CE_4\cdot_{Y_4^{(4)}}  \left([\bh_5x_{13}]-\CD_{34}\right) = \CE_4\cdot_{Y_4^{(4)}}\left(\sigma + 3\cb - \CE_1-2\CE_3+\CE_4\right)\end{equation}

\subsubsection{Spectral Divisor and Extension of Inherited Fluxes}
\label{subsubsec:extensionofinherited}

We now use the Tate divisor to construct explicit $G$-flux that extends the bundle data $\gamma$ of the generic local models described in section \ref{subsubsec:specgen}.  The Tate divisor in $Y_4^{(4)}$ is a 3-fold obtained from the proper transform of the restriction of
\begin{equation}vy^2 - x^3=0\label{tatetotal}\end{equation}
to $Y_4${\footnote{Note that restricting \eqref{tatetotal} to \eqref{defeqn} is equivalent to \eqref{TateDivisor}.  We work with the equation \eqref{tatetotal} because it makes the proper transform completely manifest without missing any components \cite{Marsano:2011hv}.}}.  Explicitly, this is
\begin{equation}\ctate:\,\,\, vy_{124}^2\delta_{23} - x_{13}^3\delta_{12}\delta_{34}^2\label{tatediveqn1}\end{equation}
When restricted to $Y_4^{(4)}$, this equation defines two irreducible divisors, one of which is a copy of the section $v=x_{13}=0$.
This piece is rather harmless so we let $\ctate$ include it{\footnote{This convention is very helpful in the discussion of non-inherited $G$-fluxes below because it means that the divisor in $W_5^{(5)}$ specified by the defining equation of $\ctate$ is irreducible and effective.  This simpifies the computation of self-intersections of non-inherited $G$-fluxes.  We don't pay a penalty for this anywhere else because all of our matter surfaces miss the zero section anyway so this piece really has no effect on anything.}} so that it is a divisor in the class
\begin{equation}\ctate = 3\sigma + 6\cb  - 2\CE_1 - \CE_2 - \CE_3 - 2\CE_4\end{equation}
The hallmark of $\ctate$ is that its restriction to $\pi^*S_2$ should yield the Higgs bundle spectral cover $\cloc$
\begin{equation}\ctate\cdot_{Y_4^{(4)}} [z] \supset \cloc\end{equation}
and it is this property that will allow us to construct $G$-fluxes that extend local model flux data.  We can study $\ctate\cdot_{Y_4^{(4)}}[z]$ quite explicitly.  It contains several components
\begin{equation}\ctate\cdot_{Y_4^{(4)}}[z] = \ctate\cdot_{Y_4^{(4)}}\CD_4 + \CD_{12}\cdot_{Y_4^{(4)}}\CD_{23} + \CD_{23}\cdot_{Y_4^{(4)}}\CD_{34} + [v]\cdot_{Y_4^{(4)}} [z]\label{ctatez}\end{equation}
We recognize the second and third terms as surfaces that arose in the decomposition of $[\bh_5]\cdot_{Y_4} [z]=0$ and both are essentially fibrations of nodes over the $\mathbf{10}$ matter curve.  The last term comes from the zero section part of $\ctate$.  The first term will give us $\cloc$ and we can see this directly by noting that when $\delta_4=0$ we can effectively set $\delta_{12}=z_1=1$.  Doing this, we find that $\ctate\cdot_{Y_4}^{(4)}\CD_4$ is given by the following 3 equations in $W_5^{(4)}$
\begin{equation}\begin{split}0 &= \delta_4 \\
&= vy_{124}^2\delta_{23} - x_{13}^3\delta_{34}^2 \\
&= \bh_0 v^2\delta_{23}^2\delta_{34} + \bh_2 v x_{13}\delta_{23}\delta_{34} + \bh_3 v y_{124}\delta_{23} + \bh_4 x_{13}^2\delta_{34} + \bh_5 x_{13}y_{124}
\end{split}\end{equation}
Since we are working along $\delta_4=0$ here, we are free to replace the $\bh_m$'s by their restrictions $b_m$ to $S_2$ and can also set $v=1$.  Further, away from the locus $b_0=0$ we have that $x_{13}$ and $y_{124}$ are both nonzero and can be effectively replaced by 1.  Doing this, the second equation tells us that $\delta_{23}=\delta_{34}^2$ which we insert into the third line to obtain
\begin{equation}0 = b_0\delta_{34}^5 + b_2\delta_{34}^3 + b_3\delta_{34}^2 + b_4\delta_{34}+b_5\label{clocemerge}\end{equation}
This is precisely the defining equation for the (noncompact) Higgs bundle spectral cover $\cloc$ of the local model.  The $x_{13}=y_{124}=0$ locus that we removed is the usual component at infinity.  As in \cite{Marsano:2011hv}, it is easy to see that the intersection of $\ctate$ with our matter surfaces yields precisely the local model matter curves \eqref{genericmattcurves} of $\cloc$.

We now construct $G$-fluxes that extend local model data by building surfaces $\CS_{\Sigma}$ inside $\ctate$ that restrict to specified curves $\Sigma$ in $\cloc$.  As we saw in section \ref{subsubsec:specgen}, the curves of interest in $\cloc$ are of two types.  The first is
\begin{equation}\ploc^*D\end{equation}
for $D$ a curve in $S_2$ and $\ploc$ the projection
\begin{equation}\ploc:\cloc\rightarrow S_2\end{equation}
If $D=\hat{D}|_{S_2}$ for some divisor $\hat{D}$ of $B_3$, it is clear that the desired surface $\CS_{\ploc^*D}$ that `extends' $\ploc^*D$ will be of the form $\ctate\cdot_{Y_4^{(4)}}\hat{D}$.

The second type of curve that we consider in $\cloc$ is the intersection of $\cloc$ with $\sigma_{loc}$
\begin{equation}\sigma_{loc}\cdot\cloc\end{equation}
where the intersection is taken in the total space of $K_{S_2}$.  The analysis leading to \eqref{clocemerge} tells us that the section $s=U/V$ in \eqref{Clocalgeneric} is identified, from a global perspective, with $\delta_{34}$ so that the surface $\CS_{\sigma_{loc}\cdot \cloc}$ will take the rough form $(\CE_3-\CE_4)\cdot_{Y_4^{(4)}}\ctate$.

In each case, we have to add some correction terms to make sure that the surface $\CS_{\Sigma}$ satisfies the `one leg on the fiber' condition by having zero intersection with all horizontal and vertical surfaces.  We must also ensure that $\CS_{\Sigma}$ has vanishing intersection with all Cartan surfaces of the form $\CD_a\cdot_{Y_4^{(4)}}D$ for all divisors $D$ in $B_3$ and Cartan divisors $\CD_a$ other than $\CD_4$.  This last condition is necessary because the Tate divisor logic assumes that $\ctate$ only meets the node $C_4$ corresponding to $\CD_4$ above $S_2$.  This is true above generic points in $S_2$ but fails above the $\mathbf{10}$ matter curve due to the additional terms in \eqref{ctatez}{\footnote{The price we pay for not including this correction term is that a traceless local model $\gamma$ would specify a $G$-flux that breaks $SU(5)_{\rm GUT}$.  In this case, we would have to turn on an explicit worldvolume $G$-flux of the form $\CD_a\cdot G'$ to compensate.  Instead of doing this, we add a correction term directly to $\CS_{\sigma_{loc}\cdot\cloc}$ when defining the map from our distinguished set of curves in $\cloc$ to surfaces in $Y_4^{(4)}$.}}.  This only affects $\CS_{\sigma_{loc}\cdot\cloc}$ and we deal with it by adding a Cartan correction term as in \cite{Marsano:2011hv}.  In the end, we construct surfaces $\CS_{\Sigma}$ in $Y_4^{(4)}$ from the indicated curves $\Sigma$ in $\cloc$ via
\begin{equation}\begin{split}\CS_{\sigma_{loc}\cdot\cloc} &= (\CE_3-\CE_4)\cdot_{Y_4^{(4)}}\ctate - \cb\cdot_{Y_4^{(4)}}\CE_2 \\
\CS_{\ploc^*D} &= \hat{D}\cdot_{Y_4^{(4)}}\left[\ctate-(3\sigma + 6\cb)\right]
\end{split}\end{equation}
We can verify that intersection numbers involving these quantities reduce to the expected results in $S_2$ \cite{Marsano:2011nn,Marsano:2011hv}
\begin{equation}\begin{split}
\CS_{\sigma_{loc}\cdot\cloc}\cdot_{Y_4^{(4)}}\CS_{\sigma_{loc}\cdot\cloc} &= -\cb \cdot_{B_3}S_2\cdot_{B_3}S_2 \\
&= (\cs - t)\cdot_{S_2} t \\
\CS_{\sigma_{loc}\cdot\cloc}\cdot_{Y_4^{(4)}}\CS_{\ploc^*D} &= -6\cb\cdot_{B_3}\hat{D}\cdot_{B_3}S_2 \\
&= -6(\cs-t)\cdot_{S_2}D \\
\CS_{\ploc^*D}\cdot_{Y_4^{(4)}}\CS_{\ploc^*D'} &= -30\hat{D}\cdot_{B_3}\hat{D}'\cdot_{B_3}S_2 \\
&= -30 D\cdot_{S_2}D'
\end{split}\end{equation}
Now, we recall that the inherited local model $\gamma$ was constructed from the combination \eqref{gammaugeneric}
\begin{equation}\gamma_{\text{inherited}} = \cloc\cdot \left(5\sigma_{loc}-\ploc^*\cb |_{S_2}\right)\end{equation}
The $G$-flux that extends this bundle data is now the Poincare dual of the holomorphic surface $\CG_{\text{inherited}}$ given by
\begin{equation}\CG_{\text{inherited}} = 5\CS_{\sigma_{loc}\cdot\cloc}-\CS_{\ploc^*\cb}\end{equation}
It is easy to verify that this is truly orthogonal to all Cartan fluxes and hence is $SU(5)_{\rm GUT}$-preserving.  As a sanity check, we can also use \eqref{chiralityfromGflux} to verify that the chiral spectrum of $\mathbf{10}$'s and $\mathbf{\overline{5}}$'s induced by $\CG_{\text{inherited}}$ reproduces the local model results.  By direct computation we find
\begin{equation}\CG_{\text{inherited}}\cdot_{Y_4^{(4)}}\CS_{\mathbf{10}} = \CG_{\text{inherited}}\cdot_{Y_4}^{(4)}\CS_{\mathbf{\overline{5}}} = -\cb \cdot_{B_3}(6\cb - 5S_2)\cdot_{B_3}S_2 = -\eta\cdot_{S_2}(\eta-5\cs)\end{equation}
which is precisely the standard local model formula \eqref{genericspec} \cite{Donagi:2009ra}.

 \subsection{Resolved Geometry and Fluxes for the 4+1 Split Spectral Cover}
\label{subsec:41resolution}

So far we have just managed to show that the analysis of \cite{Marsano:2011hv} can be repeated for the resolution procedure of \cite{Krause:2011xj}.  We now turn our attention to the case of a 4+1 split spectral divisor of the form \eqref{splitspecdiv} and its resolution using the method of \cite{Krause:2011xj}.  To achieve the splitting \eqref{splitspecdiv} we choose special forms for the sections $\bh_m$ that appear in \eqref{defeqn}
\begin{equation}\begin{split}\bh_0 &= -\eh_0^2 \alphah \\
\bh_2 &= \ah_2\eh_0 + \alphah \\
\bh_3 &= \ah_3\eh_0 + \ah_2 \eh_1^2\\
\bh_4 &= \ah_4\eh_0 + \ah_3 \eh_1\\
\bh_5 &= \ah_4 \eh_1
\end{split}\label{splitbhs}\end{equation}
The new objects here are sections of the indicated bundles
\begin{equation}\begin{array}{c|c}
\text{Section} & \text{Bundle} \\ \hline
\eh_0 & \CO(\cb -S_2) \\
\eh_1 & \CO \\
\alphah & \CO(4\cb - 3S_2) \\
\ah_m & \CO([5-m]\cb - [4-m]S_2)
\end{array}\end{equation}
Reducibility of $\ctate$ is supposed to indicate the presence of a $U(1)$ symmetry or, equivalently, the existence of a new divisor in the fully resolved Calabi-Yau that satisfies the `one leg on the fiber' condition by being orthogonal to all horizontal and vertical curves.  We can see this directly by noting that $Y_4^{(4)}$ exhibits a singularity when the $\bh_m$'s are chosen according to \eqref{splitbhs}.  Upon resolving the singularity we will introduce a new divisor class and this will give the freedom to construct a `$U(1)$ divisor'.

The existence of a singularity for $\bh_m$ as in \eqref{splitbhs} is most easily seen by writing the defining equation for $Y_4^{(4)}$ inside $W_5^{(4)}$ as
\begin{equation}\delta_{23} v \left(y_{124}+v\delta_{12}^2\delta_{23}\delta_{34}\eh_0^3z_1^3\eh_1^{-3}\right)\CA =
\left(x_{13}-v\delta_{12}\delta_{23}\eh_0^2z_1^2\eh_1^{-2}\right)\CB\label{singform}\end{equation}
where
\begin{equation}\begin{split}
\CA &= \delta_4\left(y_{124}-v\delta_{12}^2\delta_{23}\delta_{34}\eh_0^3z_1^3\eh_1^{-3}\right) - v\delta_{12}z_1^2\eh_1^{-1}\left(\ah_2\eh_1^2+\eh_0\eh_1\ah_3+\eh_0^2\ah_4\right) \\
\CB &= \ah_4 v y_{124}\eh_1 + \delta_{12}\delta_{34}\left[\delta_{12}^2\delta_{23}^2\delta_{34}\delta_4\eh_0^4\eh_1^{-4} v^2 z_1^4 + x_{13}\left(\delta_{34}\delta_4 x_{13} + (\ah_3\eh_1+\ah_4\eh_0)vz_1\right)\right. \\
&\qquad\qquad\qquad\qquad\qquad\left. + \delta_{12}\delta_{23}vz_1^2\eh_1^{-2}\left(\delta_{34}\delta_4\eh_0^2x_{13}+vz_1([\eh_0\eh_1^2\ah_2+\alphah\eh_1^4]+\eh_0^2\eh_1\ah_3+\eh_0^3\ah_4)\right)\right]
\end{split}\end{equation}
This is manifestly singular along the curve where
\begin{equation}\begin{split}
0 &= y_{124}+v\delta_{12}^2\delta_{23}\delta_{34}\eh_0^3\eh_1^{-3}z_1^3 \\
&= x_{13}-v\delta_{12}\delta_{23}\eh_0^2\eh_1^{-2}z_1^2 \\
&= \CA \\
&= \CB
\end{split}\end{equation}
There is a nice connection between the singularity and the Tate divisor that we would like to emphasize.
Recall that, to see the splitting of the Tate divisor, we looked at the original $Y_4$
\begin{equation}vy^2 = x^3 + \bh_0 z^5 v^3 + \bh_2 z^3 x v^2 + \bh_3 z^2 yv^2 + \bh_4 z x^2 v + \bh_5 xy v\end{equation}
and introduced a meromorphic section
\begin{equation}t=\frac{y}{x}\end{equation}
After stripping off factors involving the section, the restriction of $vy^2=x^3$ to the Tate equation took the form
\begin{equation}\bh_0 z^5 + \bh_2 z^3t^2 + \bh_3 z^2 t^3 + \bh_4 z t^4 + \bh_5 t^5\end{equation}
and the choices \eqref{splitbhs} ensure that this admits a linear factor
\begin{equation}\eh_0 z + t\eh_1\end{equation}
The linear factor describes one of the components of $\ctate$ but, as written, it is not defined by the vanishing of a single holomorphic section on $Y_4$.  If we follow through the (partial) resolution of $Y_4$ that yields $Y_4^{(4)}$, we can actually describe the linear component of $\ctate$ as a 3-fold in $W_5^{(4)}$ defined by
\begin{equation}\begin{split}0 &= y_{124} + v\delta_{12}^2\delta_{23}\delta_{34}\eh_0^3 \eh_1^{-3}z_1^3 \\
&= x_{13} - v\delta_{12}\delta_{23}\eh_0^2\eh_1^{-2}z_1^2\end{split}\end{equation}
The form of \eqref{singform}, though, makes clear that this is not a divisor that is inherited from $W_5^{(4)}$.  It is in fact a Weil divisor in $Y_4^{(4)}$ that is not Cartier and its presence signals the fact that $Y_4^{(4)}$ remains singular.  This had to be the case on rather general grounds; we constructed the distinct components of the Tate divisor so that one could take a linear combination as in \eqref{U1divgen} to obtain a divisor that is simultaneously nontrivial and orthogonal to all horizontal curves, vertical curves, and nodes of the $A_4$ singular fibers.  Our only Cartier divisors are horizontal divisors, vertical divisors, and $A_4$ Cartan divisors, though, so the divisor in \eqref{U1divgen} and consequently the individual Tate divisor components cannot possibly be Cartier.

The form of \eqref{singform} suggests a natural way to fully resolve $Y_4^{(4)}$ that was used in \cite{Krause:2011xj}.  We simply blow up $W_5^{(4)}$ along the 3-fold
\begin{equation}\begin{split}0 &= y_{124}+v\delta_{12}^2\delta_{23}\delta_{34}\eh_0^3\eh_1^{-3}z_1^3 \\
&= x_{13}-v\delta_{12}\delta_{23}\eh_0^2\eh_1^{-2}z_1^2
\end{split}\end{equation}
and take a proper transform to get $Y_4^{(5)}$.  In the Tate divisor language, we blow-up along the 3-fold that describes the linear component of the Tate divisor.  After we do this, the linear piece of $\ctate$ will obviously be Cartier in $Y_4^{(5)}$ because it will be the restriction of the exceptional divisor associated to the last blow-up.  Correspondingly its complement inside $\ctate$, the quartic piece, will be Cartier as well.

\subsubsection{Split Tate Divisor and $U(1)$}
\label{subsubsec:splittatedivisorandU1}

Let us describe the final resolution step and the split Tate divisor more explicitly.  When we perform this last blow-up, we obtain new sections $X$ and $Y$ on $W_5^{(5)}$ along with a unique section $\delta_5$ of $\CO(\CE_5)$ whose vanishing defines the new exceptional divisor $\CE_5$.  These satisfy
\begin{equation}\begin{split}y_{124}+v\delta_{12}^2\delta_{23}\delta_{34}\eh_0^3\eh_1^{-3}z_1^3 &= Y\delta_5 \\
x_{13}-v\delta_{12}\delta_{23}\eh_0^2\eh_1^{-2}z_1^2 &= X\delta_5
\end{split}\end{equation}
The new sections are associated to the obvious bundles
\begin{equation}\begin{array}{c|c}
\text{Section} & \text{Bundle} \\ \hline
X & \CO(\sigma+2\cb - \CE_1 - \CE_3-\CE_5) \\
Y & \CO(\sigma + 3\cb - \CE_1-\CE_2-\CE_4-\CE_5) \\
\delta_5 & \CO(\CE_5)
\end{array}\end{equation}
The defining equation of $Y_4^{(5)}$ inside $W_5^{(5)}$ is
\begin{equation}v\delta_{23}Y\CA = X\CB\end{equation}
and it is in the class
\begin{equation}3\sigma + 6\cb - 2\CE_1-\CE_2-\CE_3-\CE_4-\CE_5\end{equation}
so that it remains an anti-canonical divisor.  It is convenient to list the sets of sections that do not vanish when $\delta_5$ does
\begin{equation}\{\delta_5,\delta_{23}\}\,\,\,\{\delta_5,\delta_{12}\},\,\,\,\{\delta_5,\delta_{34}\},\,\,\,\{\delta_5,z_1\}\end{equation}

Return now to the equation \eqref{tatediveqn1} for the Tate divisor.  After the 5th blow-up, this becomes
\begin{equation}0=\delta_5\left[\delta_{12}\delta_{34}^2\delta_5^2 X^3 - \delta_{23}\delta_5 v Y^2 + 3\delta_{12}^3\delta_{23}^2\delta_{34}^2\eh_0^4\eh_1^{-4}z_1^4v^2 X + \delta_{12}^2\delta_{23}\delta_{34}\eh_0^2\eh_1^{-2}z_1^2\left(3\delta_{34}\delta_5 v X^2 + 2\delta_{23}\eh_0\eh_1^{-1} v^2 Y z_1\right)\right]\label{tatediveqn2}\end{equation}
The $\delta_5=0$ piece gives the restriction of $\CE_5$ to $Y_4^{(5)}$ which, as we said above, is the linear component $\ctate^{(1)}$ of $\ctate$
\begin{equation}\ctate^{(1)} = \CE_5\end{equation}
Correspondingly, $\ctate^{(4)}$ is given by the other factor of \eqref{tatediveqn2} and is in the class
\begin{equation}\ctate^{(4)} = 3\sigma + 6\cb - 2\CE_1 - \CE_2 - \CE_3 - 2\CE_4 - \CE_5\end{equation}
It is easy to verify that these have the expected intersections with the $A_4$ nodes.  For $\ctate^{(4)}$ these intersections are $(0,4,0,0)$ while for $\ctate^{(1)}$ they are $(0,1,0,0)$, consistent with the fact that $\ctate^{(m)}$ locally behaves like $m$ lines that meet the second node exactly once and miss all others \cite{Marsano:2011hv}.

To build a $U(1)$ divisor, we want to take the combination $\ctate^{(4)}-4\ctate^{(1)}$ \eqref{U1divgen} \cite{Dolan:2011iu,Marsano:2011nn}.  This certainly has the feature that it is orthogonal to all $A_4$ nodes but it is not quite what we want because it is not orthogonal to all horizontal and vertical curves and hence fails the `one leg on the fiber' condition.  To fix this, we have to add a correction term
\begin{equation}\begin{split}\omega &= \ctate^{(4)}-4\ctate^{(1)} - \cb - \frac{4}{3}\sigma\\
&= \frac{5\sigma}{3} + 5\cb - 2\CE_1 - \CE_2-\CE_2-2\CE_4-5\CE_5\end{split}\label{omega}\end{equation}
Note that while $\sigma/3$ is not an integral divisor in $W_5^{(5)}$, it is easy to verify that its restriction to $Y_4^{(5)}$ is integral.  The divisor $\omega$ satisfies the `one leg on the fiber' condition and is orthogonal to all $A_4$ nodes so its corresponding $(1,1)$-form gives a $U(1)$ that commutes with $SU(5)_{\rm GUT}$.
This divisor was first constructed for Calabi-Yau's that exhibit the special 4+1 splitting of $\ctate$ induced by \eqref{splitbhs} in \cite{Krause:2011xj} and we have managed here to connect it to the general procedure of the Tate divisor formalism of \cite{Marsano:2010ix,Marsano:2011nn}.

\subsubsection{Matter Surfaces}

We now turn to the structure of matter surfaces in $Y_4^{(5)}$.  As these are discussed in \cite{Krause:2011xj} we are rather brief.  For the special choice \eqref{splitbhs}, the $\mathbf{10}$ matter curve in $S_2$ remains effectively unchanged so the structure of $\mathbf{10}$ matter surfaces also carries over directly from section \ref{subsubsec:unsplitmattsurfaces}.  We will choose as our distinguished $\mathbf{10}$ matter surface
\begin{equation}\CS_{\mathbf{10}}^{(4)} = \left([\bh_5]-\CD_{23}\right)\cdot_{Y_4^{(5)}}\CD_{34} = (\cb-\CE_2+\CE_3)\cdot_{Y_4^{(5)}}(\CE_3-\CE_4)\label{10mattsurf}\end{equation}
The node associated to this surface is a curve whose homology class is specified by its intersections with the Cartan divisors as well as the $U(1)$ divisor $\omega$ \eqref{omega}.  These are easily computed and we write them with the $\omega$ charge indicated as a subscript
\begin{equation}(0,-1,0,1)_{+1}\end{equation}

Turning now to the $\mathbf{\overline{5}}$'s, we saw in the local model that the $\mathbf{\overline{5}}$ matter curve in $S_2$ is reducible
\begin{equation}\Sigma_{\mathbf{\overline{5}},\downarrow}:\,\,\,z= \hat{P}^{(44)}\hat{P}^{(41)}=0\end{equation}
with
\begin{equation}\begin{split}\hat{P}^{(44)} &= \ah_4\alphah - \ah_3(\ah_2+\ah_3\eh_0) \\
\hat{P}^{(41)} &= \ah_4\eh_0^2 + \ah_3\eh_0 + \ah_2
\end{split}\end{equation}
Correspondingly we expect distinct $\mathbf{\overline{5}}$ surfaces associated to $\hat{P}^{(44)}$ and $\hat{P}^{(41)}$.  To study these, we start by looking at the surface $z=\hat{P}^{(44)}=0$.  Most of the components of this reducible surface take the form $\CD_a\cdot_{Y_4^{(5)}}[\hat{P}^{(44)}]$ and include a node of the singular fiber corresponding to an $A_4$ root.  The component $\CD_4\cdot_{Y_4^{(5)}}[\hat{P}^{(44)}]$ is reducible, though
\begin{equation}\begin{array}{ccccccc}
\text{Surfaces:} & \CD_4\cdot_{Y_4^{(5)}}[\hat{P}^{(44)}] & \rightarrow & \CD_4\cdot_{Y_4^{(5)}}([a_4x_{13}]-[\delta_{34}]-[X]) & + & \CD_4\cdot_{Y_4^{(5)}}([\hat{P}^{(44)}]-[a_4x_{13}]+[\delta_{34}]+[X]) \\
\text{Nodes:}& (0,1,-2,1)_0 &\rightarrow & (0,1,-1,0)_{+2} & + & (0,0,-1,1)_{-2}
\end{array}\end{equation}
Here we have labeled the node associated to each surface by its Cartan charges as well as its intersection with the divisor $\omega$ that defines our $U(1)$.  We can do the same thing for $z=\hat{P}^{(41)}=0$.  Again all components are complete intersections with Cartan divisors except the ones that descend from $\CD_4\cdot_{Y_4^{(5)}}\hat{P}^{(41)}$.  In that case, we have the further splitting
\begin{equation}\begin{array}{cccccc}
\text{Surfaces:} & \CD_4\cdot_{Y_4^{(5)}}[\hat{P}^{(41)}] &\rightarrow & \CD_4\cdot_{Y_4^{(5)}} [X] &+&\CD_4\cdot_{Y_4^{(5)}}([\hat{P}^{(41)}]-[X]) \\
\text{Nodes:} & (0,1,-2,1)_0 &\rightarrow & (0,1,-1,0)_{-3} &+& (0,0,-1,1)_{+3}
\end{array}\end{equation}
Note that the $U(1)$ is able to distinguish nodes above $\hat{P}^{(44)}$ and $\hat{P}^{(41)}$ that have identical Cartan charges.  To study the spectrum of $\mathbf{\overline{5}}^{(44)}$'s and $\mathbf{\overline{5}}^{(41)}$'s we will choose distinguished matter surfaces for each
\begin{equation}\begin{split}\CS_{\mathbf{\overline{5}}}^{(44)} &= \CD_4\cdot_{Y_4^{(5)}}([a_4x_{13}]-[\delta_{34}]-[X]) \\
&= \CD_4\cdot_{Y_4^{(5)}} \left(\cb - \CE_3+\CE_4+\CE_5\right) \\
\CS_{\mathbf{\overline{5}}}^{(41)} &= \CD_4\cdot_{Y_4^{(5)}} [X] \\
&= \CD_4\cdot_{Y_4^{(5)}}\left(\sigma +2\cb - \CE_1-\CE_3-\CE_5\right)
\end{split}\label{5mattsurfs}\end{equation}
The states associated with these matter surfaces carry Cartan and $\omega$ charges given by
\begin{equation}\CS_{\mathbf{\overline{5}}}^{(44)}\leftrightarrow (0,1,-1,0)_{+2}\qquad \CS_{\mathbf{\overline{5}}}^{(41)}\leftrightarrow (0,1,-1,0)_{-3}\end{equation}

\subsubsection{Extending the Inherited Fluxes}
\label{subsubsec:extendinginherited}

We now proceed to construct $G$-fluxes that extend the inherited bundle data $\gamma^{(u)}$, $\gamma^{(p)}$, and $\gamma^{(\rho)}$ of \ref{subsubsec:chiralitygen}.  As in section \ref{subsubsec:extensionofinherited}, we first map curves in $\cloc$ to suitable surfaces in $\ctate$
\begin{equation}\begin{split}\CS_{\sigma_{loc}\cdot \cloc^{(4)}} &= (\CE_3-\CE_4)\cdot_{Y_4^{(5)}}\ctate^{(4)} - \cb\cdot{Y_4^{(5)}}\CE_2 \\
\CS_{\ploc_4^*D} &= \ctate^{(4)}\cdot_{Y_4^{(5)}}\hat{D}- (3\sigma + 6\cb)\cdot_{Y_4^{(5)}}\hat{D} \\
\CS_{\ploc_1^*D} &= \ctate^{(1)}\cdot_{Y_4^{(5)}}\hat{D}
\end{split}\end{equation}
From these, we obtain $G$-fluxes corresponding to the $\gamma$'s of \eqref{gammabuildingblocks} by including suitable subtractions to ensure orthogonality to horizontal and vertical classes
\begin{equation}\begin{split}
\CG_u &= 4\CS_{\sigma_{loc}\cdot \cloc^{(4)}} - \CS_{p_4^*\cb} - \frac{1}{3}\sigma \cdot_{Y_4^{(5)}}\cb - \cb\cdot_{Y_4^{(5)}}\cb \\
\CG_p &= \CS_{\sigma_{loc}\cdot \cloc^{(4)}} -  \CS_{p_1^*\cb} + \frac{1}{3}\sigma \cdot_{Y_4^{(5)}}\cb + \cb\cdot_{Y_4^{(5)}}\cb \\
\CG_{\rho} &=  \CS_{\sigma_{loc}\cdot \cloc^{(4)}} - 4 \CS_{\sigma_{loc}\cdot \cloc^{(4)}} + 5\cb\cdot_{Y_4^{(5)}}\rho + \frac{5}{3}\sigma\cdot_{Y_4^{(5)}}\rho \\
&= \omega\cdot_{Y_4^{(5)}}\rho
\end{split}\label{extendedinherited}\end{equation}
In the last of these, we reproduce the $U(1)$ flux of \cite{Krause:2011xj}.  The others admit natural descriptions in terms of matter surfaces; this approach has recently been discussed in \cite{Grimm:2011fx}.

It is a simple matter to check that the fluxes in \eqref{extendedinherited} are orthogonal to all surfaces obtained as complete intersections with $SU(5)_{\rm GUT}$ Cartan divisors so that they are truly $SU(5)_{\rm GUT}$-preserving.  As a sanity check, we can also verify that local model chirality results are correctly reproduced by computing the intersections of the fluxes in \eqref{extendedinherited} with our matter surfaces
$\CS_{\mathbf{10}}^{(4)}$, $\CS_{\mathbf{\overline{5}}}^{(44)}$, and $\CS_{\mathbf{\overline{5}}}^{(41)}$.  We find
\begin{equation}\begin{array}{c|ccc}
\cdot_{Y_4^{(5)}} & {\cal{G}}_u & {\cal{G}}_p & {\cal{G}}_{\rho} \\ \hline
{\cal{S}}_{\mathbf{10}}^{(4)} & -\cb(5\cb-4S_2)S_2 & -\cb(\cb-S_2)S_2 & \cb S_2\rho \\
{\cal{S}}_{\mathbf{\overline{5}}}^{(44)} & -2\cb(\cb-S_2)S_2 & \cb S_2(2\cb -S_2) & 2S_2(5\cb-3S_2)\rho \\
{\cal{S}}_{\mathbf{\overline{5}}}^{(41)} & -\cb (3\cb -2S_2)S_2 & -\cb (3\cb -2S_2)S_2 & -3(3\cb -2S_2)S_2\rho \\
\end{array}\end{equation}
where the intersections appearing in the tableÊ
entries are evaluated in $B_3$.  As expected, these reduce to intersections in $S_2$ as
\begin{equation}\begin{array}{c|ccc}
\cdot_{Y_4} & {\cal{G}}_u & {\cal{G}}_p & {\cal{G}}_{\rho} \\ \hline
{\cal{S}}_{10}^{(4)} &  -\eta_4(\eta_4-4c_1) & -c_1(\eta_4-4c_1) & \rho(\eta_4-4c_1) \\
{\cal{S}}_{\overline{5}}^{(44)} & -2c_1(\eta_4-4c_1) & \eta_4^2-7c_1\eta_4+12c_1^2 & 2\rho(2\eta_4-5c_1) \\
{\cal{S}}_{\overline{5}}^{(41)} & -8c_1^2+6c_1\eta_4-\eta_4^2 & -8c_1^2+6c_1\eta_4-\eta_4^2 & -3\rho(\eta_4-2c_1)
\end{array}\end{equation}
where the intersections appearing in these table entries are evaluated in $S_2$.  We recognize these as the local model results of \eqref{gammachiralities}.

\subsubsection{Extending the Non-inherited Fluxes}
\label{subsubsec:extendingnoninherited}

For our K3-based model, the inherited fluxes $\gamma^{(u)}$, $\gamma^{(p)}$, and $\gamma^{(\rho)}$ were not sufficient to engineer three chiral generations.  This forced us to tune $\cloc$ so that it contained additional holomorphic curves that were not inherited from the ambient space.  In that construction, we tuned the local model sections according to \eqref{tuningchoice}.  We now demonstrate that the bundle data $\gamma_{\text{noninherited}}$ of \eqref{gammanoninherited} can be extended to a globally well-defined $G$-flux provided we impose \eqref{tuningchoice} globally as in
\begin{equation}\begin{split}
\ah_4 &= \ch_4\hh_0 - \dhat_3\hh_2 \\
\ah_3 &= \ch_3\hh_0 + \dhat_3\hh_1 - \dhat_2\hh_2 \\
\ah_2 &= \ch_4\hh_0 + \dhat_2\hh_1 \\
\alphah &= \ch_1 \hh_0
\end{split}\label{globaltuningchoice}\end{equation}
where the new objects appearing here are sections of the indicated bundles on $B_3$
\begin{equation}\begin{array}{c|c}
\text{Section} & \text{Bundle} \\ \hline
\hh_0 & \CO(\xi) \\
\hh_1 & \CO(\chi) \\
\hh_2 & \CO(\chi+S_2-\cb) \\
\dhat_m & \CO([5-m]\cb - [4-m]S_2 - \chi) \\
\ch_p & \CO([5-m]\cb - [4-m]S_2-\xi)
\end{array}\end{equation}

Recall that $\gamma_{\text{noninherited}}$ was built as the difference of two interesting divisors from the local model \eqref{interestingdivisors}
\begin{equation}\begin{split}D_1: & h_0 = h_1U-h_2V=0 \\
D_2: & h_0 = d_2U+d_3V=0
\end{split}\label{interestdivisors4}\end{equation}
These, in turn, arose because $\cloc^{(4)}$ split into components when restricted to $h_0=0$.  Similarly, because we extended this structure globally through \eqref{globaltuningchoice} we expect the surface obtained by restricting $\ctate^{(4)}$ to $\hh_0=0$ to split into multiple components of which two will essentially be the global extensions of $D_1$ and $D_2$.  The easiest way to proceed is to go back to the original defining equation for $\ctate$ \eqref{tatediveqn1}
\begin{equation}vy_{124}^2\delta_{23} = x_{13}^3\delta_{12}\delta_{34}^2\label{backtoctate}\end{equation}
and, working in a patch where $y_{124}$ and $v$ are both nonzero, solve for $\delta_{23}$ and insert the result into the defining equation of our Calabi-Yau before the final blow-up, $Y_4^{(4)}$, which can be obtained from \eqref{pullbackofY4toX5}.  This produces a rather complicated expression but it simplifies considerably if we use the form \eqref{splitbhs}, \eqref{globaltuningchoice} and restrict to $\hh_0=0$
\begin{equation}\frac{x_{13}}{y_{124}^2}\left(\hat{e}_1y_{124}+\delta_{12}\delta_{34}\eh_0 x_{13}z_1\right)\left(\dhat_3y_{124}+\dhat_2\delta_{12}\delta_{34}x_{13}z_1\right)\left(-\hh_2y_{124}+\hh_1\delta_{12}\delta_{34}x_{13}z_1\right)\end{equation}
We ignore the $x_{13}$ factor since this is nonzero everywhere along $\ctate$ where $y_{124}\ne 0$.  We recognize the next factor as $\ctate^{(1)}$ and the remaining two as candidate extensions of our local model curves $D_1$ and $D_2$ \eqref{interestdivisors4}.  More specifically, this tempts us to define new surfaces from which to construct extensions of $D_1$ and $D_2$ as
\begin{equation}\begin{split}\CG_1^{(initial)} &= [\hh_0]\cdot_{W_5^{(5)}} \ctate^{(4)}\cdot_{W_5^{(5)}} [\hh_2 y_{124} - \delta_{12}\delta_{34}\hh_1 z_1 x_{13}] \\
\CG_2^{(initial)} &= [\hh_0]\cdot_{W_5^{(5)}}\ctate^{(4)}\cdot_{W_5^{(5)}}[\dhat_3y_{124}+\dhat_2\delta_{12}\delta_{34}x_{13}z_1]
\end{split}\label{noninheritedinitial}\end{equation}
where here by $\ctate^{(4)}$ we mean the divisor in $W_5^{(5)}$ that is specified by the proper transform of \eqref{backtoctate} under the final resolution step.  We obtain this from \eqref{tatediveqn2} by stripping the overall factor of $\delta_5$ and it is the divisor that restricts to the quartic component of the Tate divisor in $Y_4^{(5)}$.  We must actually be a little careful here because each of the surfaces in \eqref{noninheritedinitial} is reducible into two components with one component common to both
\begin{equation}[\hh_0]\cdot_{W_5^{(5)}}\ctate^{(4)} \cdot_{W_5^{(5)}} [y_{124}]\end{equation}
Fortunately, the non-inherited flux that we needed in the local model was constructed from the difference $D_1-D_2$ so we are ultimately interested in a $G$-flux obtained from the difference
\begin{equation}\CG_{\text{noninherited}}^{(0)} = \CG_1^{(initial)} - \CG_2^{(initial)}\end{equation}
from which the common component drops out.
We are not quite done because this is not yet orthogonal to all horizontal and vertical classes.  As usual, we can fix this with a simple correction term.  In the end, we define
\begin{equation}\begin{split}\CG_{\text{noninherited}} &= \CG_{\text{noninherited}}^{(0)}
\cdot_{W_5^{(5)}} ([\hh_2]-[\dhat_3]) + (3\cb - 2S_2 - 2\chi)\cdot_{Y_4^{(5)}}\xi \\
&=E_4\cdot_{W_5^{(5)}}\cdot \xi\cdot_{W_5^{(5)}}\cdot (3\cb - 2(S_2+\chi))
\end{split}\end{equation}
where in the last line we presented the class of $\CG_{\text{noninherited}}$ as a surface in $W_5^{(5)}$.
It is now a simple matter to verify that $\CG_{\text{noninherited}}$ is a linear combination of surfaces inside $Y_4^{(5)}$ that is orthogonal to all horizontal surfaces, vertical surfaces, and surfaces obtained as complete intersections with Cartan divisors.  As a result, it is a good $SU(5)_{\rm GUT}$-preserving $G$-flux.

As we might have expected from its connection to non-inherited curves in the local model, $\CG_{\text{noninherited}}$ includes a piece that is not the restriction to $Y_4^{(5)}$ of a 3-fold in $W_5^{(5)}$.  Nevertheless, it is straightforward to compute the intersection of $\CG_{\text{noninherited}}$ with our matter surfaces.  The matter surfaces $\CS_{\mathbf{R}}$ \eqref{10mattsurf},\eqref{5mattsurfs} descend from 3-folds in $W_5^{(5)}$ so intersections of the form $\CG_{\text{noninherited}}\cdot_{Y_4^{(5)}}\CS_{\mathbf{R}}$ still lift to complete intersections in $W_5^{(5)}$ that are easily evaluated.  We find
\begin{equation}\begin{array}{c|c}
\cdot_{Y_4^{(5)}} & \CG_{\text{noninherited}} \\ \hline
\CS_{\mathbf{10}}^{(4)} & -S_2 \xi (3\cb - 2S_2-2\chi) \\
\CS_{\mathbf{\overline{5}}}^{(44)} & 0 \\
\CS_{\mathbf{\overline{5}}}^{(41)} & -S_2\xi (3\cb-2S_2-2\chi)
\end{array}\end{equation}
where the intersections appearing in the table entries are computed in $B_3$.  These reduce to intersections in $S_2$ that can be phrased in a local model language
\begin{equation}\begin{array}{c|c}
\cdot_{Y_4^{(5)}} & \CG_{\text{noninherited}} \\ \hline
\CS_{\mathbf{10}}^{(4)} & -\xi\cdot_{S_2}(\eta_4-2c_1-2\chi) \\
\CS_{\mathbf{\overline{5}}}^{(44)} & 0 \\
\CS_{\mathbf{\overline{5}}}^{(41)} & -\xi\cdot_{S_2}(\eta_4-2c_1-2\chi)
\end{array}\end{equation}
which exactly reproduces the results that we obtained in the local model \eqref{noninherspectrum}.

We now close with a useful technical computation involving $\CG_{\text{noninherited}}$.  $G$-fluxes induce a 3-brane tadpole proportional to their self-intersection.  For most of the $G$-fluxes we have built, this is a simple thing to compute but for $\CG_{\text{noninherited}}$ it can be tricky to evaluate
because $\CG_{\text{noninherited}}$ is not inherited from a 3-fold inside the ambient space $W_5^{(5)}$.  To compute $\CG_{\text{noninherited}}\cdot_{Y_4^{(5)}}\CG_{\text{noninherited}}$ , we can proceed in steps.  First we remove the common component directly from $\CG_1^{(initial)}$ and $\CG_2^{(initial)}$ in \eqref{noninheritedinitial} to get
\begin{equation}\begin{split}\CG_1 &= \CG_1^{(initial)} - [\hh_0]\cdot_{W_5^{(5)}}\ctate^{(4)}\cdot_{W_5^{(5)}} [y_{124}] \\
&= [\hh_0]\cdot_{W_5^{(5)}}\ctate^{(4)}\cdot_{W_5^{(5)}}\left([\hh_2y_{124}+\ldots]-[y_{124}]\right) \\
\CG_2 &= \CG_2^{(initial)} - [\hh_0]\cdot_{W_5^{(5)}}\ctate^{(4)}\cdot_{W_5^{(5)}} [y_{124}] \\
&= [\hh_0]\cdot_{W_5^{(5)}} \ctate^{(4)}\cdot_{W_5^{(5)}} \left([\dhat_3y_{124}+\ldots] - [y_{124}]\right)
\end{split}\end{equation}
Because we have just removed common components from each, we have that
\begin{equation}\CG_{\text{noninherited}} = \CG_1 - \CG_2\end{equation}
We can evaluate the individual self-intersections $\CG_1\cdot_{Y_4^{(5)}}\CG_1$ and $\CG_2\cdot_{Y_4^{(5)}}\CG_2$ following \cite{Braun:2011zm} by using the general result
\begin{equation}\CS\cdot_{Y_4^{(5)}}\CS = \int_{[\CS]} c_2(N_{\CS/Y_4^{(5)}})\end{equation}
In our case, the normal bundle of $\CG_a$ in $W_5^{(5)}$ takes the form
\begin{equation}N_{\CG_a/W_5^{(5)}} = \oplus_i \CO(D_i)\end{equation}
for $D_i$ some divisor classes in $W_5^{(5)}$.  This simplifies the computation which ultimately yields
\begin{equation}\begin{split}
\CG_1\cdot_{Y_4^{(5)}}\CG_1 &= 4S_2\xi \chi \\
\CG_2\cdot_{Y_4^{(5)}}\CG_2 &= 4S_2\xi (-2\cb + S_2+\chi)
\end{split}\end{equation}
where the intersections here are performed in $B_3$.  What remains is to compute
\begin{equation}\CG_1\cdot_{Y_4^{(5)}}\CG_2\end{equation}
which is a little tricky because $\CG_1$ and $\CG_2$ are both surfaces in $W_5^{(5)}$ that sit inside the common 3-fold $Z_3$
\begin{equation}Z_3 = [\hh_0]\cdot_{W_5^{(5)}}\ctate^{(4)}\end{equation}
As a result, $\CG_1$ and $\CG_2$ meet in a curve as opposed to a collection of isolated points.  The computation we need to perform, then, is
\begin{equation}\CG_1\cdot_{Y_4^{(5)}}\CG_2 = \int_C c_1(N_{(Z_3|_{Y_4^{(5)}})/Y_4^{(5)}})\end{equation}
where $C$ is the common curve.  This evaluates to
\begin{equation}\begin{split}\CG_1\cdot_{Y_4^{(5)}}\CG_2 &= \left(\ctate^{(4)}+[\hh_0] - [Y_4^{(5)}]\right)\cdot_{W_5^{(5)}}\ctate^{(4)}\cdot_{W_5^{(5)}}[\hh_0] \\
&\qquad \cdot_{W_5^{(5)}}\left([\hh_2y_{124}+\ldots] - [y_{124}]\right)\cdot_{W_5^{(5)}}\left([\dhat_3y_{124}+\ldots]-[y_{124}]\right) \\
&= 0
\end{split}\end{equation}

Summing everything, we find that
\begin{equation}\CG_{\text{noninherited}}\cdot_{Y_4^{(5)}}\CG_{\text{noninherited}} = 4\xi \cb (2\cb - S_2)\end{equation}
which is proportional to the intersection of three effective divisors in $B_3$.

\section{F-Enriques: The Global Model}
\label{sec:globalmodel}

We are finally ready to write down our complete global model for a 3 generation $SU(5)_{\rm GUT}$ that achieves GUT-breaking with a discrete Wilson line.  Because we were careful to specify the local model in section \ref{subsec:EnriquesLocalSpec} in terms of holomorphic sections that descend from the 3-fold $B_3$,
specifying the global completion is actually very easy.  All of the hard work has already been done in section \ref{subsec:41resolution}.

\subsection{Geometry and Fluxes}

We start with the toric space $X_5$ \eqref{W5data} obtained from a pair of blow-ups in $\mathbb{P}^5$ and specify a 3-fold $B_3$ as the intersection of the quadric $Q_2^{(2)}$ from \eqref{firsttwoquadrics} with the pullback to $X_5$ of a generic quadric $Q_2^{(3)}$ on $\mathbb{P}^5$ that is invariant under the $\mathbb{Z}_2$ action \eqref{Z2P5}, \eqref{Z2W5}.  The quadric $Q_2^{(1)}$ specifies the GUT divisor in $B_3$ and is the holomorphic section that we have been calling $z$ in our general analyses
\begin{equation}z = Q_2^{(1)}(\tilde{u}_1,\tilde{v}_1,\delta_1,\tilde{u}_2,\tilde{v}_2,\delta_2,u_3,v_3)\end{equation}
Now we specify our singular Calabi-Yau 4-fold as an elliptic fibration over $B_3$ by starting with the Tate form \eqref{defeqn} and specializing according to \eqref{splitbhs} and \eqref{globaltuningchoice}.  The final result is a hypersurface $Y_{4,\uparrow}$ inside $\mathbb{P}(\CO\oplus K_{B_3}^{-2}\oplus K_{B_3}^{-3})$ defined by the equation
\begin{equation}\begin{split}vy^2 &= x^3  - \ch_1\eh_0^2\hh_0 v^3z^5 + (\ch_1\hh_0\eh_1^2 + \ch_4\eh_0\hh_0+\dhat_2\eh_0\hh_1)xv^2 z^3\\
& + (\ch_4\hh_0\eh_1 + \ch_3\eh_0\hh_0 + \dhat_2\hh_1\eh_1 + \dhat_3\eh_0\hh_1 - \dhat_2\eh_0\hh_2)yv^2z^2 \\
& + (\ch_3\hh_0 \eh_1+ \ch_4\eh_0\hh_0 + \dhat_3\hh_1\eh_1-\dhat_2\hh_2\eh_1-\dhat_3\eh_0\hh_2)x^2v z + \eh_1(\ch_4\hh_0  - \dhat_3\hh_2)xyv
\end{split}\label{finalCY}\end{equation}
This will give an $SU(5)_{\rm GUT}$ singularity over the K3 surface defined by $z=0$.  Our final Calabi-Yau geometry will be a $\zenriques$ quotient of this, $Y_4=Y_{4,\uparrow}/\zenriques$, that utilizes the geometric $\zenriques$ symmetry of $B_3$ that descends from \eqref{Z2P5} and \eqref{Z2W5} and acts freely on $S_2=K3$.  In the local model, we had to assign a `-' charge to $V$ because it is twisted by the anti-canonical bundle of $S_2$ which is $\zenriques$-odd.  This forced us to assign nontrivial $\zenriques$ charges \eqref{amcharges} \eqref{tunedbundles} to several of the holomorphic sections that defined $\cloc$.  We will carry these over to their global extensions as sections of suitable bundles on $B_3$.  With those charge assignments, we are required to extend the action of the $\zenriques$ on $B_3$ to the $\mathbb{P}^2$ fiber of $\mathbb{P}(\CO\oplus K_{B_3}^{-2}\oplus K_{B_3}^{-3})$ if we want the defining equation of our Calabi-Yau \eqref{finalCY} to be $\zenriques$ invariant.  We summarize the bundles associated to all objects in \eqref{finalCY} along with their $\zenriques$ charges in the following table
\begin{equation}\begin{array}{c|c|c}
\text{Section} & \text{Bundle} & \zenriques \\ \hline
v & \CO(\sigma) & + \\
x & \CO(\sigma + 2[2H-E_1]) & + \\
y & \CO(\sigma + 3[2H-E_1]) & - \\
\hh_0 & \CO(E_2) & + \\
\hh_1 & \CO(H-E_1) & + \\
\hh_2 & \CO(H-E_1) & - \\
\ch_3 & \CO(2H-E_1-E_2) & - \\
\ch_1, \ch_4 & \CO(2H-E_1-E_2) & + \\
\dhat_2 & \CO(H) & + \\
\dhat_3 & \CO(H) & - \\
\eh_0 & \CO & + \\
\eh_1 & \CO & - \\
\end{array}\end{equation}
It is a simple matter to check that the holomorphic $(4,0)$-form of \eqref{finalCY} is $\zenriques$-invariant so the quotient also defines a Calabi-Yau.  Locally the cotangent bundle of $Y_{4,\uparrow}$ splits into the sum of the cotangent bundle $T^*B_3$ of $B_3$ and the canonical bundle $K_{\text{fiber}}$ of the fiber so that the action of $\zenriques$ on $K_{Y_{4,\uparrow}}$ is the product of its actions on $K_{B_3}$ and $K_{\text{fiber}}$.  As it takes $y\rightarrow -y$, $\zenriques$ acts like $-1$ on $K_{\text{fiber}}$.  Furthermore, $\zenriques$ acts like $-1$ on $K_{X_5}$ and $+1$ on $N_{B_3/X_5}$ since the defining quadrics of $B_3$ are $\zenriques$-invariant.  By adjunction, then, $\zenriques$ acts like $-1$ on $K_{B_3}$ and hence like $+1$ on $K_{Y_{4,\uparrow}}$, thereby preserving the holomorphic $(4,0)$-form.

Now that we have specified the geometry it remains to choose the $G$-flux.  In section \ref{subsubsec:extendinginherited}, we found that each type of $\gamma$ in the local model corresponded to a particular $G$-flux in the resolved Calabi-Yau \eqref{extendedinherited}
\begin{equation}\begin{split}\gamma^{(u)} &\leftrightarrow \CG_u \\
\gamma^{(p)} &\leftrightarrow \CG_p \\
\gamma^{(\rho)} &\leftrightarrow \CG_{\rho} \\
\gamma_{\text{noninherited}} &\leftrightarrow \CG_{\text{noninherited}}
\end{split}\end{equation}
The $G$-flux corresponding to our $\gamma_{\text{minimal}}$ \eqref{gammaminimal} is therefore
\begin{equation}\CG_{\text{minimal}} = \CG_{\text{noninherited}}-\frac{1}{2}\CG_u\end{equation}
and yields precisely the chiral spectrum that we desire \eqref{desiredparentspectrum}
\begin{equation}\begin{array}{c|c|c}
\text{Matter} & \text{Chirality on Parent Calabi-Yau} & \text{Chirality in Quotient} \\ \hline
\mathbf{10}^{(4)}_{+1} & 6 & 3 \\
\mathbf{\overline{5}}^{(44)}_{+2} & 0 & 0 \\
\mathbf{\overline{5}}^{(41)}_{-3} & 6 & 3
\end{array}\end{equation}
Recall that the chirality in the quotient geometry is half that of the parent because $\zenriques$ acts freely on $S_2$ and hence on all of the $\mathbf{10}$ and $\mathbf{\overline{5}}$ matter curves and matter surfaces.  More generally, \eqref{moregeneralgamma} tells us that we can take
\begin{equation}\CG = \CG_{\text{minimal}} + \CG_0\end{equation}
where
\begin{equation}\CG_0 = p\left(\CG_p - 2\CG_{\text{noninherited}}\right)+\CG_{\rho}\end{equation}
with
\begin{equation}\rho = aH + bE_1 + cE_2\end{equation}
\begin{equation}a,b,c,p\in\mathbb{Z}\qquad\text{and}\qquad 6a+4b+3c+2p=0\label{pcond}\end{equation}
extends $\gamma_0$ \eqref{gamma0def} .  Note that we didn't have to do any extra work in specifying the $G$-fluxes.  Once we know the explicit dictionary relating $G$-fluxes to local model $\gamma$'s, we simply translate our results from local models.

\subsection{Tadpole and $D$-term}
\label{subsec:tadpoleandD}

Now we can put our global description of flux to work by addressing two issues that cannot be dealt with in the local model: the $U(1)$ $D$-term and the 3-brane tadpole.  We start with the flux-induced $D$-term, which takes the form
\begin{equation}D_{U(1)}\sim \int_{Y_4}G_4\wedge \omega\wedge J\end{equation}
with $\omega$ the $U(1)$ divisor \eqref{omega}.
Parametrizing the K\"ahler form $J$ as
\begin{equation}J = J_H H - J_1 E_1 - J_2E_2\end{equation}
we can directly compute the flux-induced $D$-term as in \cite{Krause:2011xj}
\begin{equation}\begin{split}D_{U(1)} &= -10\left[J_H\left(16a+8b+8c+6p\right)-J_1\left(8a+4c+4p\right) -J_2\left(8a+4b+3p\right)\right] \\
&= 5\left[ J_H(4a+8b+2c) - J_1(8a+16b+4c)-J_2(2a+4b+9c)\right]\end{split} \end{equation}
where we solved the condition in \eqref{pcond} for $p$ to obtain the second line.  We can now evaluate the $D$-term for the two flux choices \eqref{goodgammas}.  For the minimal choice $p=0$, $\rho=0$ we find that the induced $D$-term vanishes for all choices of K\"ahler moduli.  For $p=1$, $\rho=-H+E_1$ we find
\begin{equation}D_{U(1)} =  10\left(2J_H -4J_1 - J_2\right)\end{equation}
The K\"ahler cone on $B_3$ is easy to work out by requiring all curves to have positive volume.  We find the conditions
\begin{equation}J_H>J_1>0\qquad J_H > J_1+J_2\end{equation}
One linear family of $J$'s that lie in the K\"ahler cone and lead to a vanishing $D_{U(1)}$ for the choice $p=1$, $\rho=-H+E_1$ is given by
\begin{equation}J = J_0\left(5H-2E_1-2E_2\right)\end{equation}

Now, we turn to the vexing issue of the 3-brane tadpole.  This receives three contributions in general
\begin{equation}Q_{D3} = n_{D3} + n_{D3,\text{flux}}+n_{D3,\text{geometric}}\end{equation}
where
\begin{equation}\begin{split}n_{D3} &= \text{\# of space-filling D3-branes} \\
n_{D3,\text{flux}}&=\frac{1}{2}\int_{Y_4}G_4\wedge G_4 \\
n_{D3,\text{geometric}} &=- \frac{1}{24}\chi(Y_4)\end{split}\end{equation}
In general, the net 3-brane charge must vanish $Q_{D3}=0$.  We would like to accomplish this by adding $n_{D3}>0$ D3-branes but this can only be done if the flux-induced contribution $n_{D3,\text{flux}}$, which is always nonnegative, is small enough that the combination $n_{D3,\text{flux}}+n_{D3,\text{geometric}}$ is negative.

\subsubsection{Geometric Tadpole}
\label{subsubsec:geometrictadpole}

We begin with the geometric tadpole, which requires us to evaluate $\chi(Y_4)/24$.  In principle, it is easy to evaluate the Euler character of the smooth Calabi-Yau 4-fold $Y_4^{(5)}$ that we get upon resolving \eqref{finalCY}.  From section \ref{subsec:41resolution}, we know that $Y_4^{(5)}$ is a smooth hypersurface inside a relatively simple ambient space obtained by starting with a $\mathbb{P}^2$ bundle over $B_3$ and performing a series of blow-ups.  Using techniques similar to \cite{Marsano:2011hv} we compute
\begin{equation}\chi(Y_4^{(5)}) = 6\left[24\cb^3 + 2\cb c_{2,B} - 44 \cb^2 S_2 + 27 \cb S_2^2 - 5 S_2^3\right]\end{equation}
where $c_{2,B}$ is shorthand for $c_2(B_3)$.  With our $B_3$, we can readily evaluate
\begin{equation}\int_{B_3}\cb c_{2,B} = 24\end{equation}
which is equivalent to the statement that $B_3$ has Todd genus 1.  Evaluating the rest of $\chi(Y_4^{(5)})$ we find
\begin{equation}\chi(Y_4^{(5)}) = 480\end{equation}
which is nicely divisible by 24.  We are not interested in the 3-brane charge of the parent theory, though, but rather the theory on the quotient space $Y_4^{(5)}/\mathbb{Z}_2$.  We might naively expect the geometric 3-brane charge here to be
\begin{equation} n_{D3,\text{naive geometric}} = -\frac{1}{24}\chi(Y_4^{(5)}/\mathbb{Z}_2)_{\text{naive}} = -\frac{1}{48}\chi(Y_4^{(5)}) = - 10\end{equation}
This isn't quite right, though, because $\zenriques$ does not act freely on the entire base $B_3$.  There are 8 fixed points on $B_3$ above which the elliptic fiber is given by $T^2/\mathbb{Z}_2$ in the quotient.  This gives 8 singular fibers that take the form of a $\mathbb{P}^1$ with 4 orbifold singular points.

The existence of these singular fibers has a natural type IIB interpretation \cite{Denef:2005mm,Collinucci:2008zs,Collinucci:2009uh,Blumenhagen:2009up}: the image of each fixed point in the quotient is the location of an O3 plane that carries D3-brane charge $-1/4$ \cite{Denef:2005mm}.  Adding this contribution by hand produces a result that is equivalent to increasing $\chi(Y_4^{(5)}/\zenriques)$ by 6 relative to $\frac{1}{2}\chi(Y_4^{(5)})$.  This has a natural geometric interpretation because $T^2/\mathbb{Z}_2$ has orbifold Euler characteristic 6.

Summing everything, the geometrically induced 3-brane charge on the quotient space $Y_4^{(5)}/\zenriques$ is rather small
\begin{equation}n_{D3,\text{geometric}} = \frac{1}{2}\left[-\frac{1}{24}\chi(Y_4^{(5)})\right] + 8\times \left[-\frac{1}{24}\times 6\right] = -12\end{equation}

 \subsubsection{Flux-induced Tadpole}

 The flux-induced tadpole is given by the self-intersection of $\CG$ inside $Y_4^{(5)}$
 \begin{equation}n_{D3,\text{flux}} = \frac{1}{2}\int_{Y_4^{(5)}/\mathbb{Z}_2}G_4\wedge G_4 = \frac{1}{4}\left(\CG\cdot_{Y_4^{(5)}}\CG\right)\label{fluxtaddef}\end{equation}
 Since most of the terms in $\CG$ are inherited from 3-folds in the ambient space $W_5^{(5)}$, most of the computation can be reduced to complete intersections in that space.  The only tricky part comes from the self-intersection of the non-inherited flux $\CG_{\text{noninherited}}$.  Fortunately we evaluated this already in section \ref{subsubsec:extendingnoninherited}.  It is therefore straightforward to compute
 \begin{equation}\begin{split}\left(\CG_{\text{minimal}}+\CG_0\right)\cdot_{Y_4^{(5)}}\left(\CG_{\text{minimal}}+\CG_0\right) &= 32 - 160\left(a^2+a(b+c)\right) - 80b(c+p)-60p(2a+c)-92p+72p^2 \\
&= 32 + 2\left(-46 p +  \frac{160}{9} b^2 + \frac{424}{9} p^2 - \frac{80}{9} b p + 20 c^2\right) \\
&= 32 + 20\left(\frac{4b-p}{3}\right)^2 + 92p(p-1) + 40c^2
\end{split}\label{fluxD3}\end{equation}
where we substituted the solution $a=-(\frac{2}{3}b+\frac{1}{2}c+\frac{1}{3}p)$ to \eqref{pcond} in the last lines and the condition that $a$ is an integer constrains the values of $b$, $c$, and $p$.  The flux-induced contribution to the 3-brane tadpole is now $\frac{1}{4}$ of this \eqref{fluxtaddef}
\begin{equation}n_{D3,\text{flux}} = 8 + 5\left(\frac{4b-p}{3}\right)^2 + 23 p(p-1) + 10c^2\end{equation}

\subsubsection{Summing the Tadpoles}

The net 3-brane tadpole is the sum of the geometric and flux-induced contributions
\begin{equation}\begin{split}n_{D3,\text{net}} &= n_{D3,\text{geometric}}+n_{D3,\text{flux}}\\
&= -4 + 5\left(\frac{4b-p}{3}\right)^2 + 23p(p-1)+10c^2\label{nettad}\end{split}\end{equation}
To avoid an overshoot, we require that $n_{D3,\text{net}}\le 0$ so that any nonzero tadpole can be cancelled through the introduction of additional D3-branes.  From \eqref{nettad}, it is clear that we must take $c=0$ and take $p$ to be 0 or 1.  Note that when $c=0$ we can use $6a+4b+3c+2p=0$ to rewrite \eqref{nettad} in a simpler way as
\begin{equation}n_{D3,\text{net}}|_{c=0} = -4+5(a+2b)^2 + 23p(p-1)\end{equation}
The only possibility for $a+2b$ is 0.  Adding in the fact that $p$ must be 0 or 1 and enforcing the condition \eqref{pcond} $6a+4b+3c+2p=0$ then leaves us with only the minimal choice $a=b=c=p=0$.  It is amusing to note that for $a+2b=\pm 1$ the overshoot is quite minimal with only 1 anti-D3 brane required to cancel the tadpole.  Enforcing \eqref{pcond} $6a+4b+3c+2p=0$ for this choice gives one solution with $a=-1$, $b=p=1$, and $c=0$.  We tabulate both possibilities below
\begin{equation}\begin{array}{cccc|c}
a & b & c & p & n_{D3,\text{net}} \\ \hline
0 & 0 & 0 & 0 & -4 \\
-1& 1 & 0 & 1 & 1
\end{array}\end{equation}
These lead to precisely the fluxes \eqref{goodgammas} that we studied in the context of local models.  The first is the only choice that can lead to a supersymmetric compactification and the second leads to a very slight breaking of supersymmetry in which 1 anti-D3 brane is introduced into the bulk geometry.  Note that for other choices of $a,b,c,p$ the number of anti-D3-branes that must be introduced rapidly becomes quite large.  While a small number like 1 might be helpful for introducing supersymmetry-breaking, a large number can have significant and potentially damaging backreaction on our geometry.

\section{Concluding Remarks}

In this paper, we have provided a complete realization of the `local-to-global' approach to F-theory model building while implementing a previously unrealized mechanism, the discrete Wilson line, for breaking $SU(5)_{\rm GUT}\rightarrow SU(3)\times SU(2)\times U(1)_Y$.  Models of this type are of interest because they avoid the large GUT scale threshold corrections to gauge coupling unification \cite{Donagi:2008kj,Blumenhagen:2008aw}.  Starting with the basic requirement that $S_{\rm GUT}$ support a discrete Wilson line and insisting on the realization of a $U(1)_{B-L}$ symmetry to suppress dimension 4 proton decay, we constructed a large class of local models that exhibit the chiral spectrum of the MSSM.  We further argued that any model in which $SU(5)_{\rm GUT}$ is broken by a Wilson line will exhibit vector-like exotics that descend from the $SU(5)_{\rm GUT}$ adjoint.  Though local, our models were built with an eye toward the global completion in that we specified how $S_{\rm GUT}$ will sit inside the 3-fold base $B_3$ of the Calabi-Yau 4-fold.  Keeping a global perspective like this is crucial because the spectral data depends on the normal bundle of $S_{\rm GUT}$ inside $B_3$ while the available fluxes are determined by the set of line bundles on $S_{\rm GUT}$ that descend from line bundles on $B_3$.

Given our set of local models, we constructed explicit global completions by specifying compact Calabi-Yau 4-folds and $G$-fluxes.  Along the way, we met the two primary challenges to extending local models.  First, our local models required a careful tuning of the spectral data in order to increase the Picard number of $\cloc$.  We relied on fluxes derived in part from the new `non-inherited' line bundles that appeared from this tuning in order to obtain the right spectrum of chiral matter.  One must take care to ensure that the global completion admits $G$-fluxes that extend these `non-inherited' bundles.  Second, our local model exhibited a $U(1)_{B-L}$ symmetry to suppress dimension 4 proton decay.  $U(1)$'s like this are only present in the global completion if the Calabi-Yau exhibits a $(1,1)$-form with the right properties.  Both of these challenges can be addressed with the Tate divisor formalism, whose central object is a divisor $\ctate$ in the resolved 4-fold that provides an `extension' of the local model spectral cover $\cloc$.  By arranging for $\ctate$ to exhibit the same structures as $\cloc$, we ensured that Calabi-Yau had a $(1,1)$-form of the right type to yield $U(1)_{B-L}$ as well as $(2,2)$-forms of the right type to provide $G$-fluxes that extend the non-inherited bundles of $\cloc$.  Following \cite{Esole:2011sm,Marsano:2011hv,Krause:2011xj} we explicitly resolved the 4-fold and verified all of these facts, making a connection to $U(1)$-restricted Tate models \cite{Grimm:2010ez} and providing the first global extension of the non-inherited bundles of a local $SU(5)_{\rm GUT}$ model.  With the global completion in hand, we were able to compute the induced D3-brane charge and demonstrate that, among the 3 parameter family of fluxes that give rise to 3 chiral generations of quarks and leptons, a unique choice exists for which the tadpole can be cancelled by adding D3-branes.  We computed the precise spectrum for this choice and found an excess of vector-like doublet pairs in the Higgs sector.  We also studied a second flux choice where tadpole cancellation could be achieved with a single anti-D3 and found excess pairs of doublets and triplets in the Higgs sector.

Though our models represent a complete application of the current tools for understanding global completions of $U(1)$'s and $G$-fluxes, their phenomenological problems give some cause for concern.  Perhaps the most glaring issue is the general presence of vector-like exotics along $S_{\rm GUT}$ that descend from the $SU(5)_{\rm GUT}$ adjoint.  If some suitable nonperturbative dynamics cannot be identified to lift them, this will require a substantive change to the general strategy of model-building with Wilson lines.  One can attain greater flexibility by engineering a larger gauge group along $S_2$, such as $SU(6)$ or $SO(10)$, and using a combination of Wilson lines and nontrivial fluxes to break the GUT group though large GUT scale gauge threshold corrections may arise.  In models of this type, not all zero modes that descend from the adjoint must be projected out; some can be retained and identified with Standard Model fields.  This type of approach has recently been advocated in F-theory model building in \cite{Donagi:2011dv} and reflects a structure that has appeared in a number of Heterotic models \cite{Lebedev:2006kn,Lebedev:2007hv,Blaszczyk:2009in,Kappl:2010yu}.

The abundance of Higgs doublets is also troubling and is connected directly to the 3-brane tadpole.  Most of the local model fluxes that lead to 3 chiral generations are generic in the sense that the Higgs spectrum is empty for almost all choices of the spectral data.  Given such a flux, one can imagine tuning the spectral data to one of the special points where the spectrum of doublets exhibits a minimal jump from nothing to a single vector-like pair.  Such a scenario would still require a solution to the $\mu$ problem in the form of a mechanism that singles out this special point but it isn't inconceivable that this could be accomplished by a symmetry of some type.  Unfortunately, our situation is even worse in that only a flux that generates a large doublet excess survives the D3-tadpole condition.  No small deformation of the geometry can alleviate this problem so the only hope for the construction of this paper is that the extra doublets can be lifted by nonperturbative physics.  This is not very satisfactory even if the right type of instanton contributions can be found, though, because the doublet masses will not be nearly heavy enough.  Attaining a realistic Higgs sector thus requires looking to more general geometries and, specifically, new choices for the 3-fold base $B_3$.  We were not able to find any suitable alternatives but we have no reason to expect that they do not exist.  Indeed, our general approach is quite restrictive and considers only quotient spaces in which $B_3$, and in fact the entire Calabi-Yau 4-fold $Y_4$, can be lifted to a double-covering space.  In principle, many choices of 3-fold exist that do not have this property but can nevertheless serve as the base of an elliptically fibered 4-fold with an $A_4$ singularity along an Enriques surface.

Finally, our models are not completely free of proton decay problems.  Despite the realization of $U(1)_{B-L}$ for dealing with the dangerous operators at dimension 4, we have no mechanism to suppress dimension 5 proton decay.  It is by now well known that $U(1)$-based solutions can cause a variety of problems \cite{Marsano:2009gv,Marsano:2009wr,Dudas:2010zb,Marsano:2010sq,Dolan:2011iu} so the most promising approach is likely to involve a discrete symmetry such as the $\mathbb{Z}_4^{R}$ of \cite{Lee:2010gv,Lee:2011dya,Kappl:2010yu}.  The realization of discrete symmetries is very important and has been the subject of some recent work in the orientifold limit \cite{Ibanez:2012wg}.

In the end, we have provided a comprehensive example of the `local-to-global' approach to F-theory model building, from the construction of local models to their global completions, that gives a proper treatment of all aspects from the realization of $U(1)$ symmetries to the extension of all local model fluxes including those that are non-inherited.  We hope this gives a practical demonstration of the Tate divisor formalism and its ability to circumvent the problems of building global models from local ones.
We have uncovered general phenomenological problems with Wilson line models and found examples in which the Higgs sector is non-minimal.  Understanding why non-minimality like this occurs for certain choices of flux may also yield new solutions to the $\mu$ problem in F-theory models.

\section*{Acknowledgements}

We are grateful to K.~Bobkov, V.~Braun, R.~Donagi, S.~Kachru, S.~Katz, D.~Morrison, N.~Saulina, S.~Sch\"afer-Nameki, S.~Sethi, A.~Westphal, and M.~Wijnholt for valuable discussions and K.~Bobkov for collaboration at early stages of this work.
JM would like to thank the theoretical physics and algebraic geometry groups at the Ohio State University and the University of Pennsylvania for their hospitality at various stages of this work.  JM is also grateful to the Simons Center for Geometry and Physics and the organizers of the 2012 workshops on `F-theory', `String Phenomenology', and `String Theory for Mathematicians'.  TP would like to thank the Mathematics Research Institute of the Ohio State University for providing  perfect working conditions during the preparation of this paper.  The work of JM is supported by DOE grant DE- FG02-90ER-40560 and NSF grant PHY-0855039.  The work of SR is supported by DOE grant DOE/ER/01545-894.  The work of TP was partially supported by NSF RTG grant DMS-0636606 and NSF  grants DMS-0700446 and DMS-1001693. 

\appendix

\section{Cohomologies for Spectrum Computations}
\label{app:cohoms}

In this Appendix, we provide some of the details required to compute the cohomologies of section \ref{subsec:precisespectrum}.  In each case, our objective is to compute the cohomology groups in \eqref{neededcohoms} through the strategies outlined in section \ref{subsubsec:strategy}.  We proceed to discuss each flux choice in turn.

\subsection{The Minimal Flux Case of Section \ref{subsubsec:minimalspectrum}}
\label{appsubsec:minimalflux}

We start with the case of minimal flux that was discussed in section \ref{subsubsec:minimalspectrum}.  We discuss in detail the relevant cohomology computations for each matter curve in turn.

\subsubsection{$\tencurve$}
\label{appsubsubsec:minimalten}

The curve $\tencurve$ is defined by the equations
\begin{equation}U = c_4h_0 - d_3h_2=0\label{app10curveeqs}\end{equation}
inside $S_2\times\mathbb{P}^1$ and the bundle of interest is
\begin{equation}\CL_{\tencurve} = \CO_{X_5\times\mathbb{P}^1}(2H-E_1+E_2)|_{\tencurve}\otimes \CO_{\tencurve}(-2Q_{\mathbf{10}})\end{equation}
where
\begin{equation}Q_{\mathbf{10}}=\text{set of 4 points where }h_0=d_3=0\label{appQ10def}\end{equation}
The first thing to observe is that we can effectively perform all of our computations in $X_5$.  In particular, we can treat $\tencurve$ as a curve in $X_5$ and the bundle as the combination of a bundle inherited from $X_5$ and the part obtained by removing the points from $Q_{\mathbf{10}}$.  Our basic tool for computing $h^m(\tencurve,\CL_{\tencurve})$ is the exact sequence
\begin{equation}0\rightarrow \CO_{X_5}(\CD)|_{\tencurve}\otimes \CO_{\tencurve}(-2Q_{\mathbf{10}})\rightarrow \CO_{X_5}(\CD)|_{\tencurve}\xrightarrow{f_{\mathbf{10}}} \mathbb{C}^8\rightarrow 0\label{appf10def}\end{equation}
where we introduced the notation
\begin{equation}\CD = 2H-E_1+E_2\label{appCDdef}\end{equation}
so to start we need the cohomologies of the inherited bundle $\CO_{X_5}(2H-E_1+E_2)|_{\tencurve}$.  We can get these from the Koszul extension of \texttt{cohomcalg} \cite{Blumenhagen:2010pv,Jow,Blumenhagen:2010ed,cohomCalg:Implementation}.  Rather than quoting the results, though, we say a little about how they are obtained so that we can discern the behavior of the maps of cohomologies that are induced by $f_{\mathbf{10}}$ of \eqref{appf10def}.  The basic strategy of \cite{Blumenhagen:2010ed} is to write $\tencurve$ as a complete intersection in $X_5$
\begin{equation}\tencurve = \prod_{i=1}^4\CD_i\end{equation}
and determine the cohomology of $\CO_{X_5}(\CD)|_{\tencurve}$ for a divisor class $\CD$ in $X_5$ from the long exact cohomology sequences that follow from
\begin{equation}\begin{split}0\rightarrow \CO_{X_5}(\CD-\sum_{i=1}^4\CD_i)\rightarrow \oplus_{i_1<i_2<i_3}\CO_{X_5}(\CD-\CD_{i_1}-\CD_{i_2}-\CD_{i_3})\rightarrow \CI_1\rightarrow 0 \\
0\rightarrow \CI_1\rightarrow \oplus_{i_1<i_2}\CO_{X_5}(\CD-\CD_{i_1}-\CD_{i_2})\rightarrow \CI_2\rightarrow 0 \\
0\rightarrow \CI_2\rightarrow \oplus_i \CO(\CD-\CD_i)\rightarrow \CI_3\rightarrow 0 \\
0\rightarrow \CI_3\rightarrow \CO_{X_5}(\CD)\rightarrow \CO_{X_5}(\CD)|_{\tencurve}\rightarrow 0
\end{split}\label{appIdefs}\end{equation}
We can use the methods of \cite{Blumenhagen:2010pv,Jow,Blumenhagen:2010ed,cohomCalg:Implementation} to determine the cohomology groups on $X_5$ and use these to infer cohomologies of the sheaves $\CI_m$ that are implicitly defined in \eqref{appIdefs}.  In the present case, the divisor $\CD$ is defined in \eqref{appCDdef} while the $\CD_i$ are given by
\begin{equation}\begin{split}
\CD_1 &= 2H-E_1 \\
\CD_2 &=2H-E_1 \\
\CD_3 &= 2H-E_2 \\
\CD_4 &= 2H
\end{split}\end{equation}
Most of the $X_5$ cohomologies that enter here are trivial.  The only nontrivial ones are
\begin{equation}\begin{split}h^0(X_5,\CO_{X_5}(\CD-\CD_1)) &= h^0(X_5,\CO_{X_5}(E_2)) \\
&= 1 \\
h^0(X_5,\CO_{X_5}(\CD-\CD_2)) &= h^0(X_5,\CO_{X_5}(E_2)) \\
&= 1 \\
h^1(X_5,\CO_{X_5}(\CD-\CD_1-\CD_3)) &= h^1(X_5,\CO_{X_5}(-2H+2E_2)) \\
&= 1 \\
h^1(X_5,\CO_{X_5}(\CD-\CD_2-\CD_3)) &= h^1(X_5,\CO_{X_5}(-2H+2E_2)) \\
&= 1 \\
h^1(X_5,\CO_{X_5}(\CD-\CD_3)) &= h^1(X_5,\CO_{X_5}(2E_2-E_1)) \\
&= 7 \\
h^0(X_5,\CO_{X_5}(\CD)) &= h^0(X_5,\CO_{X_5}(2H-E_1+E_2)) \\
&= 11
\end{split}\label{app10seccohoms}\end{equation}
From this we conclude that
\begin{equation}h^m(X_5,\CI_1)=(0,0,0,0,0,0)\qquad h^m(X_5,\CI_2)=(0,2,0,0,0,0)\end{equation}
where the index $m$ runs from 0 to 5.  In particular
\begin{equation}H^1(X_5,\CI_2) = H^1(X_5,\CO_{X_5}(\CD-\CD_1-\CD_3)\oplus H^1(X_5,\CO_{X_5}(\CD-\CD_2-\CD_3)\end{equation}
and the others are trivial.  The long exact sequence involving $\CI_3$ now contains
\begin{equation}\begin{split}0&\rightarrow \oplus_{i=1}^2 H^0(X_5,\CO_{X_5}(\CD-\CD_i))\rightarrow H^0(X_5,\CI_3)\rightarrow \oplus_{i=1}^2H^1(X_5,\CO_{X_5}(\CD-\CD_i-\CD_3))\\
&\qquad \xrightarrow{h} H^1(X_5,\CO_{X_5}(\CD-\CD_3))\rightarrow H^1(X_5,\CI_3)\rightarrow 0\end{split}\label{hmaphere}\end{equation}
and the sequence involving our desired cohomologies splits as
\begin{equation}0\rightarrow H^0(X_5,\CI_3)\rightarrow H^0(X_5,\CO_{X_5}(\CD))\rightarrow H^0(\tencurve,\CO_{X_5}(\CD)|_{\tencurve})\rightarrow H^1(X_5,\CI_3)\rightarrow 0\label{appH0sequence}\end{equation}
and
\begin{equation}H^1(\tencurve,\CO_{X_5}(\CD)|_{\tencurve})=0\end{equation}
This is enough information to determine that
\begin{equation}h^0(\tencurve,\CO_{X_5}(\CD)|_{\tencurve}) = 14\label{appthe14}\end{equation}
but we would like to know more about the actual sections in $H^0(\tencurve,\CO_{X_5}(\CD)|_{\tencurve})$ to study how they restrict to the points in $Q_{\mathbf{10}}$ \eqref{appQ10def}.  We can shed some light on this as the procedure of \cite{Blumenhagen:2010pv,Jow} that determines the nontrivial cohomologies of \eqref{app10seccohoms} is constructive in that the `rationoms' that we count essentially provide us with explicit representatives.  For the cohomology groups of interest, the `rationoms' are (see \cite{Blumenhagen:2010pv,Jow} for a detailed discussion of the procedure)
\begin{equation}\begin{array}{c|c}
\text{Cohomology} & \text{Collection of `Rationoms'} \\ \hline
H^0(X_5,\CO_{X_5}(\CD-\CD_{1/2}) & \delta_2 \\
H^0(X_5,\CO_{X_5}(\CD-\CD_{1/2}-\CD_3) & \frac{1}{\tilde{u}_2\tilde{v}_2} \\
H^1(X_5,\CO_{X_5}(\CD-\CD_3)) & \frac{1}{\tilde{u}_2\tilde{v}_2}\left[P_1(\tilde{u}_1,\tilde{v}_1)P_1(u_3,v_3)+\delta_1P_2(\tilde{u}_1,\tilde{v}_1)\right] \\
H^0(X_5,\CO_{X_5}(\CD)) & \delta_2P_1(\tilde{u}_1,\tilde{v}_1)P_1(\tilde{u}_2\delta_2,\tilde{v}_2\delta_2,u_3,v_3) + \delta_2\delta_1P_2(\tilde{u}_1,\tilde{v}_1)
\end{array}\end{equation}
where $P_m(\ldots)$ is a homogeneous polynomial of degree $m$ in its arguments.  It is easy to count the number of rationoms in each case and verify the results of \eqref{app10seccohoms} explicitly.  We now study the  map $h$ from \eqref{hmaphere}
\begin{equation}\oplus_{i=1}^2H^1(X_5,\CO_{X_5}(\CD-\CD_i-\CD_3))\xrightarrow{h} H^1(X_5,\CO_{X_5}(\CD-\CD_3))\end{equation}
which acts as
\begin{equation}\frac{1}{\tilde{u}_2\tilde{v}_2}\begin{pmatrix}a \\ b\end{pmatrix}\xrightarrow{h} \frac{1}{\tilde{u}_2\tilde{v}_2}\left[a Q_2^{(1)} + b Q_{\Sigma}\right]\label{apphaction}\end{equation}
for complex numbers $a$ and $b$.  Here $Q_2^{(1)}$ and $Q_{\Sigma}$ are the equations that define the divisors $\CD_1$ and $\CD_2$.  These are the quadric from \eqref{firsttwoquadrics} that defines $S_2$ inside $B_3$ and the equation that, along with $U=0$, defines $\tencurve$ inside $S_2\times\mathbb{P}^1$ \eqref{app10curveeqs}.  The right hand side of \eqref{apphaction} becomes trivial in cohomology if the term in $[\,]$'s vanishes or acquires an overall factor of $\tilde{u}_2$ or $\tilde{v}_2$.  This can only happen if $Q_2^{(1)}$ and $Q_{\Sigma}$ satisfy a nontrivial linear relation modulo $\tilde{u}_2$ and $\tilde{v}_2$.  For clarity, we write the most general form of $Q_2^{(1)}$ and $Q_{\Sigma}$
\begin{equation}\begin{split}Q_2^{(1)} &= \tilde{u}_1 P_1(\tilde{u}_1\delta_1,\tilde{u}_2\delta_2,u_3) - \tilde{v}_1 P_1(\tilde{v}_1\delta_1,\tilde{v}_2\delta_2,v_3) \\
Q_{\Sigma} &= \delta_2P_2(\tilde{u}_1,\tilde{v}_1) + P_1(\tilde{u}_1,\tilde{v}_1)P_1(\tilde{u}_1\delta_1,\tilde{u}_2\delta_2,u_3,\tilde{v}_1\delta_1,\tilde{v}_2\delta_2,v_3)
\end{split}\end{equation}
where as before we use $P_m(\ldots)$ to refer to any homogeneous polynomial of the given degree in its arguments.  Of course specific choices for the polynomials in $Q_2^{(1)}$ and $Q_{\Sigma}$ will have to be made.  For a generic such choice, though, it is easy to see that they will not satisfy any linear relation modulo $\tilde{u}_2$ and $\tilde{v}_2$.  This means that
\begin{equation}\text{ker }h=0\end{equation}
from which it follows that
\begin{equation}\begin{split}H^0(X_5,\CI_3) &= \oplus_{i=1}^2 H^0(X_5,\CO_{X_5}(\CD-\CD_i)) \\
H^1(X_5,\CI_3) &= \text{cok }h\end{split}\end{equation}
We now turn to the sequence \eqref{appH0sequence} that determines $H^0(\tencurve,\CO_{X_5}(\CD)|_{\tencurve})$.  We see that $H^0(\tencurve,\CO_{X_5}(\CD)|_{\tencurve})$ is a sum of $\text{cok }h$ and the cokernel of the map
\begin{equation}H^0(X_5,\CI_3)= \oplus_{i=1}^2 H^0(X_5,\CO_{X_5}(\CD-\CD_i)) \rightarrow H^0(X_5,\CO_{X_5}(\CD))\end{equation}
We can therefore describe the elements of $H^0(\tencurve,\CO_{X_5}(\CD)|_{\tencurve})$ schematically as follows
\begin{equation}H^0(\tencurve,\CO_{X_5}(\CD)|_{\tencurve})\leftrightarrow\left\{\begin{array}{lll}
\delta_2P_1(\tilde{u}_1,\tilde{v}_1)P_1(\tilde{u}_2\delta_2,\tilde{v}_2\delta_2,u_3,v_3) & +\delta_2\delta_1P_2(\tilde{u}_1,\tilde{v}_1) & \text{with nontriv restriction} \\
& & \text{ to }Q_2^{(1)}\text{ and }Q_{\Sigma} \\
+8 & +3 & -2 \\
&& \\
P_1(\tilde{u}_1,\tilde{v}_1)P_1(u_3,v_3) & +\delta_1P_2(\tilde{u}_1,\tilde{v}_1) & \text{with nontriv restriction} \\
& & \text{ to }Q_2^{(1)}\text{ and }Q_{\Sigma} \\
& & \text{ modulo }\tilde{u}_2\text{ and }\tilde{v}_2 \\
+4 & +3 & -2
\end{array}\right.\end{equation}
This explicitly displays the 14 elements of $H^0(\tencurve,\CO_{X_5}(\CD)|_{\tencurve})$.  We are now interested in the rank of the map $f_{\mathbf{10}}$ in
\begin{equation}0\rightarrow H^0(\tencurve,\CO_{X_5}(\CD)|_{\tencurve}\otimes \CO_{\tencurve}(-2Q_{\mathbf{10}}))\rightarrow H^0(\tencurve,\CO_{X_5}(\CD)|_{\tencurve})\xrightarrow{f_{\mathbf{10}}} \mathbb{C}^8\label{f10again}\end{equation}
obtained by restriction and taking first derivatives.  We recall that $Q_{\mathbf{10}}$ is given by the 4 points where $h_0=d_3=0$ with $h_0$ a section of $\CO_{X_5}(E_2)$ and $d_3$ a section of $\CO_{X_5}(H)$ which is $\zenriques$-odd.  Up to scaling we have that $h_0=\delta_2$ and $d_3=P_1(\delta_1\tilde{v}_1,\delta_2\tilde{v}_2,v_3)$.  Of the 14 sections of $H^0(\tencurve,\CO_{X_5}(\CD)|_{\tencurve})$, 9 are identically zero when $\delta_2=0$ and, when we restrict to $\delta_2=0$, a linear combination of one of the 5 remaining sections will be proportional to $d_3$.  This leaves us with exactly 4 that restrict nontrivially to all of the points in $Q_{\mathbf{10}}$ which is borderline with what we need.  When we consider first derivatives, 8 of the 9 sections that were proportional to $\delta_2$ will have nontrivial restriction.  Taken together, this means that the map $f_{\mathbf{10}}$ is indeed of maximal rank 4+4=8 and hence from \eqref{appthe14}, \eqref{f10again} we have that
\begin{equation}H^0(\tencurve,\CL_{\tencurve}) = 14-8=6\qquad H^1(\tencurve,\CL_{\tencurve}) = 0\end{equation}
which is the result quoted in section \ref{subsubsec:minimalspectrum}.

\subsubsection{$\fiveonecurve$}
\label{appsubsubsec:minimalfiveone}

The case of $\fiveonecurve$ follows almost immediately.  As a curve in $X_5\times\mathbb{P}^1$ it is equivalent to $\tencurve$ and the inherited bundle of interest is also the same.  The only difference from the previous case lies in the points subtracted.  In particular, we have
\begin{equation}\CL|_{\fiveonecurve} = \CO_{X_5\times\mathbb{P}^1}(\CD)|_{\fiveonecurve}\otimes\CO_{\fiveonecurve}(-2Q_{\mathbf{\overline{5}}})\end{equation}
where
\begin{equation}Q_{\mathbf{\overline{5}}} = \text{set of 4 points where }h_0 = d_2+d_3=0\end{equation}
The reasoning is now exactly the same as before except we should replace $d_3$, which was an antisymmetric section of $\CO_{X_5}(H)$, with $d_2+d_3$, which is a sum of an antisymmetric and symmetric section of $\CO_{X_5}(H)$.  It is easy to see that nothing changes from the previous argument and we have
\begin{equation}H^0(\fiveonecurve,\CL_{\fiveonecurve}) = 6\qquad H^1(\fiveonecurve,\CL_{\fiveonecurve}) = 0\end{equation}

\subsubsection{$\fivefourcurve$}
\label{appsubsubsec:minimalfivefour}

We now turn to the cohomologies of $\CL_{\fivefourcurve}$ on $\fivefourcurve$.  As discussed in the text, this is a little simpler because the bundle is inherited from $X_5\times\mathbb{P}^1$.  An additional complication, though, is that we need to keep track of a pair of $\mathbb{Z}_2$ involutions.  The first is the involution $\ztau$ $\tau$ that acts as a reflection on the $\mathbb{P}^1$.  This interchanges the two sheets of $\fivefourcurve$ and the physical zero modes correspond to cohomology elements that are $\ztau$-odd.  Once we have zero modes, however, the 4-dimensional fields to which they give rise depend on their parity with respect to the involution $\zenriques$, which we defined on the toric space $X_5$ \eqref{Z2P5} \eqref{Z2W5} and descends to an Enriques involution on $S_2=K3$.  Zero modes that are $\zenriques$-odd give rise to doublets and zero modes that are $\zenriques$-even give rise to triplets.  Other than the added complication of these $\mathbb{Z}_2$'s, the computation is straightforward following the Koszul techniques of \cite{Blumenhagen:2010ed}.

The defining equations of $\fivefourcurve$ inside $X_5\times\mathbb{P}^1$ are the defining equations of $S_2$ inside $X_5$ combined with
\begin{equation}a_3V^2+e_1\alpha U^2 = a_4V^2+(a_2+a_3e_0)U^2 = 0\end{equation}
so that
\begin{equation}\fivefourcurve = \prod_{i=1}^5\CD_i\end{equation}
with
\begin{equation}\begin{split}\CD_1 &= 2H-E_1 \\
\CD_2 &= 2H-E_2 \\
\CD_3 &= 2H \\
\CD_4 &= 2\sigma+2H-E_1 \\
\CD_5 &= 2\sigma + 2H-E_1
\end{split}\end{equation}
The cohomologies of interest are
\begin{equation}h^m_-(\fivefourcurve,\CO_{X_5\times\mathbb{P}^1}(\CD)|_{\fivefourcurve})\end{equation}
where the $-$ means to take the odd cohomology with respect to $\mathbb{Z}_2^{\tau}$ and $\CD$ is the divisor class
\begin{equation}\CD = 4H-2E_1-2\sigma\end{equation}
Following \cite{Blumenhagen:2010ed} we use the long exact cohomology sequences that follow from the following set of exact sequences
\begin{equation}\begin{split}0\rightarrow \CO_{X_5}(\CD-\sum_{i=1}^5\CD_i)\rightarrow \oplus_{i_1<i_2<i_3<i_4}\CO_{X_5}(\CD-\CD_{i_1}-\CD_{i_2}-\CD_{i_3}-\CD_{i_4})\rightarrow \CI_1\rightarrow 0 \\
0\rightarrow \CI_1\rightarrow \oplus_{i_1<i_2<i_3}\CO_{X_5}(\CD-\CD_{i_1}-\CD_{i_2}-\CD_{i_3})\rightarrow \CI_2\rightarrow 0 \\
0\rightarrow \CI_2\rightarrow \oplus_{i_1<i_2} \CO(\CD-\CD_{i_1}-\CD_{i_2})\rightarrow \CI_3\rightarrow 0 \\
0\rightarrow \CI_3\rightarrow \oplus_i \CO_{X_5}(\CD-\CD_i)\rightarrow \CI_4\rightarrow 0 \\
0\rightarrow \CI_4\rightarrow \CO_{X_5}(\CD)\rightarrow \CO_{X_5}(\CD)|_{\fivefourcurve}\rightarrow 0
\end{split}\label{appHiggscurveIdefs}\end{equation}
Note that unlike the case of $\tencurve$ and $\fiveonecurve$ we must work in the full $X_5\times \mathbb{P}^1$ here so there is one more sequence compared to \eqref{appIdefs}.  We pass to long exact cohomology sequences involving the odd-graded cohomology groups with respect to $\ztau$.  The nonvanishing cohomologies that enter are
\begin{equation}\begin{split}
h^1_-(X_5\times\mathbb{P}^1,\CO_{X_5\times\mathbb{P}^1}(\CD-\CD_1) &= h^1_-(X_5\times\mathbb{P}^1,\CO_{X_5\times\mathbb{P}^1}(2H-E_1-2\sigma) \\
&= 11 =6+5\\
h^1_-(X_5\times\mathbb{P}^1,\CO_{X_5\times\mathbb{P}^1}(\CD-\CD_2) &= h^1_-(X_5\times\mathbb{P}^1,\CO_{X_5\times\mathbb{P}^1}(2H-2E_1+E_2-2\sigma) \\
&= 3 =2+1\\
h^1_-(X_5\times\mathbb{P}^1,\CO_{X_5\times\mathbb{P}^1}(\CD-\CD_3) &= h^1_-(X_5\times\mathbb{P}^1,\CO_{X_5\times\mathbb{P}^1}(2H-2E_1-2\sigma) \\
&= 3 = 2+1\\
h^1_-(X_5\times\mathbb{P}^1,\CO_{X_5\times\mathbb{P}^1}(\CD-\CD_{4/5}) &= h^1_-(X_5\times\mathbb{P}^1,\CO_{X_5\times\mathbb{P}^1}(2H-E_1-4\sigma) \\
&= 22 = 12+10\\
h^1_-(X_5\times\mathbb{P}^1,\CO_{X_5\times\mathbb{P}^1}(\CD-\CD_1-\CD_{4/5}) &= h^1_-(X_5\times\mathbb{P}^1,\CO_{X_5\times\mathbb{P}^1}(-4\sigma) \\
&= 2 = 2+0\\
h^1_-(X_5\times\mathbb{P}^1,\CO_{X_5\times\mathbb{P}^1}(\CD-\CD_4-\CD_5) &= h^1_-(X_5\times\mathbb{P}^1,\CO_{X_5\times\mathbb{P}^1}(-6\sigma) \\
&= 3 = 3+0\\
h^6_-(X_5\times\mathbb{P}^1,\CO_{X_5\times\mathbb{P}^1}(\CD-\CD_1-\CD_2-\CD_3-\CD_4-\CD_5) &= h^6_-(X_5\times\mathbb{P}^1,\CO_{X_5\times\mathbb{P}^1}(-6H+E_1+E_2-6\sigma) \\
&= 3=0+3\\
h^1_-(X_5\times\mathbb{P}^1,\CO_{X_5\times\mathbb{P}^1}(\CD) &= h^1_-(X_5\times\mathbb{P}^1,\CO_{X_5\times\mathbb{P}^1}(4H-2E_1-2\sigma) \\
&= 51= 27+24
\end{split}\label{higgscohoms}\end{equation}
In each case we have written the result as a sum $a+b$ where $a$ denotes the number of elements that are even under $\zenriques$ and $b$ the number that are odd.
The long exact sequences that follow from \eqref{appHiggscurveIdefs} tell us that
\begin{equation}\begin{split}H^2_-(X_5\times\mathbb{P}^1,\CI_4) &= H^3_-(X_5\times\mathbb{P}^1,\CI_3) = H^4_-(X_5\times\mathbb{P}^1,\CI_2) = H^5_-(X_5\times\mathbb{P}^1,\CI_1) \\
&= H^6_-(X_5\times\mathbb{P}^1,\CO_{X_5\times\mathbb{P}^1}(\CD-\sum_{i=1}^5\CD_i))\end{split}\end{equation}
while
\begin{equation}H^1_-(X_5\times\mathbb{P}^1,\CI_3) = \oplus_{i_1<_2}H^1_-(X_5\times\mathbb{P}^1,\CO_{X_5\times\mathbb{P}^1}(\CD-\CD_{i_1}-\CD_{i_2}))\end{equation}
Further since $H^0(X_5\times\mathbb{P}^1,\CO_{X_5\times\mathbb{P}^1}(\CD))=0$ we have that $H^0(X_5\times\mathbb{P}^1,\CI_4)=0$ and
\begin{equation}0\rightarrow H^1_-(X_5\times\mathbb{P}^1,\CI_3)\rightarrow \oplus_iH^1_-(X_5\times\mathbb{P}^1,\CO_{X_5\times\mathbb{P}^1}(\CD-\CD_i)\rightarrow H^1_-(X_5\times\mathbb{P}^1,\CI_4)\rightarrow 0\end{equation}
or
\begin{equation}H^1_-(X_5\times\mathbb{P}^1,\CI_4) = \left[\oplus_i H^1_-(X_5\times\mathbb{P}^1,\CO_{X_5\times\mathbb{P}^1}(\CD-\CD_i))\right] \left/ \left[\oplus_{i_1<i_2}H^1_-(X_5\times\mathbb{P}^1,\CO_{X_5\times\mathbb{P}^1}(\CD-\CD_{i_1}-\CD_{i_2}))\right]\right.\end{equation}
The sequence that determines the cohomologies of interest is now
\begin{equation}\begin{split}0&\rightarrow H^0_-(\fivefourcurve,\CO_{X_5\times\mathbb{P}^1}(\CD)|_{\fivefourcurve})\rightarrow H^1_-(X_5\times\mathbb{P}^1,\CI_4)\xrightarrow{j} H^1_-(X_5\times\mathbb{P}^1,\CO_{X_5\times\mathbb{P}^1}(\CD))\\
&\qquad \rightarrow H^1_-(\fivefourcurve,\CO_{X_5\times\mathbb{P}^1}(\CD)|_{\fivefourcurve})\rightarrow H^2_-(X_5\times\mathbb{P}^1,\CI_4)\rightarrow 0\end{split}\end{equation}
Since $h^1_-(X_5\times\mathbb{P}^1,\CI_4)=54$, $h^2_-(X_5\times\mathbb{P}^1,\CI_4)=3$ and $h^1_-(X_5\times\mathbb{P}^1,\CO_{X_5\times\mathbb{P}^1}(\CD))=51$ we have that
\begin{equation}h^0_-(\fivefourcurve,\CO_{X_5\times\mathbb{P}^1}(\CD)|_{\fivefourcurve}) = h^1_-(\fivefourcurve,\CO_{X_5\times\mathbb{P}^1}(\CD)|_{\fivefourcurve}) = \text{ker } j = 3+\text{coker }j\end{equation}
where $j$ is a linear map from a 54-dimensional space to a 51-dimensional one.  This kernel always has dimension at least 3.  This means we are guaranteed to get at least 3 vector-like pairs of particles.  What kind of particles depends on the parity of the zero modes under $\zenriques$
Note that it is enough to determine the parities of the elements of $H^1_-(\fivefourcurve,\CO_{X_5\times\mathbb{P}^1}(\CD)|_{\fivefourcurve})$.  This is because $\zenriques$ acts freely on $\fivefourcurve$ so that the indices over the even states and odd states vanish independently.  It is easy to see that all 3 elements of $H_-^2(X_5\times\mathbb{P}^1,\CI_4)=H^1_-(\fivefourcurve,\CO_{X_5\times\mathbb{P}^1}(\CD)|_{\fivefourcurve})$ are odd so the 3 vector-like pairs that are guaranteed are all doublets.

This is not the end of the story, though, because a more careful study of the map $j$ reveals that its kernel generically has dimension 4.  To see this, let us consider the construction of a matrix representative of $j$.
We start by understanding elements of $H^1_-(X_5\times\mathbb{P}^1,\CO_{X_5\times\mathbb{P}^1}(\CD))$.  We can characterize these elements as follows \cite{Blumenhagen:2010pv,Jow}
\begin{equation}
H^1_-(X_5\times\mathbb{P}^1,\CO_{X_5\times\mathbb{P}^1}(\CD))\leftrightarrow\frac{1}{UV}\left\{\begin{array}{l}P_2(\tilde{u}_1,\tilde{v}_1)P_2(\tilde{u}_2\delta_2,\tilde{v}_2\delta_2,u_3,v_3) \\
\delta_1P_3(\tilde{u}_1,\tilde{v}_1)P_1(\tilde{u}_2\delta_2,\tilde{v}_2\delta_2,u_3,v_3) \\
\delta_1^2P_4(\tilde{u}_1,\tilde{v}_1)
\end{array}\right.\end{equation}
We now turn to elements of $\oplus_iH^1_-(X_5\times\mathbb{P}^1,\CO_{X_5\times\mathbb{P}^1}(\CD-\CD_i)$
{\small{\begin{equation}\oplus_iH^1_-(X_5\times\mathbb{P}^1,\CO_{X_5\times\mathbb{P}^1}(\CD-\CD_i))\leftrightarrow\frac{1}{UV}\left\{
\begin{array}{lr}
\left[P_1(\tilde{u}_1,\tilde{v}_1)P_1(\tilde{u}_2\delta_2,\tilde{v}_2\delta_2,u_3,v_3) + \delta_1P_2(\tilde{u}_1,\tilde{v}_1)\right] & i=1\\
\delta_2P_2(\tilde{u}_1,\tilde{v}_1) & i=2\\
P_2(\tilde{u}_1,\tilde{v}_1) & i=3\\
P_1\left(\frac{1}{U^2},\frac{1}{V^2}\right)\left[P_1(\tilde{u}_1,\tilde{v}_1)P_1(\tilde{u}_2\delta_2,\tilde{v}_2\delta_2,u_3,v_3) + \delta_1P_2(\tilde{u}_1,\tilde{v}_1)\right] & i=4\\
P_1\left(\frac{1}{U^2},\frac{1}{V^2}\right)\left[P_1(\tilde{u}_1,\tilde{v}_1)P_1(\tilde{u}_2\delta_2,\tilde{v}_2\delta_2,u_3,v_3) + \delta_1P_2(\tilde{u}_1,\tilde{v}_1)\right] & i=5\\
\end{array}\right.\end{equation}}}
To obtain $H^1_-(X_5\times\mathbb{P}^1,\CI_4)$, we must quotient these by $\oplus_{i<j}H^1_-(X_5\times\mathbb{P}^1,\CO_{X_5\times\mathbb{P}^1}(\CD-\CD_i-\CD_j))$ which can be characterized by
\begin{equation}\begin{split}H^1_-(X_5\times\mathbb{P}^1,\CO_{X_5\times\mathbb{P}^1}(\CD-\CD_1-\CD_{4/5}))&\leftrightarrow \frac{1}{UV}P_1(U^{-2},V^{-2}) \\
H^1_-(X_5\times\mathbb{P}^1,\CO_{X_5\times\mathbb{P}^1}(\CD-\CD_4-\CD_5)) &\leftrightarrow \frac{1}{UV}P_2(U^{-2},V^{-2})\end{split}\end{equation}
where the mapping $\oplus_{i<j}H^1_-(X_5\times\mathbb{P}^1,\CO_{X_5\times\mathbb{P}^1}(\CD-\CD_i-\CD_j))\rightarrow \oplus_kH^1_-(X_5\times\mathbb{P}^1,\CO_{X_5\times\mathbb{P}^1}(\CD-\CD_k)$ is the obvious one defined in terms of the defining equations for the divisors $\CD_i$.  All elements of $\oplus_{i<j}H^1_-(X_5\times\mathbb{P}^1,\CO_{X_5\times\mathbb{P}^1}(\CD-\CD_i-\CD_j))$ map to zero in $H^1_-(X_5\times\mathbb{P}^1,\CO_{X_5,\times\mathbb{P}^1}(\CD))$ so to study the cokernel of $j$ it is enough to construct a matrix representative of $\oplus_i H^1_-(X_5\times\mathbb{P}^1,\CO_{X_5\times\mathbb{P}^1}(\CD-\CD_i))$.  The defining equations of the $\CD_i$ are
\begin{equation}\begin{split}Q_2^{(1)} &= \tilde{u}_1 f_1^{(1)}(\tilde{u}_1,\delta_1,\tilde{u}_2\delta_2,u_3) - \tilde{v}_1 g_1^{(1)}(\tilde{v}_1\delta_1,\tilde{v}_2\delta_2,v_3) \\
Q_2^{(2)} &= \tilde{u}_2 f_1^{(1)}(\tilde{u}_1\delta_1,\tilde{u}_2\delta_2,u_3) - \tilde{v}_2 g_1^{(2)}(\tilde{v}_1\delta_1,\tilde{v}_2\delta_2,v_3) \\
Q_2^{(3)} &= h_2(\tilde{u}_1\delta_1,\tilde{u}_2\delta_2,u_3,\tilde{v}_1\delta_1,\tilde{v}_2\delta_2,v_3) \\
Q_{\Sigma}^{(4)} &= a_3V^2 + \alpha U^2 \\
Q_{\Sigma}^{(5)} &= a_4V^2 + (a_2+a_3e_0)U^2
\end{split}\end{equation}
where $a_2,a_4$ are holomorphic sections of $\CO_{X_5}(2H-E_1)$ that are $\zenriques$-even while $a_3$ and $\alpha$ are holomorphic sections of $\CO_{X_5}(2H-E_1)$ that are $\zenriques$-odd.  The map $\oplus_i H^1_-(X_5\times\mathbb{P}^1,\CO_{X_5\times\mathbb{P}^1}(\CD-\CD_i))\rightarrow H^1_-(X_5\times\mathbb{P}^1,\CO_{X_5\times\mathbb{P}^1}(\CD))$ is the obvious one that consists of multiplication by the relevant $Q^{(i)}$'s.  We should be a little careful how the $Q_{\Sigma}^{(4/5)}$ act, though.  Given an element of $H^1_-(X_5\times\mathbb{P}^1,\CO_{X_5\times\mathbb{P}^1}(\CD-\CD_4))$ of the form
\begin{equation}\frac{1}{UV}\left(\frac{m}{U^2}+\frac{n}{V^2}\right)P(\tilde{u}_1,\tilde{v}_1,\tilde{u}_2\delta_2,\tilde{v}_2\delta_2,u_3,v_3)\end{equation}
its image in $H^1_-(X_5\times\mathbb{P}^1,\CO_{X_5\times\mathbb{P}^1}(\CD))$ is
\begin{equation}\frac{1}{UV}\left(na_3 + m\alpha\right)P(\tilde{u}_1,\tilde{v}_1,\tilde{u}_2\delta_2,\tilde{v}_2\delta_2,u_3,v_3)\end{equation}
as the terms with positive powers of $U$ and $V$ are trivial in cohomology.  With this observation, it is easy to see that we can construct a matrix representative of the map $\hat{j}$
\begin{equation}\hat{j}:\oplus_i H^1_-(X_5\times\mathbb{P}^1,\CO_{X_5\times\mathbb{P}^1}(\CD-\CD_i))\rightarrow H^1_-(X_5\times\mathbb{P}^1,\CO_{X_5\times\mathbb{P}^1}(\CD))\end{equation}
by `translating' the problem to $\mathbb{P}^5$ as follows.  Consider the vector spaces $V_1$ and $V_2$ defined as
\begin{equation}\begin{split}V_1 &= \{\text{quadratic polynomials in }u_i,v_j\text{ of degree at least }1\text{ in }u_1,v_1\} \\
V_2 &= \{\text{quadratic polynomials in }u_1,v_1\}
\end{split}\end{equation}
The dimensions of $V_1$ and $V_2$ are trivial
\begin{equation}\begin{array}{c|c|cc}
\text{Space} & \text{dim} & \text{dim}^+& \text{dim}^- \\ \hline
V_1 & 11 & 6 & 5 \\
V_2 & 3 & 2 & 1
\end{array}\end{equation}
and we list the dimensions of the subspaces that are even and odd under $\zenriques$ for later use.
Our domain $V$ consists of five copies of $V_1$ and two copies of $V_2$
\begin{equation}V = V_1^{\oplus 5}\oplus V_2^{\oplus 2}\end{equation}
and has dimension 61.  We will write an element of $V$ schematically as
\begin{equation}\begin{pmatrix}p^{(1)}_{2|1}(u_1,v_1;u_2,v_2,u_3,v_3)\\
p^{(2)}_{2|1}(u_1,v_1;u_2,v_2,u_3,v_3) \\
p^{(3)}_{2|1}(u_1,v_1;u_2,v_2,u_3,v_3) \\
p^{(4)}_{2|1}(u_1,v_1;u_2,v_2,u_3,v_3) \\
p^{(5)}_{2|1}(u_1,v_1;u_2,v_2,u_3,v_3) \\
q^{(1)}_2(u_1,v_1) \\
q^{(2)}_2(u_1,v_1)
\end{pmatrix}\end{equation}
where the notation $p^{(a)}_{2|1}(x_a;y_b)$ means a polynomial of degree 2 that has degree at least 1 in the $x_a$.  The map $\hat{j}$ is contraction with a vector of polynomials that takes the form
\begin{equation}\begin{pmatrix}
r^{(1),+}_{2|1}(u_1,v_1;u_2,v_2,u_3,v_3) \\
r^{(2),+}_{2|1}(u_1,v_1;u_2,v_2,u_3,v_3) \\
r^{(3),+}_{2|1}(u_1,v_1;u_2,v_2,u_3,v_3) \\
r^{(4),-}_{2|1}(u_1,v_1;u_2,v_2,u_3,v_3) \\
r^{(5),-}_{2|1}(u_1,v_1;u_2,v_2,u_3,v_3) \\
s^{(1),+}_{2|1}(u_2,v_2;u_1,v_1,u_3,v_3) \\
s^{(2),+}_2(u_1,v_1,u_2,v_2,u_3,v_3)
\end{pmatrix}\label{rsdefined}\end{equation}
where the $\pm$ index denotes whether the given polynomial is even or odd with respect to $\zenriques$.  The polynomials $r^{(1)}$, $s^{(1)}$, and $s^{(2)}$ are determined by the quadrics $Q_2^{(1)}$, $Q_2^{(2)}$, and $Q_2^{(3)}$ respectively.  The polynomials $r^{(2)}$, $r^{(3)}$, $r^{(4)}$, and $r^{(5)}$ are determined by the sections of $\CO_{X_5\times\mathbb{P}^1}(2H-E_1)$ that enter into the defining equations of $\fivefourcurve$.  The even ones are determined by $a_4$ and $a_2$ while the odd ones are determined by $\alpha$ and $a_3${\footnote{Note that in constructing the map in this way we have effectively replaced the action of $a_2+a_3e_0$ on $V_1$ by $a_2$.  Because we take $e_0=1$, $a_3$ appears twice so this represents a basis change that makes the preservation of $\mathbb{Z}_2^{(Enriques)}$-parity manifest without worrying about the fact that $e_0$ is taken to be odd when we perform the quotient.}}.

What we have obtained is the map $\hat{j}$
\begin{equation}\hat{j}:V\rightarrow W\qquad \text{dim }V=61\qquad \text{dim }W=51\end{equation}
It will be important in what follows to keep track of the $\zenriques$-even and $\zenriques$-odd subspaces of $W$, denoted $W^+$ and $W^-$, respectively
\begin{equation}\text{dim }W^+=27\qquad \text{dim }W^-=24\label{appdimWplusminus}\end{equation}
The space $V_1^{\oplus 5}$ decomposes into a direct sum of two subspaces that map individually to $W^+$ and $W^-$ under $\hat{j}$
\begin{equation}V_1^{\oplus 5} = \hat{V}_1^{(+)}\oplus \hat{V}_1^{(-)}\end{equation}
\begin{equation}\hat{j}:\hat{V}_1^{(+)}\rightarrow W^+\qquad \hat{j}:\hat{V}_1^{(-)}\rightarrow W^-\end{equation}
It is not true that all elements in $\hat{V}^{(+)}$ are $\zenriques$-even since two of the $r^{(a)}$'s in \eqref{rsdefined} are $\zenriques$-odd.  It is easy to see that
\begin{equation}\text{dim }\hat{V}_1^{(+)} = 28\qquad\text{dim }\hat{V}_1^{(-)}=27\end{equation}
Now, we know that the kernel of $\hat{j}$ has dimension at least 10.  The obvious 10-dimensional part of the kernel sits in $V_1^{\oplus 5}$ and is spanned by the columns of a simple matrix that we write schematically as
\begin{equation}\begin{pmatrix}r^{(2)} & r^{(3)} & r^{(4)} & r^{(5)} & 0 & 0 & 0 & 0 & 0 & 0  \\
-r^{(1)} & 0 & 0 & 0 & r^{(3)} & r^{(4)} & r^{(5)} & 0 & 0 & 0 \\
0 & -r^{(1)} & 0 & 0 & -r^{(2)} & 0 & 0 & r^{(4)} & r^{(5)} & 0 \\
0 & 0 & -r^{(1)} & 0 & 0 & -r^{(2)} & 0 & -r^{(3)} & 0 & r^{(5)} \\
0 & 0 & 0 & -r^{(1)} & 0 & 0 & -r^{(2)} & 0 & -r^{(3)} & -r^{(4)} \\
\end{pmatrix}\end{equation}
From this it is easy to see that
\begin{equation}\text{dim Im}(\hat{V}_1^{(+)}\xrightarrow{\hat{j}}W^+) = 22\qquad \text{dim Im}(\hat{V}_1^{(-)}\xrightarrow{\hat{j}}W^-)=23
\end{equation}
Now we turn to the action of $\hat{j}$ on $V_2^{\oplus 2}$ which we can also decompose into subspaces that map to $W^+$ and $W^-$
\begin{equation}V_2^{\oplus 2} = \hat{V}_2^{(+)}\oplus\hat{V}_2^{(-)}\end{equation}
\begin{equation}\hat{j}:\hat{V}_2^{(+)}\rightarrow W^+\qquad \hat{j}:\hat{V}_2^{(-)}\rightarrow W^-\end{equation}
The kernel of $\hat{j}:V_2^{\oplus 2}\rightarrow W$ is trivial in general so it is easy to see that
\begin{equation}\text{dim Im}(\hat{V}_2^{(+)}\xrightarrow{\hat{j}}W^+) = 4\qquad \text{dim Im}(\hat{V}_2^{(-)}\xrightarrow{\hat{j}}W^-)=2\end{equation}
What we have shown then is that if we write
\begin{equation}V = V^{(+)}\oplus V^{(-)}\end{equation}
where
\begin{equation}\hat{j}:V^{(+)}\rightarrow W^+\qquad \hat{j}:V^{(-)}\rightarrow W^-\end{equation}
then
\begin{equation}\text{dim Im}(V^{(+)}\xrightarrow{\hat{j}}W^+) = 26\qquad \text{dim Im}(V^{(-)}\xrightarrow{\hat{j}}W^-) = 25\end{equation}
From \eqref{appdimWplusminus} we conclude that the map $\hat{j}$ must have a cokernel of dimension at least 1.  Further, that cokernel sits in the subspace of $W$ that is $\zenriques$-odd.  It is easy to construct explicit representatives of $\hat{j}$ for suitable choices of the defining equations of $S_2$ and $\fivefourcurve$ for which the cokernel of $\hat{j}$ is in fact exactly 1.

Now, recall that
\begin{equation}h^m_-(\fivefourcurve,\CL_{\fivefourcurve}) = 3+\text{cok }j = 3+\text{cok }\hat{j}\end{equation}
where the elements corresponding to the `3' come from $h^2_-(X_5\times\mathbb{P}^1,\CI_4)$ and are all $\zenriques$-odd.  We conclude that $h^m_-(\fivefourcurve,\CL_{\fivefourcurve})$ has 4 elements for $m=0,1$ that are all $\zenriques$-odd and hence that $\fivefourcurve$ supports 4 vector-like pairs of doublets.  This is quite remarkable as we didn't impose any symmetry to guarantee massless vector-like pairs.

\subsection{The `Non-Minimal' Flux of Section \ref{subsubsec:nonminflux}}
\label{app:nonmin}
We turn now to the `non-minimal' flux choice of section \ref{subsubsec:nonminflux}, treating each matter curve in turn.

\subsubsection{$\tencurve$}
\label{app:nonminimalten}

The curve $\tencurve$ is defined by the equations
\begin{equation}U = c_4h_0 - d_3h_2=0\end{equation}
inside $S_2\times\mathbb{P}^1$ and the bundle of interest is
\begin{equation}\CL_{\tencurve} = \CO_{X_5\times\mathbb{P}^1}(H+E_2)|_{\tencurve}\otimes \CO_{\tencurve}(-2Q_{\mathbf{10}}')\end{equation}
where
\begin{equation}Q_{\mathbf{10}}' = \text{set of 2 points where }h_0 = h_2=0\end{equation}
As in section \ref{appsubsubsec:minimalten}, we can effectively perform all of our computations in $X_5$ here.  The basic tool for computing $h^m(\tencurve,\CL_{\tencurve})$ is the exact sequence
\begin{equation}0\rightarrow \CO_{X_5}(\CD)|_{\tencurve}\otimes \CO_{\tencurve}(-2Q_{\mathbf{10}}')\rightarrow \CO_{X_5}(\CD)|_{\tencurve}\xrightarrow{f'_{\mathbf{10}}}\mathbb{C}^4\rightarrow 0\label{nonmin10sequence}\end{equation}
where in this subsection
\begin{equation}\CD = H+E_2\end{equation}
We use the Koszul extension of \texttt{cohomcalg} to study cohomologies of the inherited bundle $\CO_{X_5}(H+E_2)|_{\tencurve}$ and investigate the behavior of the map $f_{\mathbf{10}}'$.  We again write
\begin{equation}\tencurve = \prod_{i=1}^4\CD_i\end{equation}
with
\begin{equation}\begin{split}\CD_1 &= 2H-E_1 \\
\CD_2 &= 2H-E_1 \\
\CD_3 &= 2H-E_2 \\
\CD_4 &= 2H
\end{split}\end{equation}
and use the long exact cohomology sequences derived from \eqref{appIdefs}.  Only two of the $X_5$ cohomologies that enter here are nontrivial.  They are
\begin{equation}\begin{split}h^1(X_5,\CO_{X_5}(\CD-\CD_3)) &= h^1(X_5,\CO_{X_5}(-H+2E_2) \\
&= 4 \\
h^0(X_5,\CO_{X_5}(\CD)) &= h^0(X_5,\CO_{X_5}(H+E_2) \\
&= 6
\end{split}\label{nonmin10twocohoms}\end{equation}
It follows that
\begin{equation}\begin{split}H^0(\tencurve,\CO_{X_5}(\CD)|_{\tencurve}) &= H^1(X_5,\CO_{X_5}(\CD-\CD_3))\oplus H^0(X_5,\CO_{X_5}(\CD)) \\
H^1(\tencurve,\CO_{X_5}(\CD)|_{\tencurve}) &= 0
\end{split}\label{nonmin10Dcohoms}\end{equation}
it is easy to describe the elements of $H^0(\tencurve,\CO_{X_5}(\CD)|_{\tencurve})$ schematically as
\begin{equation}H^0(\tencurve,\CO_{X_5}(\CD)|_{\tencurve})\leftrightarrow\left\{\begin{array}{l}
\frac{1}{\tilde{u}_2\tilde{v}_2}P_2(\tilde{u}_1\delta_1,\tilde{v}_1\delta_1,u_3,v_3) \\
\\
\delta_2 P_1(\tilde{u}_1\delta_1,\tilde{v}_1\delta_1,\tilde{u}_2\delta_2,\tilde{v}_2\delta_2,u_3,v_3)
\end{array}\right.\end{equation}
This explicitly displays all 10 of the elements of $H^0(\tencurve,\CO_{X_5}(\CD)|_{\tencurve})$.  We now want to study their restriction to the locus $h_0=h_2=0$.  For this, recall that $h_0\sim \delta_2$ and $h_2$ is an Enriques-odd section of $\CO_{X_5}(H-E_1)$ and hence proportional to $\tilde{v}_1$.  It is easy to see that the rank of $f_{\mathbf{10}}'$ is indeed maximal (i.e. 2+2=4) as three sections will have nonzero restriction to these points when the defining equations of $S_2$ (which supplement $h_0=h_2=0$ in defining the points $Q_{\mathbf{10}}'$ inside $X_5$) are generic and 7 have zeroes of degree at most 1.  From \eqref{nonmin10twocohoms}, \eqref{nonmin10Dcohoms}, \eqref{nonmin10sequence} this means that
\begin{equation}H^0(\tencurve,\CL_{\tencurve}) = 10-4=6\qquad H^1(\tencurve,\CL_{\tencurve}) = 0
\end{equation}
which is the result quoted in section \ref{subsubsec:nonminflux}.

\subsubsection{$\fiveonecurve$}
\label{app:nonminmatt}

Even though $\fiveonecurve$ is equivalent to $\tencurve$ as a curve in $X_5\times\mathbb{P}^1$, we must address this case in its own right because the bundle of interest $\CL_{\fiveonecurve}$ is different from $\CL_{\tencurve}$.  As with $\tencurve$, we can think of $\fiveonecurve$ as a curve in $X_5$ and work directly there.  The basic tool for computing $h^m(\fiveonecurve,\CL_{\fiveonecurve})$ is the exact sequence
\begin{equation}0\rightarrow \CO_{X_5}(\CD)|_{\fiveonecurve}\otimes \CO_{\fiveonecurve}(-2Q_{\mathbf{\overline{5}}}')\rightarrow \CO_{X_5}(\CD)|_{\fiveonecurve}\xrightarrow{f'_{\mathbf{\overline{5}}}}\mathbb{C}^4\rightarrow 0\label{nonminmattsequence}\end{equation}
where in this subsection
\begin{equation}\CD = 3H-3E_1+E_2\end{equation}
We use the Koszul extension of \texttt{cohomcalg} \cite{Blumenhagen:2010pv,Jow,Blumenhagen:2010ed,cohomCalg:Implementation} to study cohomologies of the inherited bundle $\CO_{X_5}(3H-3E_1+E_2)|_{\fiveonecurve}$ and investigate the behavior of the map $f_{\mathbf{\overline{5}}}'$.  We write
\begin{equation}\fiveonecurve = \prod_{i=1}^4\CD_i\end{equation}
with
\begin{equation}\begin{split}\CD_1 &= 2H-E_1 \\
\CD_2 &= 2H-E_1 \\
\CD_3 &= 2H-E_2 \\
\CD_4 &= 2H
\end{split}\end{equation}
and use the long exact cohomology sequences derived from \eqref{appIdefs}.  Only three of the $X_5$ cohomologies that enter here are nontrivial.  They are
\begin{equation}\begin{split}
h^4(X_5,\CO_{X_5}(\CD-\CD_1-\CD_2-\CD_3-\CD_4)) &= h^4(X_5,\CO_{X_5}(-5H-E_1+2E_2) \\
&= 2 \\
h^1(X_5,\CO_{X_5}(\CD-\CD_3)) &= h^1(X_5,\CO_{X_5}(H-3E_1+2E_2) \\
&= 4 \\
h^0(X_5,\CO_{X_5}(\CD)) &= h^0(X_5,\CO_{X_5}(3H-3E_1+E_2) \\
&= 4
\end{split}\label{nonminmattcohoms}\end{equation}
It follows that
\begin{equation}\begin{split}H^0(\tencurve,\CO_{X_5}(\CD)|_{\tencurve}) &= H^4(X_5,\CO_{X_5}(\CD-\CD_1-\CD_2-\CD_3-\CD_4)\oplus H^1(X_5,\CO_{X_5}(\CD-\CD_3))\\
&\qquad \oplus H^0(X_5,\CO_{X_5}(\CD)) \\
H^1(\tencurve,\CO_{X_5}(\CD)|_{\tencurve}) &= 0
\end{split}\label{nonminmattDcohoms}\end{equation}
It is now easy to describe the elements of $H^0(\fiveonecurve,\CO_{X_5}(\CD)|_{\fiveonecurve})$ schematically as
\begin{equation}H^0(\fiveonecurve,\CO_{X_5}(\CD)|_{\fiveonecurve})\leftrightarrow\left\{\begin{array}{l}
\frac{1}{u_3v_3\delta_1\delta_2\tilde{u}_2\tilde{v}_2}P_1\left(\frac{1}{\tilde{u}_2},\frac{1}{\tilde{v}_2}\right) \\
\\
\frac{1}{\tilde{u}_2\tilde{v}_2}P_3(\tilde{u}_1,\tilde{v}_1) \\
\\
\delta_2P_3(\tilde{u}_1,\tilde{v}_1)
\end{array}\right.\end{equation}
From this depiction of the 10 elements of $H^0(\fiveonecurve,\CO_{X_5}(\CD)|_{\fiveonecurve})$, we can study the behavior of $f_{\mathbf{\overline{5}}}'$.  The points $Q_{\mathbf{\overline{5}}}'$ are defined by the vanishing of $h_0=h_1+h_2=0$.  Since $h_1$ and $h_2$ are $\zenriques$-even and $\zenriques$-odd sections of $\CO_{X_5}(H-E_1)$, respectively, these equations take the form $\delta_2 = \tilde{u}_1+r\tilde{v}_1=0$ for some $r$, supplemented of course by the defining equations of $S_2$ in $X_5$.  Four of the sections are identically zero when $\delta_2=0$ but this leaves more than enough that are nonzero to ensure that the map $f_{\mathbf{\overline{5}}}'$ has maximal rank $2+2=4$.  From \eqref{nonminmattcohoms}, \eqref{nonminmattDcohoms}, \eqref{nonminmattsequence} this means that
\begin{equation}H^0(\fiveonecurve,\CL_{\fiveonecurve}) = 10-4=6\qquad H^1(\fiveonecurve,\CL_{\fiveonecurve})=0\end{equation}
which is the result quoted in section \ref{subsubsec:nonminflux}.

\subsubsection{$\fivefourcurve$}
\label{app:nonminhiggs}

We finally turn to the Higgs curve $\fivefourcurve$.  As emphasized in section \ref{appsubsubsec:minimalfivefour}, we must keep track of both the $\zenriques$ and $\ztau$ gradings.  We write
\begin{equation}\fivefourcurve = \prod_{i=1}^1\CD_i\end{equation}
with
\begin{equation}\begin{split}\CD_1 &= 2H-E_1 \\
\CD_2 &= 2H-E_2 \\
\CD_3 &= 2H \\
\CD_4 &= 2\sigma+2H-E_1 \\
\CD_5 &= 2\sigma + 2H-E_1
\end{split}\end{equation}
and seek the cohomologies
\begin{equation}h^m_-(\fivefourcurve,\CO_{X_5\times\mathbb{P}^1}(\CD)|_{\fivefourcurve})\end{equation}
where the $-$ means to take the odd cohomology with respect to $\ztau$ and $\CD$ is the divisor class
\begin{equation}\CD = 2H\end{equation}
We again follow \cite{Blumenhagen:2010ed} and use the long exact cohomology sequences that follow from \eqref{appHiggscurveIdefs}.  The nonvanishing $X_5$ cohomologies that enter are
\begin{equation}\begin{split}h^6_-(X_5\times\mathbb{P}^1,\CO_{X_5\times\mathbb{P}^1}(\CD-\CD_1-\CD_2-\CD_3-\CD_4-\CD_5)) &=
h^6_-(X_5\times\mathbb{P}^1,\CO_{X_5\times\mathbb{P}^1}(-8H+3E_1+E_2-4\sigma)) \\
&= 6 \\
h^2_-(X_5\times\mathbb{P}^1,\CO_{X_5\times\mathbb{P}^1}(\CD-\CD_4-\CD_5)) &=
h^2_-(X_5\times\mathbb{P}^1,\CO_{X_5\times\mathbb{P}^1}(-2H+2E_1-4\sigma))\\
&= 2 \\
h^2_-(X_5\times\mathbb{P}^1,\CO_{X_5\times\mathbb{P}^1}(\CD-\CD_1-\CD_{4/5}))
&= h^2_-(X_5\times\mathbb{P}^1,\CO_{X_5\times\mathbb{P}^1}(-2H+2E_1-2\sigma)) \\
&= 1 \\
h^1_-(X_5\times\mathbb{P}^1,\CO_{X_5\times\mathbb{P}^1}(\CD-\CD_{4/5}))
&= h^1_-(X_5\times\mathbb{P}^1,\CO_{X_5\times\mathbb{P}^1}(E_1-2\sigma)) \\
&= 1
\end{split}\end{equation}
These results are actually enough to completely fix the dimensions of $H^m_-(\fivefourcurve,\CO_{X_5\times\mathbb{P}^1}(\CD)|_{\fivefourcurve})$ to
\begin{equation}h^m_-(\fivefourcurve,\CO_{X_5\times\mathbb{P}^1}(\CD)|_{\fivefourcurve})= 6\qquad m=0,1\end{equation}
so we get 6 vector-like pairs of fields.  It is actually easy to see how many $\zenriques$-even and $\zenriques$-odd elements we have since
\begin{equation}H^1_-(\fivefourcurve,\CO_{X_5\times\mathbb{P}^1}(\CD)|_{\fivefourcurve}) = H^6_-(X_5\times\mathbb{P}^1,\CO_{X_5\times\mathbb{P}^1}(\CD-\CD_1-\CD_2-\CD_3-\CD_4-\CD_5))\end{equation}
The `rationoms' associated to this group are
\begin{equation}\frac{1}{UV\tilde{u}_1\tilde{v}_1\tilde{u}_2\tilde{v}_2u_3v_3}P_1\left(\frac{1}{U^2},\frac{1}{V^2}\right)P_2\left(\frac{1}{\tilde{u}_1},\frac{1}{\tilde{v}_1}\right)\end{equation}
from which we deduce that 4 of the 6 elements are $\zenriques$-even and 2 are $\zenriques$-odd.  We therefore have that
\begin{equation}\begin{split}h^{0,+}_-(\fivefourcurve,\CL_{\fivefourcurve}) &= 4 \\
h^{0,-}_-(\fivefourcurve,\CL_{\fivefourcurve}) &= 2 \\
h^{1,+}_-(\fivefourcurve,\CL_{\fivefourcurve}) &= 4 \\
h^{1,-}_-(\fivefourcurve,\CL_{\fivefourcurve}) &= 2
\end{split}\end{equation}
This corresponds to 4 vector-like pairs of triplets and 2 vector-like pairs of doublets.

\newpage

\bibliographystyle{JHEP}
\renewcommand{\refname}{Bibliography}
\addcontentsline{toc}{section}{Bibliography}

\providecommand{\href}[2]{#2}\begingroup\raggedright\endgroup

\end{document}